\documentclass[aps,pra,twocolumn,nofootinbib]{revtex4-1}

\usepackage{graphicx}
\usepackage{rotating}
\usepackage{amsmath}
\usepackage{amssymb}
\usepackage{amsfonts}
\usepackage{amsthm}
\usepackage{hyperref}
\usepackage{color}
\usepackage[caption=false]{subfig}
\usepackage{bbold}
\usepackage{comment}
\usepackage{xspace}
\usepackage{bm}
\newcommand{\Tr}{\operatorname{Tr}}
\newcommand{\ket}[1]{\left | #1 \right \rangle}

\newcommand{\bs}{boson sampling\xspace}

\begin{document}

\title{
No imminent quantum supremacy by boson sampling
}
\author{Alex Neville$^{1}$}
\author{Chris Sparrow$^{1,2}$}
\author{Rapha\"el Clifford$^{3}$}
\author{Eric Johnston$^{1}$}
\author{Patrick M. Birchall$^{1}$}
\author{Ashley Montanaro$^{4}$}

\author{Anthony Laing$^{1}$}
\email{anthony.laing@bristol.ac.uk}
\affiliation{
$^{1}$Quantum Engineering and Technology Laboratories, School of Physics and Department of Electrical and Electronic Engineering, University of Bristol, UK\\
$^{2}$Department of Physics, Imperial College London, UK\\
$^{3}$Department of Computer Science, University of Bristol, UK\\
$^{4}$School of Mathematics, University of Bristol, UK\\
}
\date{\today}
\begin{abstract}
It is predicted that quantum computers will dramatically outperform their conventional counterparts.
However, large-scale universal quantum computers are yet to be built.
Boson sampling is a rudimentary quantum algorithm tailored to the platform of photons in linear optics, 
which has sparked interest as a rapid way to demonstrate this quantum supremacy.
%
Photon statistics are governed by intractable matrix functions known as permanents,
which suggests that sampling from the distribution obtained by injecting photons into a linear-optical network
could be solved more quickly by a photonic experiment than by a classical computer.
The contrast between the apparently awesome challenge faced by any classical sampling algorithm
and the apparently near-term experimental resources required for a large boson sampling experiment
has raised expectations that quantum supremacy by boson sampling is on the horizon.
Here we present classical boson sampling algorithms
and theoretical analyses of prospects for scaling boson sampling experiments,
showing that near-term quantum supremacy via boson sampling is unlikely.
While the largest boson sampling experiments reported so far are with 5 photons,
our classical algorithm, based on Metropolised independence sampling (MIS), allowed the boson sampling problem
to be solved for 30 photons with standard computing hardware.
We argue that the impact of experimental photon losses means that demonstrating quantum supremacy by boson sampling would require a step change in technology.
\end{abstract}
\maketitle
\begin{figure*}[t]
\centering
\center{\includegraphics[clip,width=0.8\textwidth]{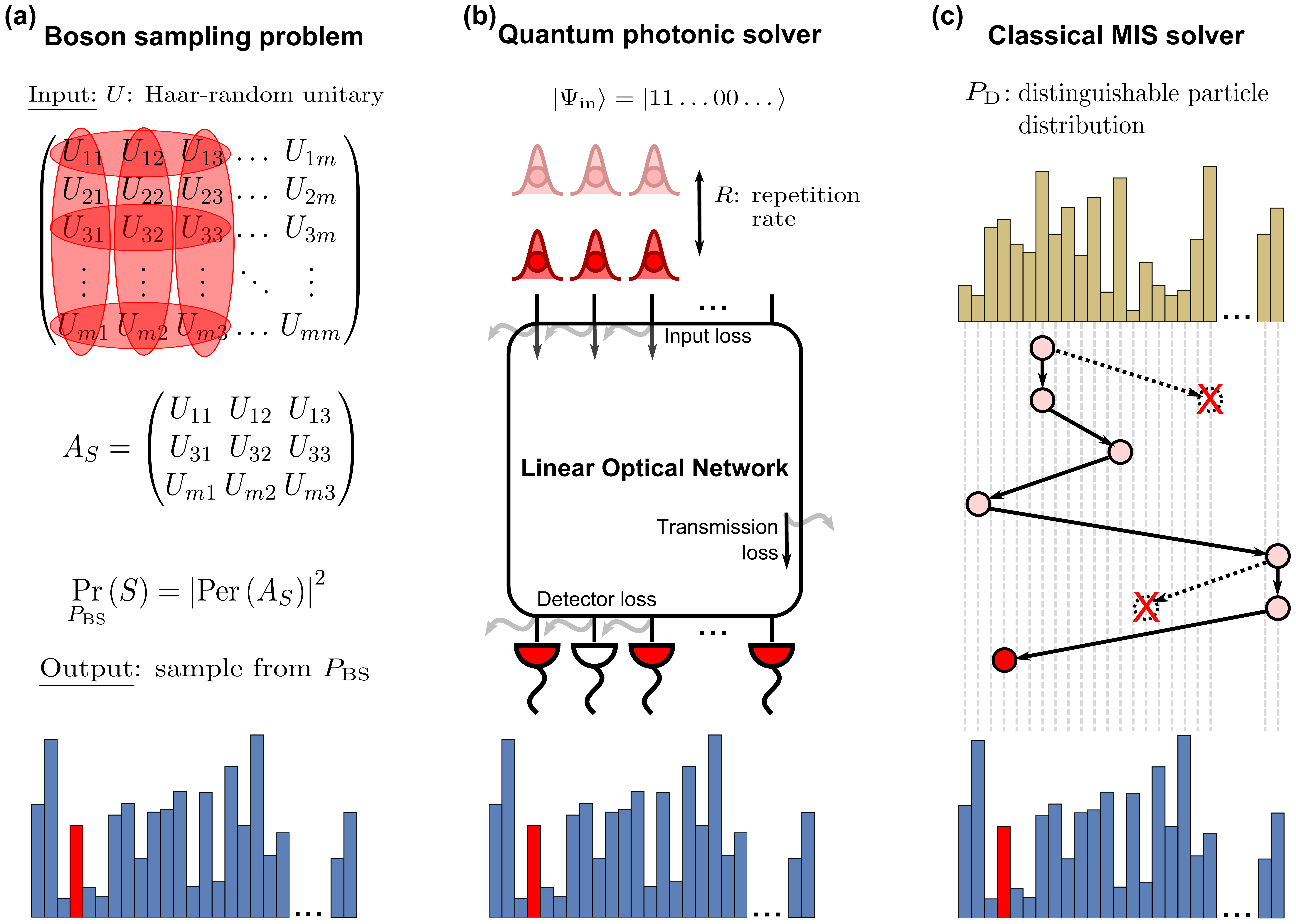}}
\caption{
Photonic and classical approaches to the boson sampling problem.
(a) Definition of the problem. Output a sample from the distribution defined by the modulus squared permanents of submatrices of a Haar-random unitary matrix $U$.
(b) Photonic experiments solve the problem by propagating single photons through a linear optical network followed by single photon detection and can be broadly parameterised by $R$, the $n$-photon generation rate, and $\eta$, the transmission probability for a single photon taking into account input, coupling, transmission and detection losses.
(c) A classical boson sampling algorithm based on Metropolised independence sampling using the distinguishable particles transition probabilities as the proposal distribution.
The algorithm computes 100 complex and real permanents to produce a single output pattern, and enabled classical boson sampling for $30$ bosons on a laptop. 
}
\label{fgMain}
\end{figure*}
%
It is believed that new types of computing machines will be constructed to exploit quantum mechanics
for an exponential speed advantage in solving certain problems compared with classical computers~\cite{Aaronson2011,Broome2013,Tillmann2013,Spring2013,Crespi2013,lund17, Latmiral2016,Wang2016,Shor1997,Lloyd1996}.
Recent large state and private investments in developing quantum technologies have increased interest in this challenge.
However, it is not yet experimentally proven that a large computationally useful quantum system can be assembled,
and such a task is highly non-trivial given the challenge of overcoming the effects of errors in these systems.

Boson sampling is a simple task which is native to linear optics and has captured the imagination of quantum scientists
because it seems possible that the anticipated supremacy of quantum machines
could be demonstrated by a near-term experiment.
%
%
The advent of integrated quantum photonics \cite{Politi2008, OBrien2009} has enabled large, complex, stable and programmable optical circuitry \cite{Shadbolt2012, carolan2015, Harris2015},
while recent advances in photon generation \cite{Xing2016, Wang2016b, Somaschi2016, Spring2017} and detection~\cite{Lita2008,Hadfield2009} have also been impressive.
The possibility to generate many photons, evolve them under a large linear optical unitary transformation, then detect them,
seems feasible, so the role of a boson sampling machine as a rudimentary but legitimate computing device is particularly appealing.
Compared to a universal digital quantum computer, the resources required for experimental boson sampling appear much less demanding.
This approach of designing quantum algorithms to demonstrate computational supremacy with near-term experimental capabilities has inspired a raft of proposals suited to different hardware platforms~\cite{bremner16, bremner16b, boixo16, BermejoVega2017}.

Based on a simple architecture, the boson sampling problem is similarly straightforward to state. 
A number $n$ of indistinguishable noninteracting bosons (e.g.\ photons)
should be injected into $n$ input ports
of a circuit comprised of a number $m$ of linearly coupled bosonic modes.
The circuit should be configured so that the transformation between input and output ports
is described by a uniformly (``Haar'') random unitary matrix.
The probability for the $n$ bosons to be detected at given set of $n$ output ports
is equal to the square of the absolute value of the permanent of the transfer matrix
that describes the transformation.

While choosing a number of modes $m \sim n^5\log^2n$ guarantees that the distribution of any $n\times n$ sub-matrix is
approximately equal to that of a matrix of elements drawn independently from the complex normal distribution~\cite{Aaronson2011},
the less impractical scaling of $m \sim n^2$ is typically targeted.
This polynomial relation between $n$ and $m$ is also important because it ensures
a not too large probability that two or more of the bosons arrive at the same output port, i.e.\ bunch; 
the conjectured hardness only applies to collision-free events, i.e.\ no bunching.
Because approximating the permanent of a random matrix is conjectured to be computationally hard~\cite{Aaronson2011},
calculating any transition probability is intractable;
the collection of all of the possible collision-free transition probabilities ($m$ choose $n$)
constitutes an exponentially large probability distribution, where each element is exponentially hard to calculate.
Running an ideal boson sampler would solve the problem of producing samples from this distribution.

\begin{figure*}[t]
  \centering
\center{\includegraphics[clip,width=1\textwidth]{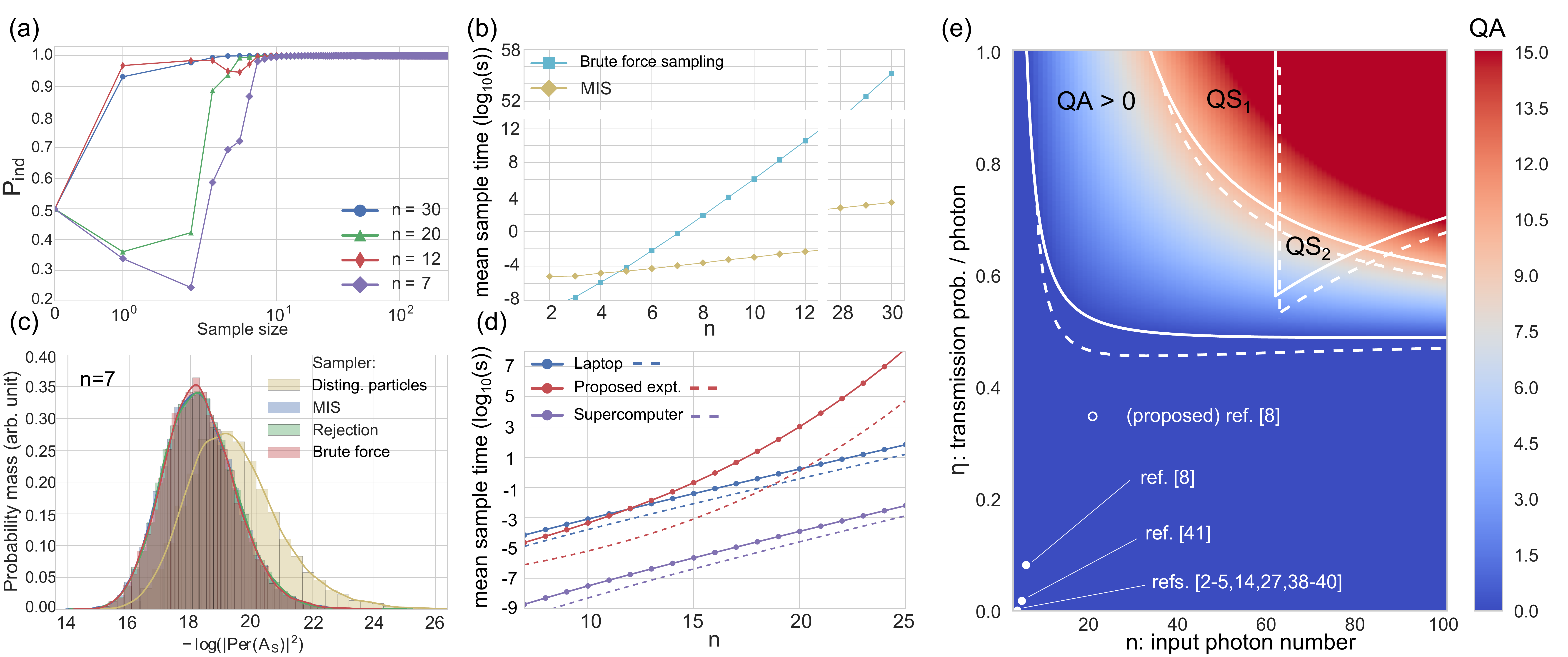}}
\caption{
Results and projections for classical boson sampling.
(a) A likelihood ratio test with the boson sampling and distinguishable particle distributions as the hypotheses for sample sizes of up to 250 for $n=7,12,20$ and $30$ bosons. $P_\mathrm{ind}$ is the probability that the data are drawn from the boson sampling and not the distinguishable particle distribution.
(b) Mean time to get a sample value using a laptop via the MIS and brute force approach to classical boson sampling, averaged over samples of size 100.
(c) Verification of sampler by comparing the distribution of $-\log(|\text{Per} A_{S}|^{2})$ for a sample size of 20000 to other boson samplers (rejection sampler and brute force sampler) and a distinguishable particle sampler.
(d) Mean time to get a sample using a laptop, supercomputer and the proposed experiment in Ref. \cite{Wang2016}.
Dashed lines represent the time to get sample in a variant of boson sampling where $2$ photons are lost. 
(e) Quantum advantage, QA, as a function of $n$ and $\eta$ assuming the classical time scaling of a supercomputer and an experimental rate $R=10$GHz. Lines separate the regions of no quantum advantage, positive quantum advantage and quantum supremacy (as measured by criterion QS$_1$ or QS$_2$). Dashed lines demonstrate adjusted regions when up to $2$ photons can be lost (optimised to maximise QA).
Solid circles represent existing experimental demonstrations and the empty circle represents a proposed future experiment.
}
\label{fgMain2}
\end{figure*}

Importantly, a strong case for the classical hardness of boson sampling can be made
even when the distribution being sampled from is only approximately correct~\cite{Aaronson2011}:
assuming certain conjectures from computational complexity theory, there can be no efficient classical algorithm to sample from any distribution within a small distance from the boson sampling distribution.
%
%
%
%

Current estimates for the regime in which photonic experiments could achieve quantum supremacy have been between 20 and 30 photons \cite{Aaronson2011, preskill12, Bentivegna15} and, recently, as low as 7 \cite{Latmiral2016}.
However, our classical algorithm, based on Metropolised independence sampling (MIS),
while necessarily inefficient for large $n$, was able to output a sample for $n=30$ bosons in half an hour on a standard laptop,
and would enable an $n=50$ sample to be produced in under 10 days on a supercomputer~\cite{wu16}.
%
MIS~\cite{liu96,liu04} is a specific Markov Chain Monte Carlo (MCMC) method.
For an instance of the problem,
our aim is to construct a Markov chain where each state in the chain signifies a boson sampling detection event.
New states in the chain are proposed from a classical mockup of the boson sampling distribution:
the distribution of distinguishable particles,
where probabilities of detection events are equal to permanents of real and positive transition matrices,
and sampling becomes efficient \cite{Aaronson2013}.
Proposed states are then accepted or rejected by comparison with the actual probability to observe that event for indistinguishable bosons.
This means that, at each step, only a relatively small number of permanents must be calculated;
a calculation of the full and exponentially large boson sampling distribution is not required.

More precisely, let $P_\mathrm{D}$ be the distinguishable particle distribution with probability mass function $g(x)$, over the set of tuples of length $m$ with elements in $\{0,1\}$ which sum to $n$.
And let $P_\mathrm{BS}$ be the boson sampling distribution over these tuples, with probability mass function $f(x)$.
Then starting at a random (according to $P_\mathrm{D}$) tuple $x$, propose a new random tuple $x^\prime$. 
The transition from $x$ to $x^\prime$ is accepted with probability
\begin{equation}\label{short:metropolis}
T\left(x^\prime | x\right) =\min\left(1,\frac{f(x^\prime)}{f(x)} \frac{g(x)}{g(x^\prime)}\right).
\end{equation}
Repeating this procedure generates a Markov chain, which will converge such that thereafter, the chain is sampling from $P_\mathrm{BS}$.

Not all states in the Markov chain are retained as detection events.
The time taken for the Markov chain to converge means that a number of tuples at the beginning of the chain must be discarded, known as the `burn in' period, $\tau_{\mathrm{burn}}$.
For the size of simulations covered here, empirical tests find that a burn in period of 100 is sufficient for convergence to have occurred.
In addition, autocorrelation between states in the chain can occur, for example because two consecutive states in the chain will be identical whenever a proposed new state $x^\prime$ is not accepted. 
We empirically find for the size of problem we tackle that autocorrelation is suppressed with a thinning procedure that retains only every $100$th state (see appendix Sec.\ \ref{sec:assessment}).
Generally, the burn in period and thinning interval are reduced by a greater overlap between target and proposal distributions,
as measured by (\ref{short:metropolis}), the transition probability.
We find that a proposal distribution of the distinguishable particle transition probabilities has a high acceptance rate of $\sim40\%$, a sign that the distributions overlap well.
In Fig.\ \ref{fgMain} we present a schematic of the MIS based approach to boson sampling, alongside schematics of a quantum photonic approach and the boson sampling problem itself.

The classical resources used to produce the thinned Markov chain are
far fewer than those required by the brute force approach of calculating all permanents in the full distribution.
The asymptotically fastest known algorithm for exactly computing the permanent of an $n \times n$ matrix is Ryser's algorithm~\cite{ryser63}, whose runtime when implemented efficiently is $\mathcal{O}(n 2^n)$.
%
Generating the first tuple in a sample requires
the computation of $\tau_{\mathrm{burn}}$ $n \times n$ real valued matrix permanents
and $\tau_{\mathrm{burn}}$ $n \times n$ complex valued matrix permanents.
Each subsequent sample requires $\tau_{\mathrm{thin}}$ $n \times n$ real valued and complex valued matrix permanents to be computed.
The relative scaling of the approaches to classical boson sampling using a standard laptop is shown in Fig.\ \ref{fgMain2}(b),
setting both $\tau_{\mathrm{burn}}$ and $\tau_{\mathrm{thin}}$ equal to $100$.
The MIS sampler is over 50 orders of magnitude faster for the $n=30$ case than the brute force computation of the entire distribution.

We used this algorithm on a standard laptop to produce samples of size 20,000 for up to 20 bosons, and used
a local server, which allowed around 30 times more chains to be run in parallel, to produce 250 samples for 30 bosons in 900 modes in less than five hours.
As in the experimental case,
a central challenge is to provide evidence
for sampling from the correct distribution.
Here we addressed this using standard statistical techniques.
The likelihood ratio tests \cite{Bentivegna2014} in Fig.\ \ref{fgMain2}(a) show a rapid growth in confidence in the hypothesis that these samples are from the indistinguishable boson distribution rather than the distinguishable particle distribution for $n=7,12,20$ and $30$.

Further verification results are shown in Fig.\ \ref{fgMain2}(c) for the case of 7 bosons in 49 modes.
For each tuple in a sample size of 20,000 produced by our classical algorithm,
we calculate $-\log(|\text{Per} (A_{S})|^{2})$,
where $A_{S}$ is the matrix associated to each tuple,
and produce a probability mass histogram.
The same function and associated histogram
is plotted for a sample of 20,000 tuples chosen from the distinguishable particle distribution
(note that $\text{Per} (|A_{S}|^{2})$ gives the probability to observe a transition of distinguishable particles).
Fig.\ \ref{fgMain2}(c) shows the clear difference between the two distributions,
which we analyse with a (bootstrapped) 2-sample Kolmogorov-Smirnov (KS) test~\cite{Praestgaard1995,Sekhon2011}.
We are able to reject the null hypothesis,
that the two samples are chosen from the same distribution,
at a significance level of $0.001$.
Conversely, a similar KS test between distributions from
our MIS algorithm and a rejection sampling algorithm,
and between
our MIS algorithm and the brute force approach of calculating all permanents in the full distribution,
both found large $p$-values (see appendix Sec.\ \ref{sec:assessment}).


We next compare our classical approach with plausible experimental parameters.
It is worth noting that asymptotically, experimental boson sampling will have a slower runtime than our algorithm.
This is because photon losses scale exponentially badly with $n$ \cite{Aaronson2011, Rohde2014}.
The runtime for an experiment with a transmission probability resulting from fixed loss (generation, coupling, detection) $\eta_\mathrm{f}$ and a transmission probability $\eta_0$, resulting from loss per unit optical circuit depth scales as $\mathcal{O}\big((\frac{1}{\eta_\mathrm{f}})^n (\frac{1}{\eta_0})^{dn}\big)$ for an optical circuit depth of $d$, which is worse than Ryser's algorithm if $d$ grows with $n$ for any $\eta_\mathrm{f},\eta_0 < 1$.
However, the region of interest for quantum supremacy is likely to be restricted to $n < 100$, where low-loss experiments still have the potential to produce large speedups.
Assuming that our MIS sampler continues to perform equally well for larger instance sizes,
we can compare its runtime with current and future experiments.
The classical and quantum runtimes for an instance of size $n$ bosons in $m=n^2$ modes can be estimated as
\begin{align}
&c_t(n) = a 100 n 2^n  \\
&q_t(n,\eta) = \frac{e}{R\eta^n}
\end{align}
where $a$ is the time scaling of the classical computer (for computing one real and one complex permanent), the factor of $e$ is an approximation to the probability of obtaining a collision-free event~\cite{Arkhipov2012},
$R$ is the experimental source repetition rate,
and $\eta=\eta_\mathrm{f} \eta_0^d$ is the experimental transmission probability of a single photon including the efficiencies of photon generation, coupling, circuit propagation and detection
(note that $R$ and $\eta$ will generally be a function of $n$).
We define the \emph{quantum advantage} (QA) as the improvement in quantum runtime versus classical runtime measured in orders of magnitude,
\begin{equation}
\text{QA}(n,\eta) = \max\Big[0, \log_{10}\Big(\frac{c_t}{q_t}\Big)\Big].
\end{equation}

We now consider two plausible notions of quantum supremacy.
First, we can define supremacy as
a speedup so large that it is unlikely to be overcome
by algorithmic or hardware improvements to the classical sampler,
for which we choose a speedup of ten orders of magnitude.
Secondly, we may wish to define supremacy as the point at which
a computational task is performed in a
practical runtime on a quantum device, for which we choose under a week,
but in an impractical runtime on a classical device, for which we choose over a century.

These criteria can be summarised as
\begin{align}
&\text{QS}_1: \text{QA} > 10 \\
&\text{QS}_2: q_t < 1 \text{week} , c_t > 100 \text{yrs}.
\end{align}
In order to make concrete estimates of future runtimes, we need to fix $a$ and $R$. Choosing $a = 3n\times10^{-15} s$ as the time scaling for computing one real and one complex matrix permanent recently reported for the supercomputer \emph{Tianhe 2}~\cite{wu16} and $R = 10$GHz, which is faster than any experimentally demonstrated photon source to our knowledge, we can plot QA against $n$ and $\eta$. 

We first note that current approaches using spontaneous parametric down conversion (SPDC) photon pairs are generally inefficient with $\eta < 0.002$ \cite{Broome2013,Tillmann2013,Spring2013,Crespi2013,spagnolo2014,Carolan2014,carolan2015,Bentivegna15}.
Recently, improved rates have been demonstrated with quantum dot photon sources \cite{Loredo2017, He2016, Wang2016}. 
The current leading experimental demonstration however is still restricted to $\eta \approx 0.08$ for $n=5$ \cite{Wang2016} where $q_t \approx 10^9 c_t$.
This calculation includes the rate used in the experiment ($76n^{-1}\text{MHz}$) and includes a suppression factor caused by a lower collision-free event rate using a linear instead of quadratic mode scaling.
In Wang \emph{et al.} \cite{Wang2016} a number of realistic, near-term experimental improvements are suggested to reach 20 photon boson sampling.
Using these projections we find that $\eta$ is increased to $\approx 0.35$, which would be a major experimental breakthrough.
However, as shown in Fig.\ \ref{fgMain2}(d), in this case we predict that the classical runtime would still be over six orders of magnitude faster.
Fig.\ \ref{fgMain2}(e) shows the regions of quantum advantage and quantum supremacy with current and projected experiments.

In Ref.~\cite{Aaronson16}, the authors showed that the boson sampling problem can be modified to allow for a fixed number of lost photons at the input of the circuit whilst retaining computational hardness.
In appendix Sec.\ \ref{sec:loss}, we show that if the overall losses in the experiment are path-independent then this is equivalent to loss at the input.
The MIS sampler can be readily adapted to this scenario by adding an initial step which generates a uniformly random input subset, followed by the usual MIS method for this input state.
The dashed contours and lines in Fig.\ \ref{fgMain2}(e) and (d) take into account the adjusted classical and quantum runtimes when up to two lost photons are allowed.
Although allowing loss helps the experiments to compete, the complexity of realistic experimental regimes such as losing a constant fraction of photons remains unknown and it is easy to see that losing too many photons eventually allows the problem to become efficiently solvable classically.
%

We conclude that a near term experimental demonstration of quantum supremacy by boson sampling would require a technological step change, reaching photon numbers of over 50 and ultra-low loss interferometers with thousands of modes.
A convincing demonstration would additionally need to solve general instances of the problem to a high degree of precision which will require full programmability and increased control, both of which are likely to add loss.
%
%
Although the boson sampling algorithm could be run on a fault-tolerant quantum computer, this approach would lose the appealing simplicity of the original proposal.
However, it may be the case that a more limited, partially error-corrected device could demonstrate quantum supremacy, and theoretical and experimental schemes which can overcome photon loss should be investigated.
We finally note that our algorithm is unoptimised and we expect significant improvements can be made, pushing the supremacy threshold further still from the current experimental reality.

\begin{acknowledgements}
		\noindent The authors would like to thank Nicola Maraviglia and Pete Shadbolt for helpful discussions.
		AN is grateful for support from the Wilkinson Foundation.
		This work was supported by the Engineering and Physical Sciences Research Council (EPSRC) and QUCHIP (H2020-FETPROACT-3-2014: Quantum simulation). Fellowship support from EPSRC is acknowledged by RC (EP/J019283/1), AM (EP/L021005/1), and AL (EP/N003470/1).
\end{acknowledgements}
\bibliographystyle{apsrev4-1}
%

\clearpage

\onecolumngrid
\begin{center}
	\bf APPENDIX
\end{center}


\section{Introduction}
The \bs problem has, since its introduction by Aaronson and Arkhipov in 2011~\cite{Aaronson2011}, been the subject of great interest within theoretical and experimental quantum physics.
Informally, \bs is the problem of sampling from the distribution obtained by transmitting photons through a linear-optical network.
No efficient classical algorithm for this task is known, but by its nature, \bs is suited to being carried out on a specialised linear optical quantum device which could be substantially simpler than a fully universal linear-optical quantum computer.
As such, \bs is considered a leading candidate for a problem which could see enhanced performance when solved by a quantum device compared with a classical device -- so-called \emph{quantum supremacy}~\cite{preskill12,lund17} -- in the near future.

Estimates for the size of experiment required to demonstrate quantum supremacy via \bs have evolved over time. Aaronson and Arkhipov initially predicted this to be in the region of 20-30 photons in a 400-900 mode linear optical network~\cite{Aaronson2011}; Preskill suggested ``about 30'' photons~\cite{preskill12}; and Bentivegna et al.~\cite{Bentivegna15} suggested that, by making some minor modifications to the problem, 30 photons and 100 modes would suffice. Recently Latmiral et al.~\cite{Latmiral2016} have reported that the number of photons and modes required to achieve quantum supremacy could be as small as 7 and 50 respectively.

This estimate is based on a comparison with the simplest possible classical technique for simulating \bs: evaluating the entire probability distribution.
However, there exist well-known classical techniques for exact or approximate sampling that do not require the entire distribution of interest to be determined.
Here we assess the requirements on a linear-optical implementation in order to achieve quantum supremacy in light of these.

We find that, using projected loss parameter estimates with a promising experimental \bs setup~\cite{Wang2016}, quantum supremacy (via the standard \bs problem or in a modified version with loss~\cite{Aaronson16} ) is not achieved.
%

\section{The \bs problem}

The original \bs problem, as described by Aaronson and Arkhipov~\cite{Aaronson2011}, is based on sampling from the probability mass function defined by the linear scattering of multiple bosons (in practice, photons) prepared in a Fock state and measured in the Fock basis.

Given $U\in SU(m)$, let $A$ be the column-orthonormal, $m\times n$ matrix formed by taking the first $n$ columns of $U$. 
Also let $\Phi_{m,n}$ be the set of all possible tuples of length $m$ of non-negative integers whose elements sum to $n$. 

For some tuple $S = \left(s_1,\dots,s_m\right) \in \Phi_{m,n}$, let $A_S$ be the sub-matrix of $A$ with $s_i$ copies of row $i$.
Boson sampling is the problem of sampling from the probability distribution $P_\mathrm{BS}$ over $\Phi_{m,n}$, for a given input $U$, with probabilities defined as:
\begin{equation}\label{BS_dist}
\Pr(S) = \frac{\left| \mathrm{Per}\left( A_S\right) \right| ^{2}}{s_1!\dots s_n!} 
\end{equation}
where $\mathrm{Per}(X)$ is the permanent of an $n\times n$ matrix $X=\left(x\right)_{ij}$, defined by:
\begin{equation}\label{perm}
\mathrm{Per}(X) = \sum\limits_{\sigma\in \mathcal{S}_n} \prod\limits_{i=1}^{n} x_{i,\sigma\left(i\right)}
\end{equation}
where $\mathcal{S}_n$ is the group of permutations of integers 1 to $n$.

The matrix permanent is similar to the more commonly encountered matrix determinant, but the definition lacks an alternating sign.
Although this difference is seemingly minor, computing the permanent of a complex-valued matrix falls in the \#P-hard computational complexity class~\cite{valiant79,aaronson11a}, and is in general vastly more demanding than computing the determinant.
Indeed, the fastest algorithms known for computing the permanent run in time exponential in $n$.

Using the fact that computing the permanent is a \#P-hard problem, Aaronson and Arkhipov~\cite{Aaronson2011} have shown that the existence of an efficient, exact classical algorithm for \bs would collapse the polynomial hierarchy to its third level, a complexity-theoretic consequence considered extremely unlikely.
However, the \bs distribution could (in principle) be sampled from using a simple linear-optical circuit with $n$ photons in $m$ modes, where the circuit corresponds to the desired unitary $U$.
To resolve the apparent contradiction with \#P-hardness, note that their result only shows that a classical sampler from the \bs distribution could be used to compute the permanent within the polynomial hierarchy, not that a quantum sampler could.

In addition to these results on the exact form of the problem, Aaronson and Arkhipov proposed two additional (yet plausible) conjectures which together would imply that an approximate form of \bs could not be carried out efficiently on a classical computer.
In approximate \bs, the task is to output a sample from an arbitrary probability distribution $P$ within total variation distance $\epsilon$ of the real \bs distribution $P_\mathrm{BS}$, for some small constant $\epsilon$.
This is a particularly important result for experimental implementations of \bs, as it allows for the inevitable imperfections in a real \bs device, up to a point~\cite{Rohde2012, Arkhipov15}. 

An important feature of the approximate \bs problem is that the sub-unitary matrices $A_S$ must look like Gaussian random matrices. 
In order for this to be true, it must hold that $m\gg n$ (the currently best proven bound is $m=\mathcal{O}(n^5 \log^2 n)$~\cite{Aaronson2011}, although it is generally accepted that $m=\mathcal{O}(n^2)$ should be sufficient).
A side effect of this condition is that, due to the bosonic birthday paradox~\cite{Arkhipov2012}, $\mathrm{Pr}\left(s_i>1\right)$ is small, and hence that one can  consider the problem restricted to the ``collision free subspace" (CFS) of the overall Hilbert space of the bosonic states, where there is no bunching at the output ($s_i \le 1$ for all $i$).
So $\Pr(S) = \left| \mathrm{Per}\left( A_S\right) \right| ^{2}$.
Then our task becomes to sample from the restriction of the \bs distribution to this subspace, suitably renormalised by a constant close to 1.
This is the problem on which we focus in this work.

In the setting of linear optics, an instance of the \bs problem for random $U$ can be implemented by injecting the $n$-photon Fock state
\begin{equation}\label{input_state}
\ket{\Psi_{in}}=| \underbrace{1,\dots,1}_n, \underbrace{0,\dots,0}_{m-n} \rangle
\end{equation}
into a linear optical network of beam-splitters and phase-shifters with transfer matrix chosen Haar-randomly, and a single-photon detector coupled to each output mode.
These detectors need not be able to discern the number of output photons in each mode, as one can restrict the problem to the CFS.
A sample is obtained by recording the $n$-fold coincident detection signals at the output.

Schemes for implementing arbitrary unitary linear optical transfer matrices using beam-splitters and phase-shifters are well known~\cite{Reck1994, Clements2016} and have been realised successfully~\cite{carolan2015, Harris2015}.
The current record for implementing the ``classic'' version of \bs is 5 photons in 9 modes~\cite{Wang2016}.

For large $n$, producing the input state $\ket{\Psi_{in}}$ is a major experimental challenge.
A ``scattershot'' variant of \bs has therefore been proposed which is easier to implement experimentally, and yet satisfies similar complexity-theoretic properties to classic \bs~\cite{lund14,scottblog}.
This approach is useful when using photon pair sources such as spontaneous parametric down conversion (SPDC) sources, where each source has a small probability of generating a photon pair per pulse of a pump laser.
In the scattershot setup, $m$ sources are pumped concurrently, and some subset of these sources may generate a photon pair.
One photon in each pair is sent to a detector in order to herald the presence of the other, which is sent to an input mode of the linear optical circuit (which may now be any of the $m$ modes, rather than one of the first $n$ modes).

In principle, and with perfectly efficient single photon detectors, this allows one to select the instances where $2n$ detection events occur ($n$ herald signals and $n$ post-circuit signals) and generate samples from a \bs distribution with a known input configuration.
The benefit of this approach is that, as the number of pairs generated is distributed binomially, one can tune the probability that each source generates a photon pair so as to maximise the probability of an $n$ photon-pair outcome, and this is much greater than if one were limited to $n$ sources.
Scattershot \bs has been implemented with 6 sources and 13 modes~\cite{Bentivegna15}.

From the point of view of classical simulation, the scattershot \bs problem is not significantly harder than the classic \bs problem. The only difference is that, rather than sampling from a distribution defined by a fixed $m \times n$ submatrix $A$ of an $n \times n$ unitary matrix $U$, one first chooses $A$ itself at random, according to a probability distribution which is easy to sample from classically. We therefore focus on classic \bs here, but stress that all our results also apply to scattershot \bs.

\section{Classical sampling techniques for \bs}

We now describe some techniques that can be used to implement \bs on a classical computer.
All the classical algorithms we consider will ultimately be based on computing permanents, and our goal will be to minimise the number of permanents computed per sample obtained.
The asymptotically fastest known algorithm for exactly computing the permanent of an $n \times n$ matrix is Ryser's algorithm~\cite{ryser63}, whose runtime when implemented efficiently is $\mathcal{O}(n 2^n)$.
The Balasubramanian--Bax/Franklin--Glynn algorithm achieves the same asymptotic performance but may be preferred from the perspective of numerical stability~\cite{wu16}.
For the values of $n$ we consider here, numerical stability is unlikely to be a significant issue~\cite{wu16}.

Although this exponential runtime may seem discouraging, note that this is substantially faster than the na\"ive approach of evaluating the sum (\ref{perm}) directly, which would take time $\mathcal{O}(n!)$.
On a standard personal computer, the permanent of an $n \times n$ matrix can be computed in under 1 second for $n \approx 25$ (see~\cite{wu16,Latmiral2016} and Sec.~\ref{sec:permexps} below for more detailed numerical experiments), and it is predicted in~\cite{wu16} that a supercomputer could solve the case $n=50$ in under 2 hours.
In addition, some proposed verification techniques for \bs experiments (e.g.\ likelihood-ratio tests) rely themselves on computing permanents~\cite{Aaronson2011,Aaronson2013}.

\subsection{Brute force exact sampling}

Perhaps the most obvious way to tackle the \bs problem on a classical computer is to compute all of the probabilities in the CFS, and sample from the probability mass function associated with these probabilities~\cite{Latmiral2016}.
It becomes apparent that this scheme is immensely computationally demanding when one considers how the dimension of the CFS scales with the size of the problem. 
Indeed, this approach requires computing $\binom{m}{n}$ permanents of $n\times n$ complex valued matrices before a single sample can be output.
Using the lower bound $\binom{m}{n} \ge (m/n)^n$, for $m \ge n^2$ at least $n^n$ permanents must be computed; and even for $n=10$ we need to compute more than $17 \times 10^{12}$ permanents.
Thus our computation quickly becomes swamped by the number of permanents to compute, rather than the complexity of computing the permanent itself.

We therefore seek an approach which relies on computing a number of permanents which scales more favourably with $n$ and $m$ than $\binom{m}{n}$.

\subsection{Rejection sampling with a uniform proposal}


Rejection sampling is a general approach for exactly sampling from a desired distribution $P$ with probability mass function $f(x)$, given the ability to sample from a distribution $Q$ with probability mass function $g(x)$, where $f(x) \le \lambda g(x)$ for some $\lambda$ and all $x$.
The algorithm proceeds as follows:
\begin{enumerate}
	\item Generate a sample $x$ from $Q$.
	\item With probability $\frac{f(x)}{\lambda g(x)}$, output $x$. Otherwise, go to step 1.
\end{enumerate}
It is easy to show that the sample eventually output by the rejection sampling algorithm is distributed precisely according to $P$. The probability that an accepted sample is generated is
\[ \sum_x g(x) \frac{f(x)}{\lambda g(x)} = \frac{1}{\lambda}, \]
so the expected number of samples from $Q$ required to output a sample from $P$ is just $\lambda$.
The simplest case in which we can apply the algorithm is where $Q$ is the uniform distribution on $N$ elements, and we have the upper bound $f(x) \le \mu$ for some $\mu$ and all $x$.
Then the expected number of uniform samples required to obtain a sample from $P$ is $\mu N$, which will be minimised when $\mu = \max_x f(x)$.

%
%
Here we take $P$ to be the \bs distribution restricted to the CFS, and $Q$ to be the uniform distribution on the CFS (so $N=\binom{m}{n}$).
Note that $P$ is subnormalised, so is not quite a probability distribution.
However, the rejection sampling algorithm is blind to this subnormalisation (as this is effectively the same as increasing $\lambda$), so will generate samples from the renormalised distribution.

Each iteration of rejection sampling requires the computation of one permanent, corresponding to $f(x)$.
However, we seem to have a problem: in order to use rejection sampling most efficiently, it is required to know the maximum value of $f(x)$, which corresponds to the largest permanent of all $n\times n$ submatrices $A$ of a given $m \times m$ unitary matrix $U$.
One might expect that computing this would be computationally hard.
Without any bound on this quantity, we would be forced to use the trivial bound $\mu = 1$, corresponding to $\binom{m}{n}$ permanent computations being required to obtain one sample from $P$.

Fortunately, as we are only attempting to perform \emph{approximate} \bs, we only require a good estimate of $\mu = \max_x f(x)$.
Recall that it was argued in~\cite{Aaronson2011} that sampling from a distribution within total variation distance $\epsilon$ of the real \bs distribution $P_\mathrm{BS}$ should be computationally difficult, for some small constant $\epsilon$.
Imagine that our guess $\widetilde{\mu}$ for $\mu$ is too small, such that $\sum_{x,f(x) > \widetilde{\mu}} f(x) = \epsilon > 0$.
Then if $x$ is sampled uniformly at random and $f(x) > \widetilde{\mu}$, step 2 of the rejection sampling algorithm will fail.
If we modify the rejection sampling algorithm to simply produce a new uniform sample in this case and repeat, it is easy to see that we can view the modified algorithm as sampling from the truncated distribution $P_{\operatorname{low}}$ with probability mass function
\[ \widetilde{f}(x) = \begin{cases} \frac{f(x)}{\sum_{x, f(x) \le \widetilde{\mu}} f(x)} & \text{if $f(x) \le \widetilde{\mu}$}\\ 0 & \text{if $f(x) > \widetilde{\mu}$} \end{cases}. \]
Then the total variation distance between $P_{\operatorname{low}}$ and $P$ is
\[ \frac{1}{2} \sum_x |f(x) - \widetilde{f}(x)| = \sum_{x, f(x) > \widetilde{\mu}} f(x). \]
So if the probability mass of $P$ above $\mu$ is at most $\epsilon$, we have sampled from a distribution within distance $\epsilon$ of $P$.
We have found that we are able to use a simple random restart hill climbing algorithm to provide a suitable estimate of $\mu$ with $\mathcal{O}(m^2n)$ computations of $n\times n$ matrix permanents.

Our random restart hill climbing algorithm works as follows. 
We start by randomly sampling a submatrix, represented by the tuple $S$, from the uniform distribution.
For one pass, we greedily try replacing each row in the sampled submatrix by each row from $A$ in turn, accepting only if this increases $\Pr(S)$ while also making sure to avoid selecting the same row twice.  
We perform repeat passes until there is no improvement of the probability over a complete pass. 
At this point we randomly resample a new starting submatrix and repeat from the beginning. The total number of permanent calculations for one pass is $n(m-n)$. 

This method is not guaranteed to find a global maximum. However, in our experiments for $n \leq 7$ where we can still compute the full probability mass function exactly, we found the estimates for the maximum probability to be exactly equal to the global maximum in the overwhelming majority of cases. 
In the range $8 \leq n \leq 12$ where we no longer are able to compute the exact maximum probability, the bounds from our hill-climbing algorithm also allowed us to sample using rejection sampling efficiently and then compare our results with our Metropolised independence sampler (qv).  
This provided further evidence for both sampling techniques.   




\subsection{Metropolised independence sampling}
Markov Chain Monte Carlo (MCMC) methods have been long been the standard tool within statistics for the task of sampling from complicated probability distributions. 
We employ for our problem of \bs a special case of the MCMC procedure known as Metropolised independence sampling~\cite{liu96,liu04} (MIS), which enables us to sample efficiently from the target distribution.


The general principle of MCMC methods is to construct a Markov chain which has the target distribution (in our case, the \bs distribution of tuples) as its stationary distribution. 
The overall MCMC method requires us only to specify a proposal probability distribution, in our case over tuples $(s_1,\dots,s_m)$. 
At each turn we sample from this probability distribution and accept the new state according to the prespecified rules of the MCMC method.
In the general case the proposal distribution can depend on the current state of the chain but in MIS these proposals are entirely independent.
By choosing a proposal distribution which is not too far from our target \bs distribution, this simplification turns out to be particularly useful for us.
Empirically we show that it not only ensures fast convergence but a simple and efficient thinning procedure almost completely eliminates dependence between successive samples.

Let $Y$ be some proposal distribution over the tuples in $\Phi_{m,n}$ with probability mass function $g(x)$. 
As before, let $P$ be an instance of the CFS restricted (subnormalised) \bs distribution over these tuples, with probability mass function $f(x)$.  Starting at some random (according to $Y$) tuple $x$, propose a new random tuple $x^\prime$. 
We accept the proposal and transition from $x$ to  $x^\prime$ with probability
\begin{equation}\label{metropolis}
T\left(x^\prime | x\right) =\min\left(1,\frac{f(x^\prime)}{f(x)} \frac{g(x)}{g(x^\prime)}\right)
\end{equation}
where we note that any normalisation factor multiplying probabilities would cancel at this stage.
Repeating this generates a Markov chain with stationary distribution equal to the target distribution, $P$. 
If one can argue that the chain has converged then each sample from the generated chain after this point will be from $P$.
Each state in the Markov chain requires generating a single tuple sample from $Y$, one evaluation of $h(x^\prime)$ and one evaluation of $g(x^\prime)$, which entails one computation of an $n\times n$ complex valued matrix permanent.

It should be noted that it takes some time for the Markov chain to converge to its stationary distribution, and so some number of tuples at the beginning of the chain must be discarded (referred to as ``burn in'') in order for samples to come from the target distribution.
In addition to this, for example as there is a probability that the state of the Markov chain stays the same between iterations, there exists some autocorrelation in the chain. 
In order to generate independent samples from the Markov chain, one usually throws away a number of states (referred to as ``thinning'') at least equal to the autocorrelation period of the chain before harvesting a sample state. 
Both burn period $\tau_{\mathrm{burn}}$ and thinning interval $\tau_{\mathrm{thin}}$ depend both on the target distribution and the choice of proposal distribution $Y$. 
In general, the greater the overlap between target and proposal distributions (as measured by the acceptance probability (\ref{metropolis})), the smaller $\tau_{\mathrm{burn}}$ and $\tau_{\mathrm{thin}}$ need be.

A candidate for the proposal distribution is, as above in the case of rejection sampling, the uniform distribution over the tuples. 
However, we have found that a more suitable proposal distribution is the distribution $P_\mathrm{D}$ of distinguishable particles for a given input $U$, as it has a greater overlap with the target distribution.
Indeed, our experiments have found that for $n = 20$, the acceptance rate when using $P_\mathrm{D}$ as the proposal distribution is roughly 40\%.

This distribution is tantalisingly similar to the usual \bs distribution (\ref{BS_dist}):
\begin{equation}\label{BS_dist_dist}
\Pr_{P_\mathrm{D}}(S) = \frac{\mathrm{Per}\left( |A_S|^2 \right)}{s_1!\dots s_n!},
\end{equation}
where for a complex matrix $A$ with elements $A_{ij}$, $|A|^2$ denotes the matrix with elements $|A_{ij}|^2$.
However, there is a classical algorithm which can sample from this distribution in time $O(mn)$~\cite{Aaronson2013}.

Using $P_\mathrm{D}$ as the proposal distribution, one evaluation of $f(x^\prime)$ requires computing one $n \times n$ \emph{real} valued matrix permanent. 
So, in order to generate the first tuple in a sample, $\tau_{\mathrm{burn}}$ $n \times n$ real valued matrix permanents and $\tau_{\mathrm{burn}}$ $n \times n$ complex valued matrix permanents must be computed. Each subsequent sample requires $\tau_{\mathrm{thin}}$ $n \times n$ real valued and complex valued matrix permanents to be computed.

Testing (see Section \ref{sec:assessment}) has suggested that, using the distinguishable particle proposal distribution, it is sufficient for $\tau_{\mathrm{burn}}$ and $\tau_{\mathrm{thin}}$ to grow very slowly with $n$, and for neither to grow above 100 for $n\leq 25$.

Adapting our MIS-based method to carry out the scattershot \bs problem is simple.
For each sample that we wish to output, we can first sample (efficiently) from the uniform distribution on $n$-fold input modes (which fixes the columns of $U$ contributing to submatrices), before running the algorithm in the way described above for a single sample.
In this case, $\tau_{\mathrm{thin}}$ becomes meaningless and we are only interested in $\tau_{\mathrm{burn}}$, as we start a new chain for each sample.

Alternatively, our proposal distribution can be changed to include the uniform distribution over $n$-fold input modes, meaning that each state in a given Markov chain can correspond to a different input configuration.

Although in this work we do not examine the more general situation of \bs where there can be more than one boson in an output mode, we anticipate that relaxing the CFS restriction will not increase the run time of the MIS method. 
In fact, it is possible that the average run time could be decreased with this relaxation, as there exists an algorithm for computing the permanent which is exponential in matrix rank, rather than matrix size~\cite{Barvinok1996}.
However, due to there usually existing very large permanents of sub-unitary matrices with many repeated rows, relaxing the CFS restraint has an adverse effect on the average run time of our rejection sampling method.

\section{Numerical results}



For concreteness and convenience, we will examine the performance of our samplers for the case where the number of modes scales exactly as the square of the number of photons, i.e.\ $m=n^2$. This regime is generally expected to be hard to simulate classically~\cite{Aaronson2011,Latmiral2016}. 

\subsection{Computing matrix permanents}
\label{sec:permexps}

In order to make scaling based arguments for the expected run time of our sampling methods, we need to both confirm that we are able to compute permanents of $n\times n$ matrices in $\mathcal{O}(n2^n)$ time and compute the associated constant of proportionality.

\begin{figure}[t]
	\centering
	\includegraphics[width=0.5\linewidth]{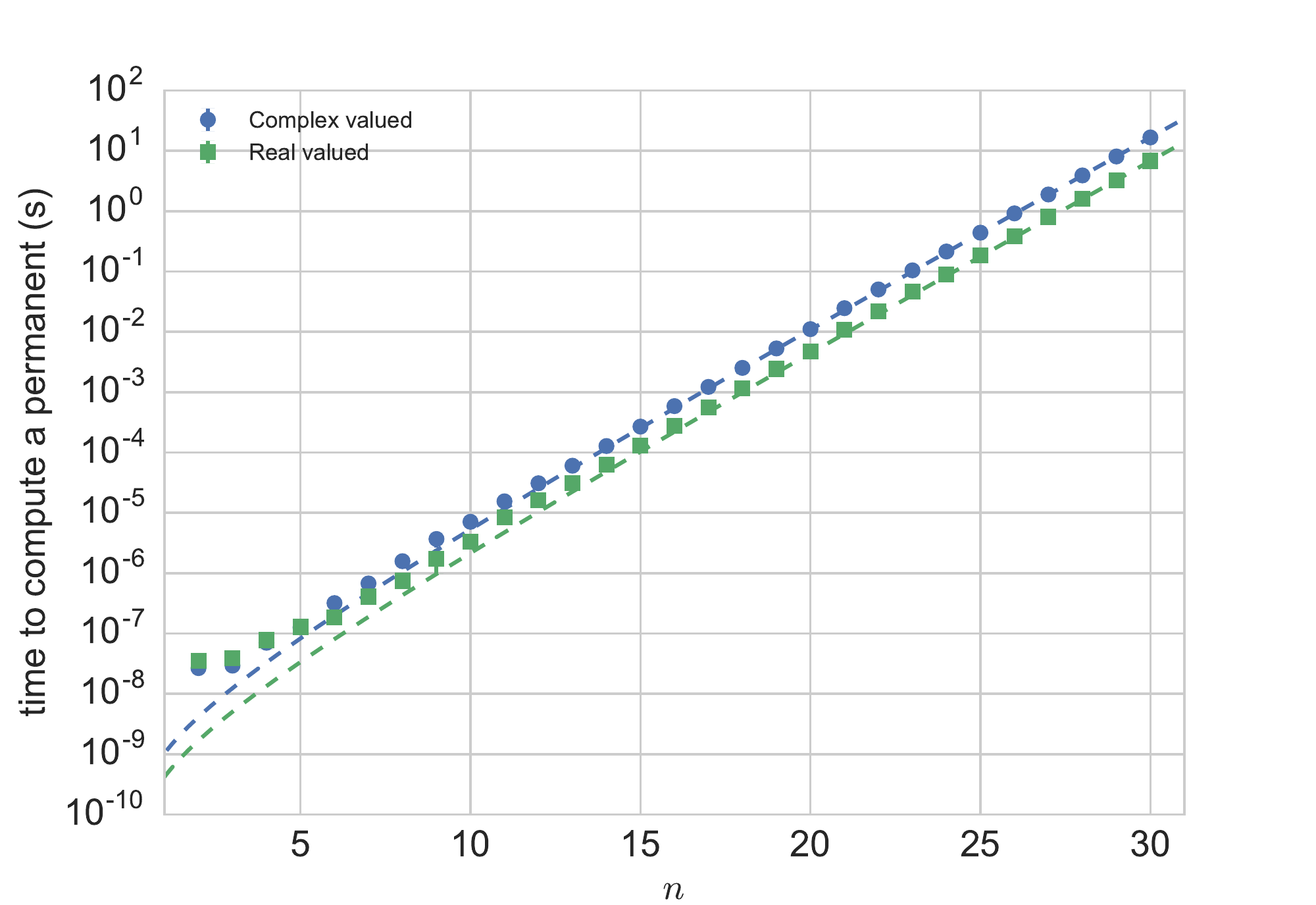}
	\caption{Numerical results for the mean time to compute $n\times n$ matrix permanents on a laptop with a 2.30GHz Intel(R) Core(TM) i5-5300U CPU.
		Blue points correspond to computing permanents of complex valued matrices and green squares to real valued matrices.
		Values correspond to the mean time recorded over 20 random matrices, with 12000000, 120000, 12000, 1200 and 12 repeats for values of $n$ in ranges $2\leq n \leq 3$, $4\leq n \leq 7$, $8\leq n \leq 15$, $16\leq n \leq 20$ and $21\leq n \leq 30$ respectively.
		Dashed lines correspond to a fit to the function $t(n)=cn2^n$, with $c$ calculated as $5.180\times10^{-10}$ (blue) and $2.106\times10^{-10}$ (green).
		Error bars displayed (where visible) represent the standard error in the mean time.
	}
	
	\label{fig:matrix-permanents}
\end{figure}

Timing results are presented in fig.\ \ref{fig:matrix-permanents}, where we see the expected scaling for large $n$.
We attribute the difference between the fitting function $t(n)=cn2^n$ and the data points for small $n$ to extra operations in our permanent computation code (checks, memory allocation etc.), on top of an implementation of Ryser's algorithm.
From the fit we were able to calculate $c$ for the case of computing permanents of complex and real valued matrices on a personal computer\footnote{\label{personal-computer}Specifically, numerical experiments in this paper were performed on a Dell Latitude E5450 laptop with 2.30GHz Intel Core i5-5300U CPU} finding that $c=5.180\times10^{-10}$ and $c=2.106\times10^{-10}$ respectively.
In the case of computing permanents of real valued matrices we note that all values are in fact nonnegative, and therefore it may be beneficial to use the fully-polynomial randomized approximation scheme of Jerrum, Sinclair and Vigoda~\cite{Jerrum2004}.

Importantly, these values (and approximate values for more powerful machines) allow us to predict the average run time for our sampling methods for any value of $n$.

\subsection{Rejection sampling and random restart hill climbing}

In addition to how long it takes to compute a single matrix permanent, we are also interested in how many matrix permanent evaluations are required before a sample value is output.

In the case of rejection sampling with a uniform proposal distribution, we must compute the upper bound $\mu$ using our random restart hill climbing algorithm before implementing the rejection sampling algorithm.
We emphasise here that $\mu$ need only be computed once per input $U$, and so for large enough sample sizes, the computational contribution of computing $\mu$ gets washed out.
For concreteness, we focus here on outputting a sample consisting of 100 tuples, and present the number of required permanent computations averaged over these 100 tuples.

\begin{figure}[t]
	\centering
	\captionsetup[subfloat]{farskip=0pt,captionskip=0pt}
	\subfloat{\includegraphics[width=0.5\linewidth]{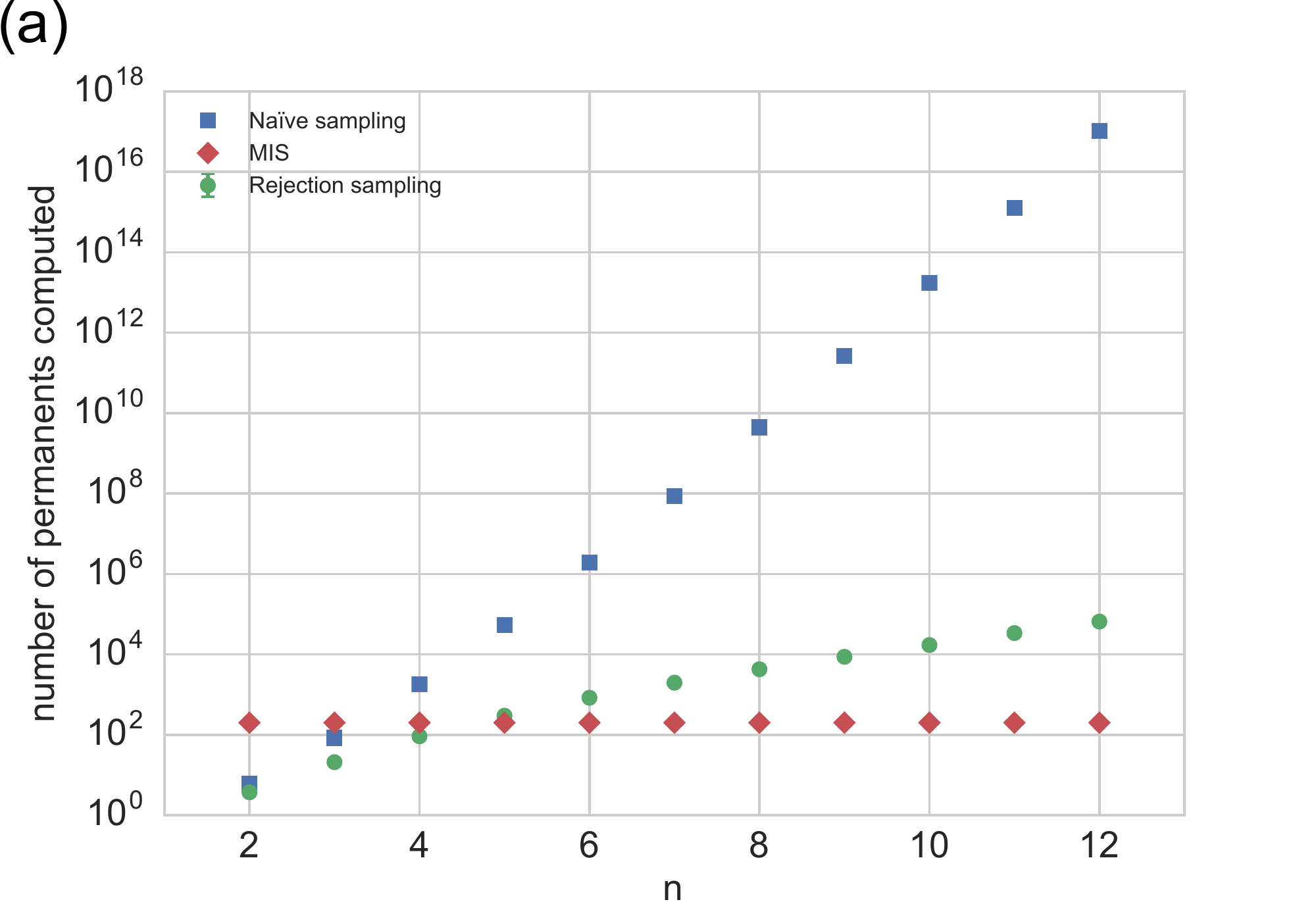}}
	\subfloat{\includegraphics[width=0.5\linewidth]{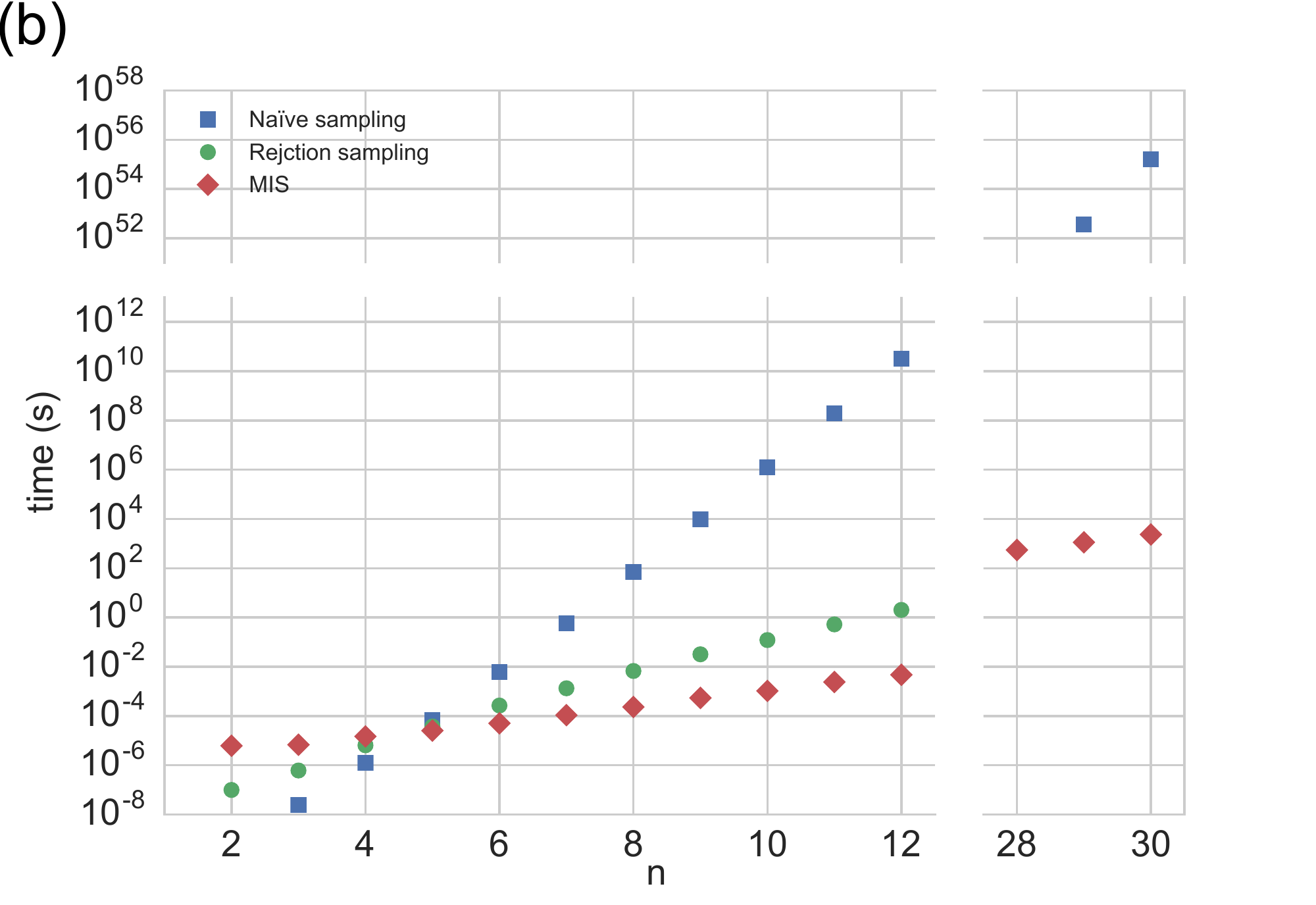}}
	\caption{(a): Numerical results for the mean number of $n\times n$ permanent computations required to output a single tuple, averaged over a sample of 100 tuples, with $m=n^2$. Red diamonds correspond to MIS with $\tau_{\mathrm{burn}} = \tau_{\mathrm{thin}} = 100$, green circles to rejection sampling with $\mu$ (an approximation of $\max_S |\mathrm{Per}\left(A_S\right)|^2$) computed using our random restart hill climbing algorithm with $4m$ restarts, and blue squares correspond to the number of permanent computations for brute force sampling (i.e.\ simply $\binom{\,n^2}{n} / 100$).
		Each rejection sampling point is calculated by finding the mean number of permanents computed before accepting a proposed tuple over 10000 accepted tuples (100 repeats for 100 Haar-random input unitaries $U$), including those computed in performing the random restart hill climbing algorithm.
		Error bars, where visible, represent one standard error in the mean number.
		(b): These values converted to mean time (in seconds) to output a single tuple using timing data from fig.\ \ref{fig:matrix-permanents}, and extended to $n=30$ photons. 
	}
	
	\label{fig:rejection-numperms}
\end{figure}

In fig.\ \ref{fig:rejection-numperms} we see that, by using rejection sampling instead of brute force sampling, one can reduce the number of computations required at a rate which increases rapidly with $n$.
In fact, a significant portion of the computation for all values of $n$ investigated here is spent getting an estimate of $\mu$.
It should be noted, however, that this is an artefact of our choice of number of restarts implemented in our random restart hill climbing algorithm.
We chose the number of restarts to be $4m$, based on empirical evidence that this results in the probability mass above this estimate being small.
For $600$ random instances of the problem with $2 \leq n \leq 7$, we found that the largest submatrix permanent was found $598$ times.
When the largest value was not found, the probability mass above the estimate was found to be small ($<2 \times 10^{-5}$).

As an example for reference, on a personal computer the average time to produce one sample for $n=12$ (averaged over 100 samples) was 2.03s.

%

\subsection{Metropolised independence sampling}
\label{sec:mis}

Determining the theoretical runtime of MIS is straightforward: $\tau_{\mathrm{thin}}$ and $\tau_{\mathrm{burn}}$, together with the time to compute a matrix permanent, completely determine the time to produce a sample, as discussed above.
We include these bounds in fig.\ \ref{fig:rejection-numperms}, taking $\tau_{\mathrm{thin}} = \tau_{\mathrm{burn}} = 100$, and amortised over 100 samples.
Observe that for $n=20$, each tuple can be output in 1.58s on a personal computer; even for $n=25$, only 62.5s are required.

\subsection{Assessment of our sampling methods}
\label{sec:assessment}


\begin{figure}[tp]
	\centering
	\captionsetup[subfloat]{farskip=0pt,captionskip=0pt}
	\subfloat{\includegraphics[width=0.42\columnwidth]{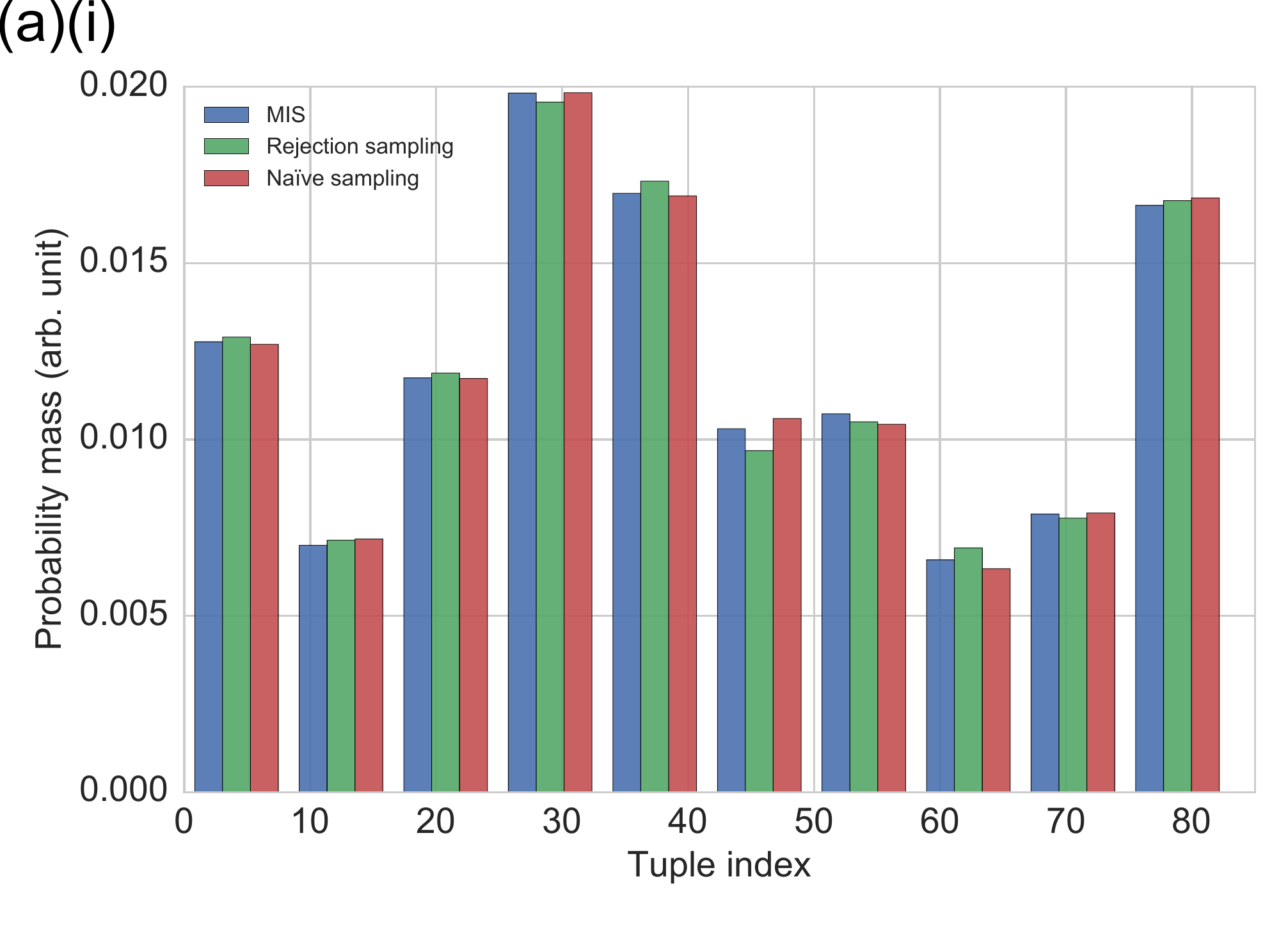}}
	\subfloat{\includegraphics[width=0.42\columnwidth]{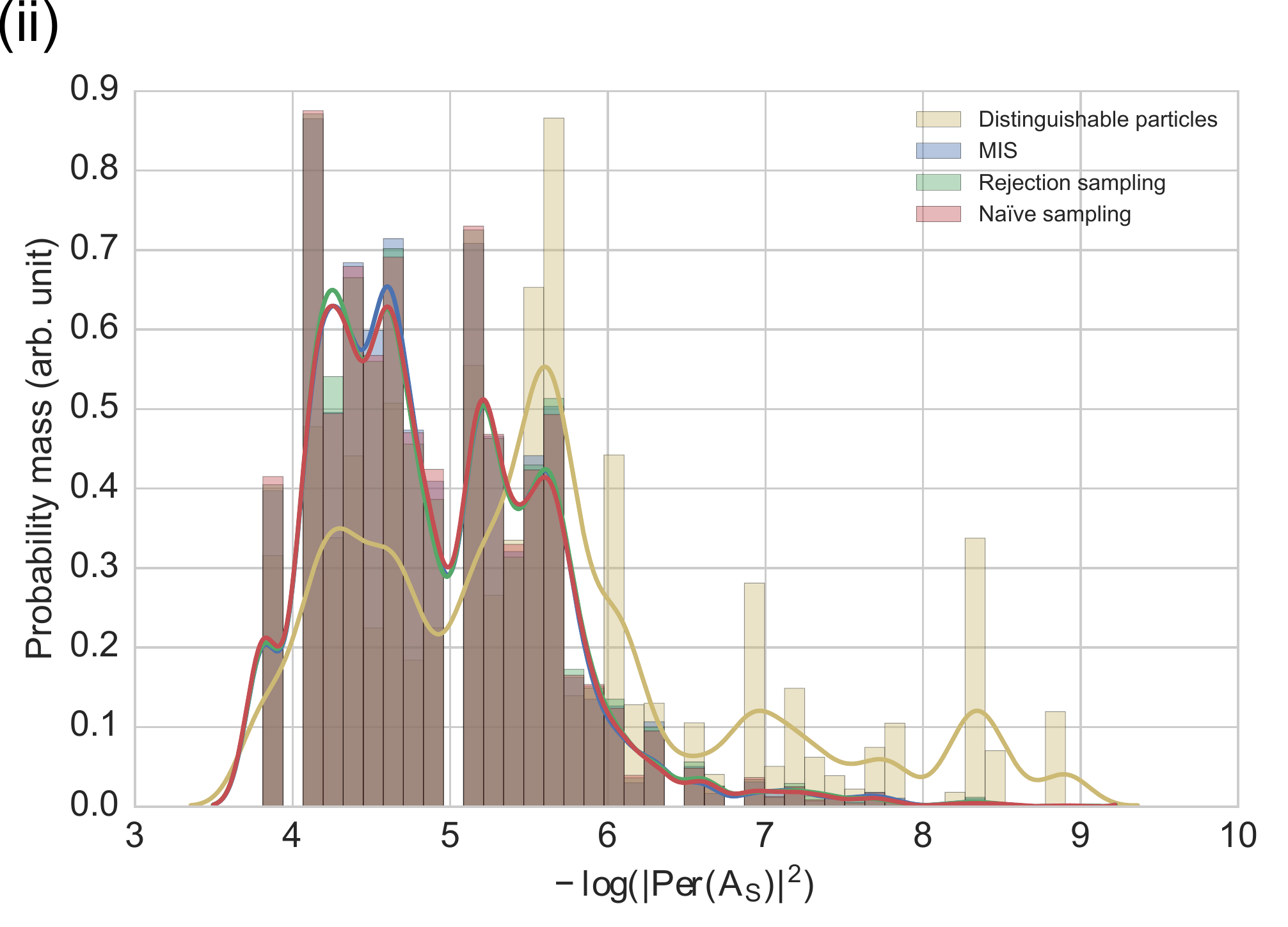}}
	\linebreak
	\subfloat{\includegraphics[width=0.42\columnwidth]{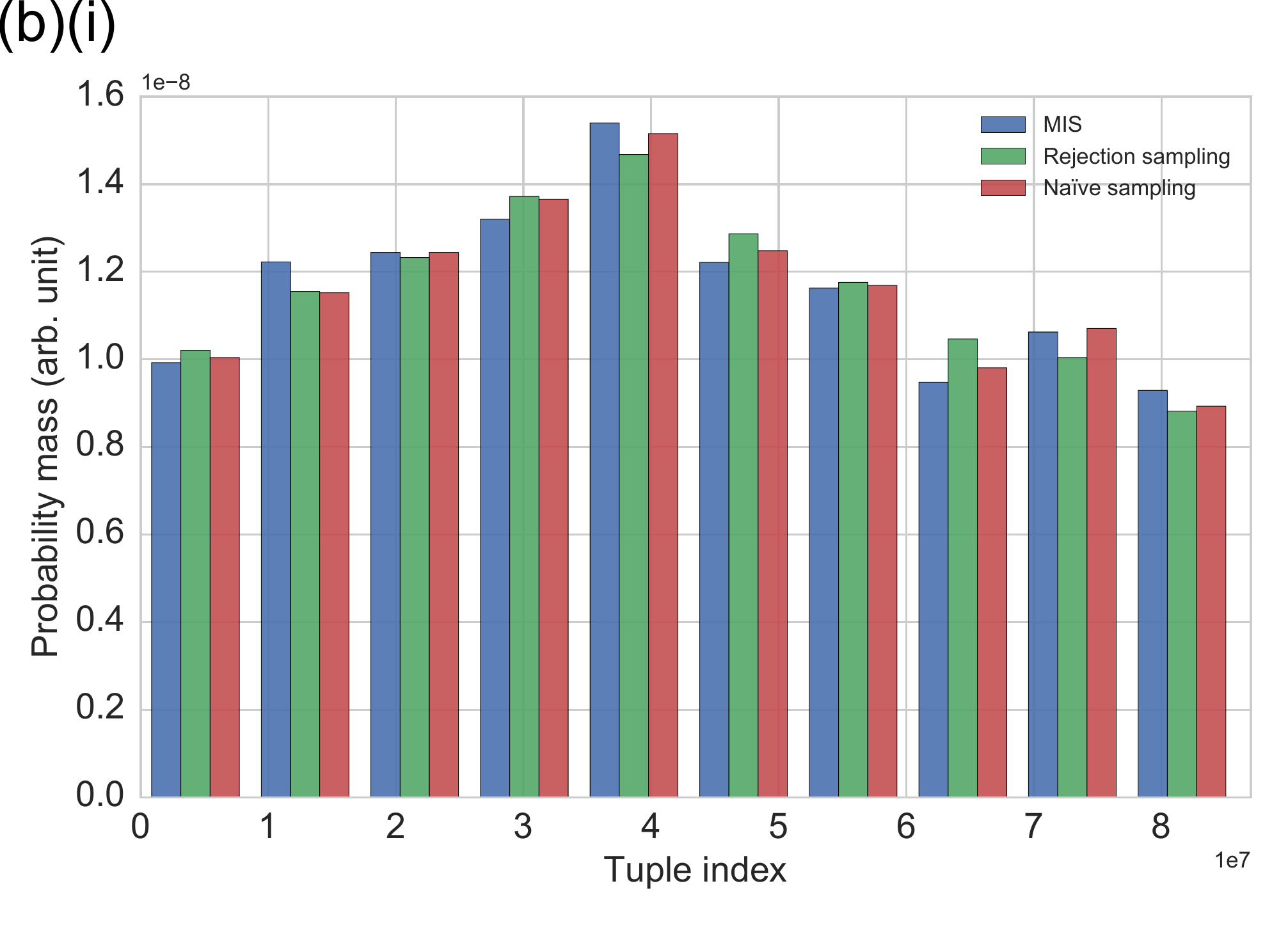}}
	\subfloat{\includegraphics[width=0.42\columnwidth]{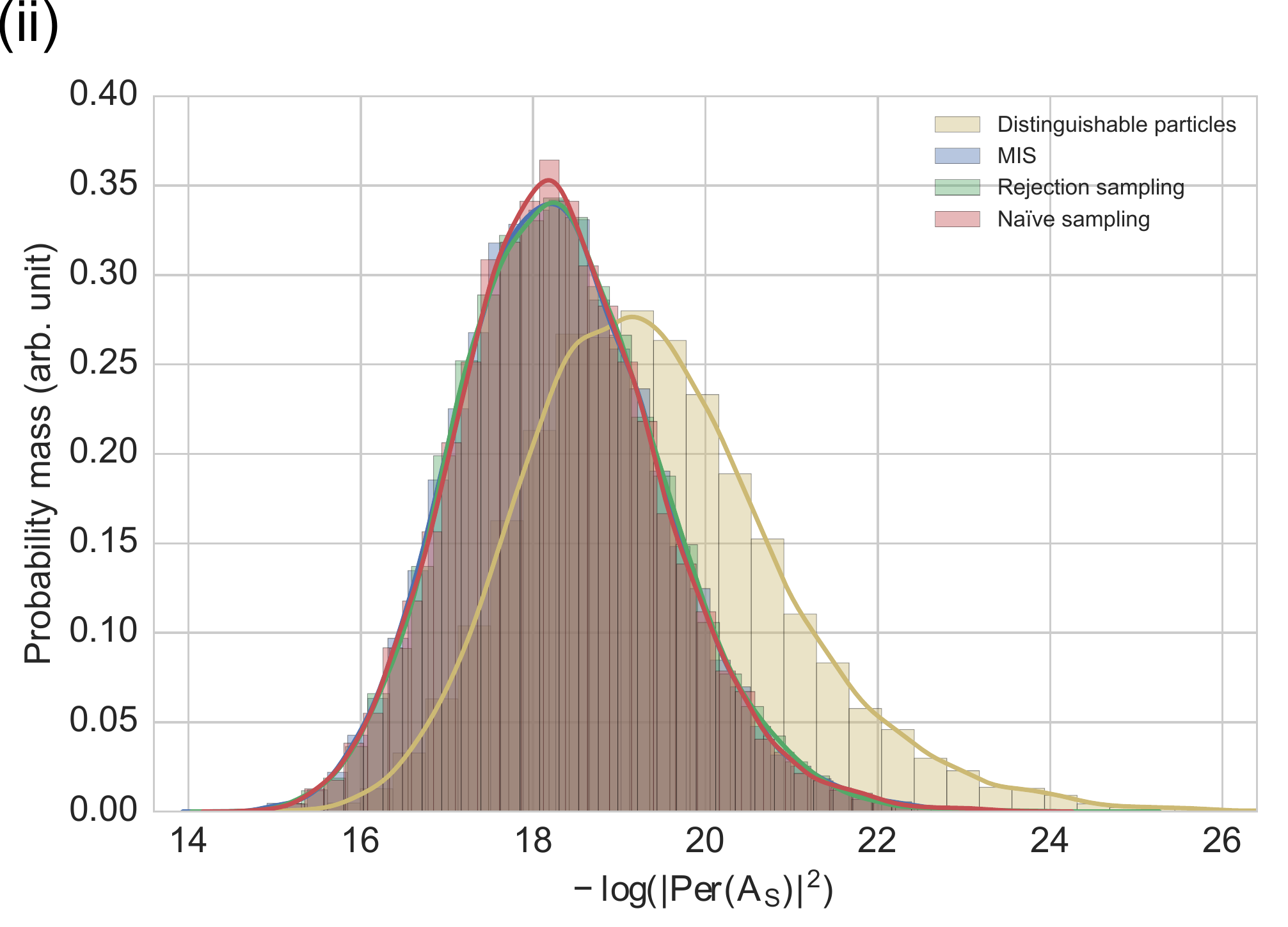}}
	\linebreak
	\subfloat{\includegraphics[width=0.42\columnwidth]{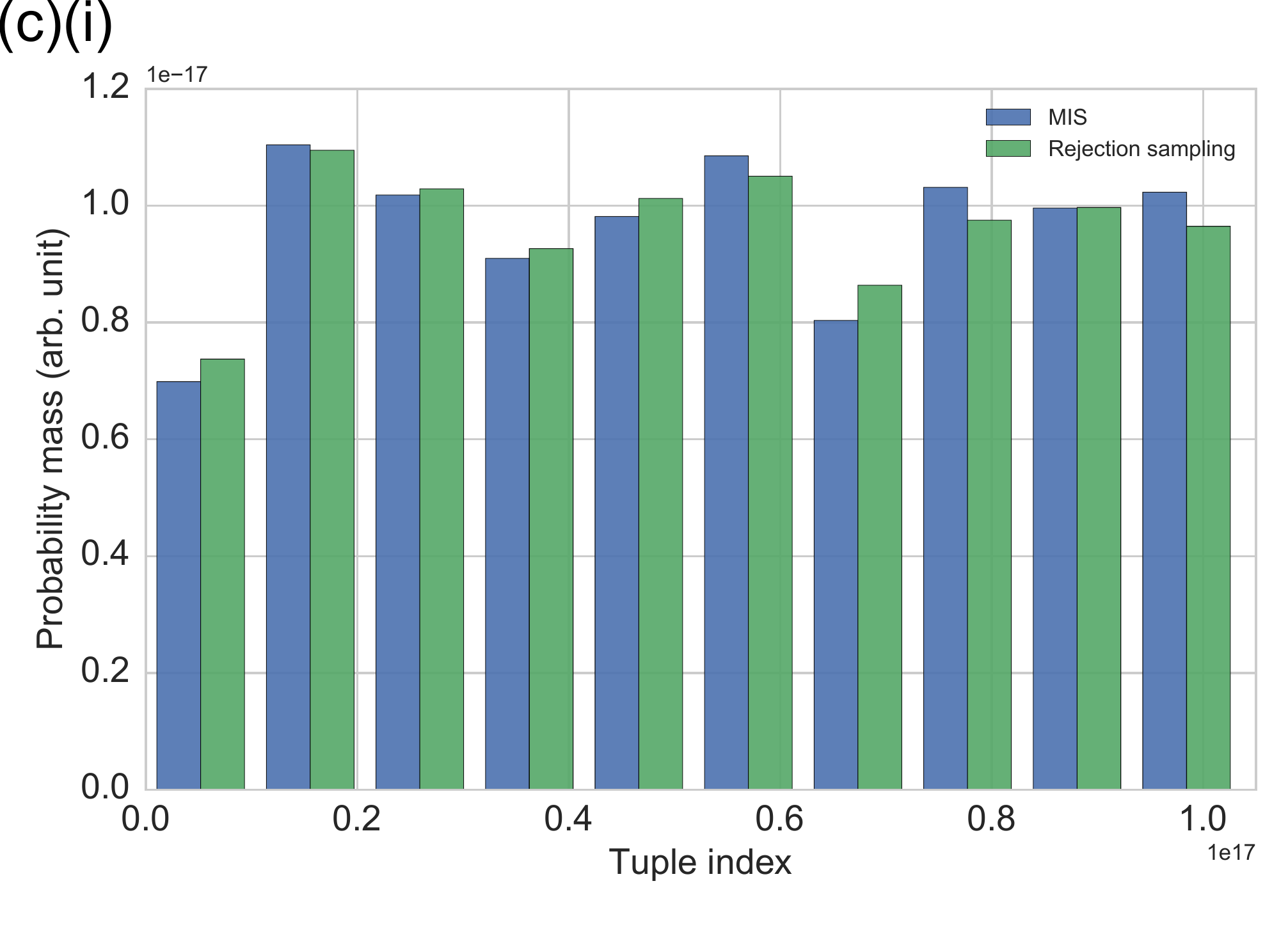}}
	\subfloat{\includegraphics[width=0.42\columnwidth]{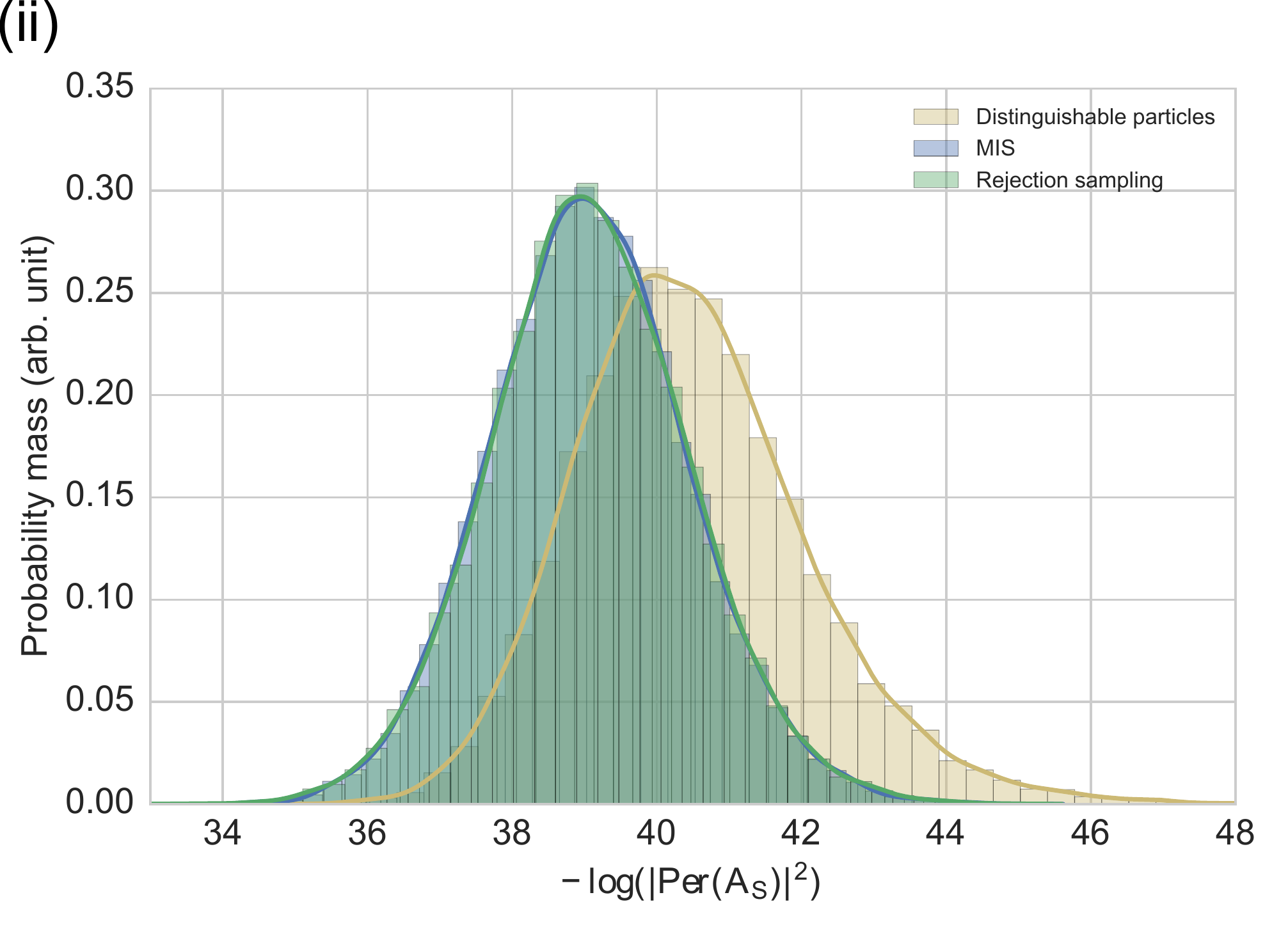}}
	\linebreak
	\subfloat{\includegraphics[width=0.42\columnwidth]{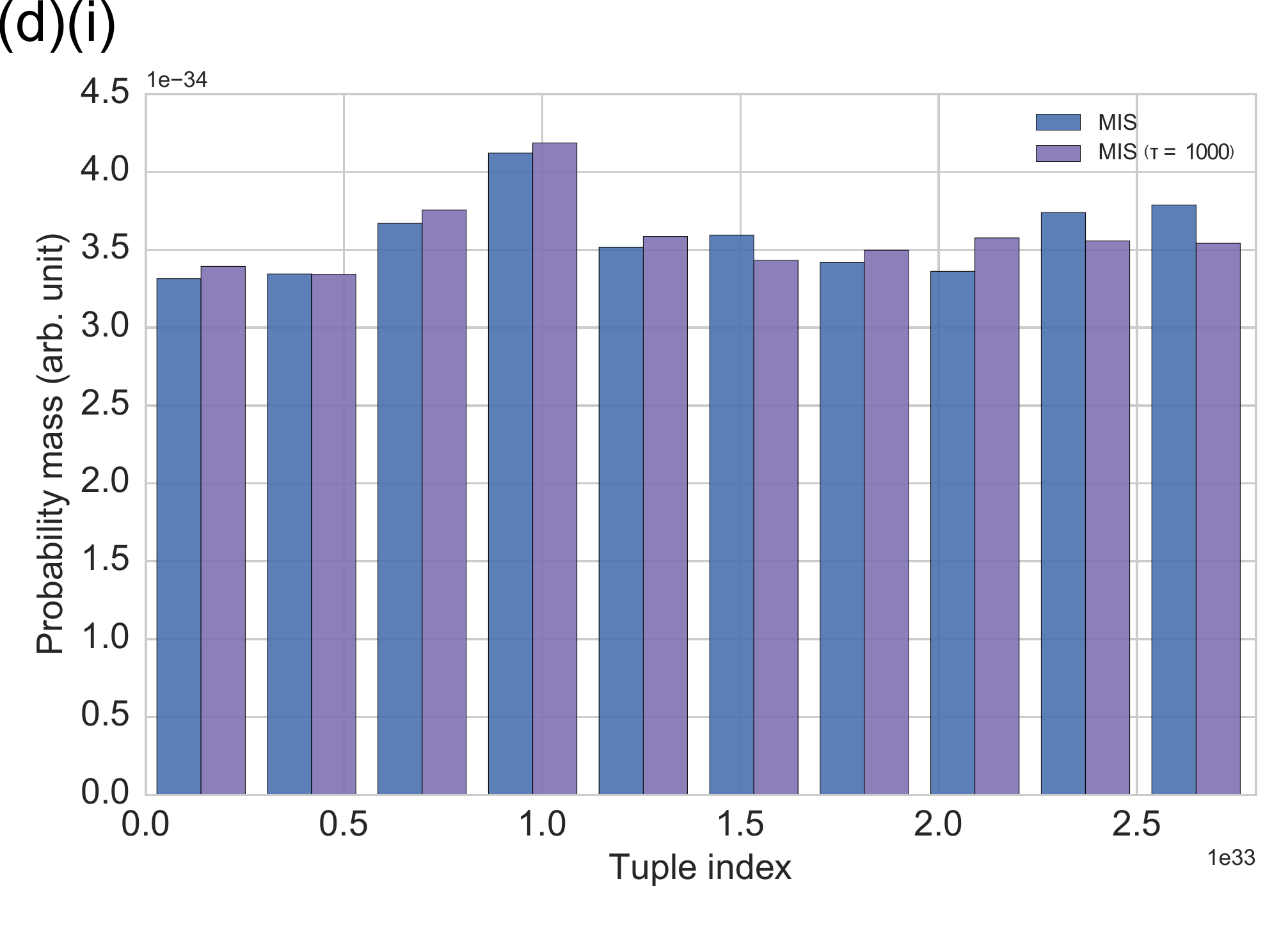}}
	\subfloat{\includegraphics[width=0.42\columnwidth]{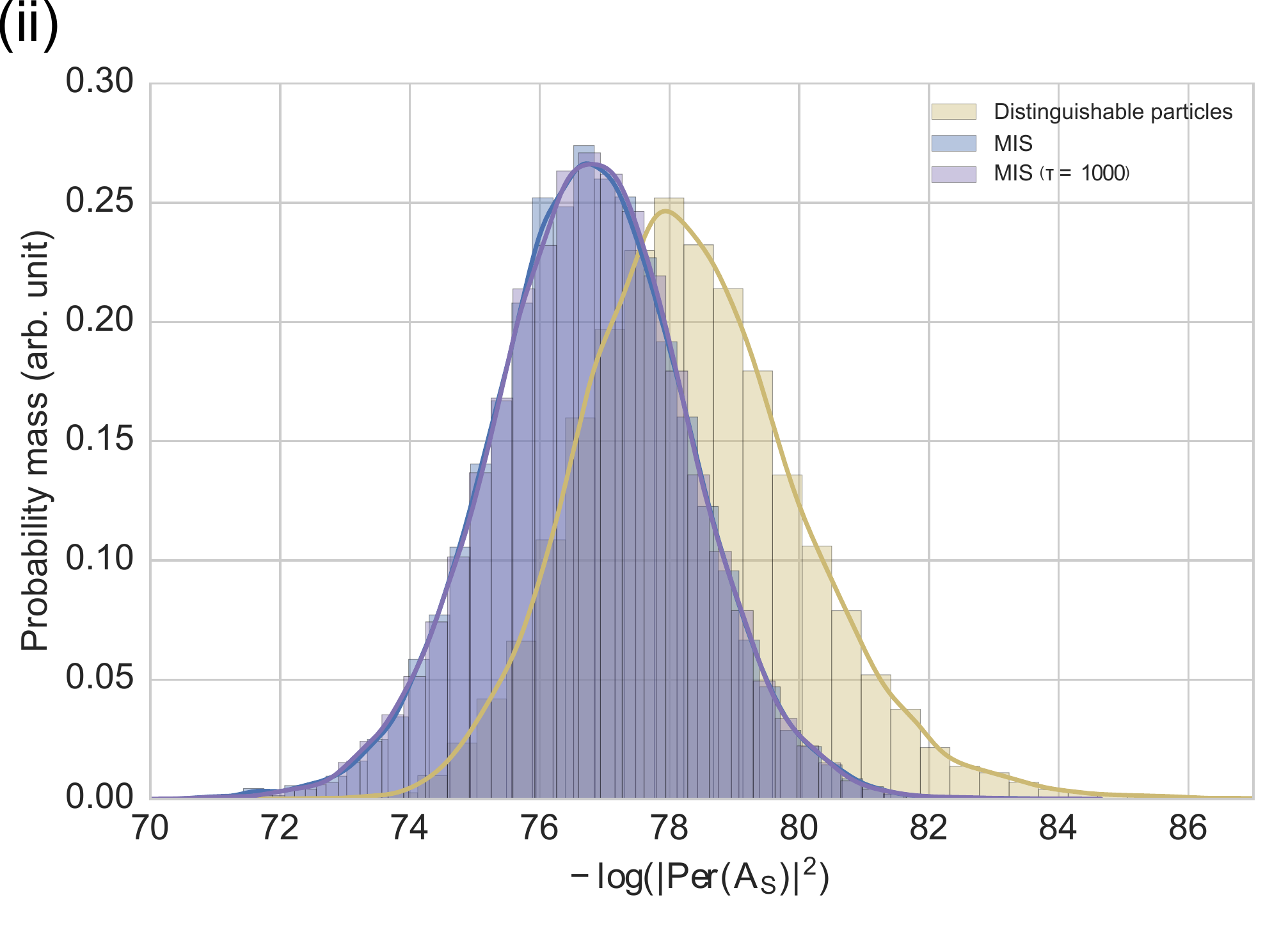}}
	
	\caption{(a): (i): Distribution of 20000 indexed and binned tuples obtained using MIS (blue), rejection sampling (green) and na\"ive brute-force sampling (red) for a Haar-random instance of the \bs problem with $n=3$ and $m=9$.
		(ii): Distribution of $-\log(|\mathrm{Per}(A_S)|^2)$ for each tuple $S$ sampled, along with the distribution of these values computed for a sample from the distinguishable particle distribution (yellow).
		Solid lines are obtained via kernel density estimation, and are included as a visual aid.
		(b): Distributions for a Haar-random instance of the \bs problem with $n=12$ and $m=144$ (colours as in A).
		(c): $n=12$ and $m=144$.
		Note that no sampling was done using the na\"ive brute-force sampling method, due to its inefficiency at this problem size.
		(d): $n=20$ and $m=400$ using MIS with $\tau=100$ (blue) and $\tau=1000$ (purple).
	}
	\label{fig:sample-distributions}
\end{figure}

\begin{figure}[tp]
	\captionsetup[subfloat]{farskip=0pt,captionskip=0pt}
	\centering	\subfloat{\includegraphics[width=0.5\columnwidth]{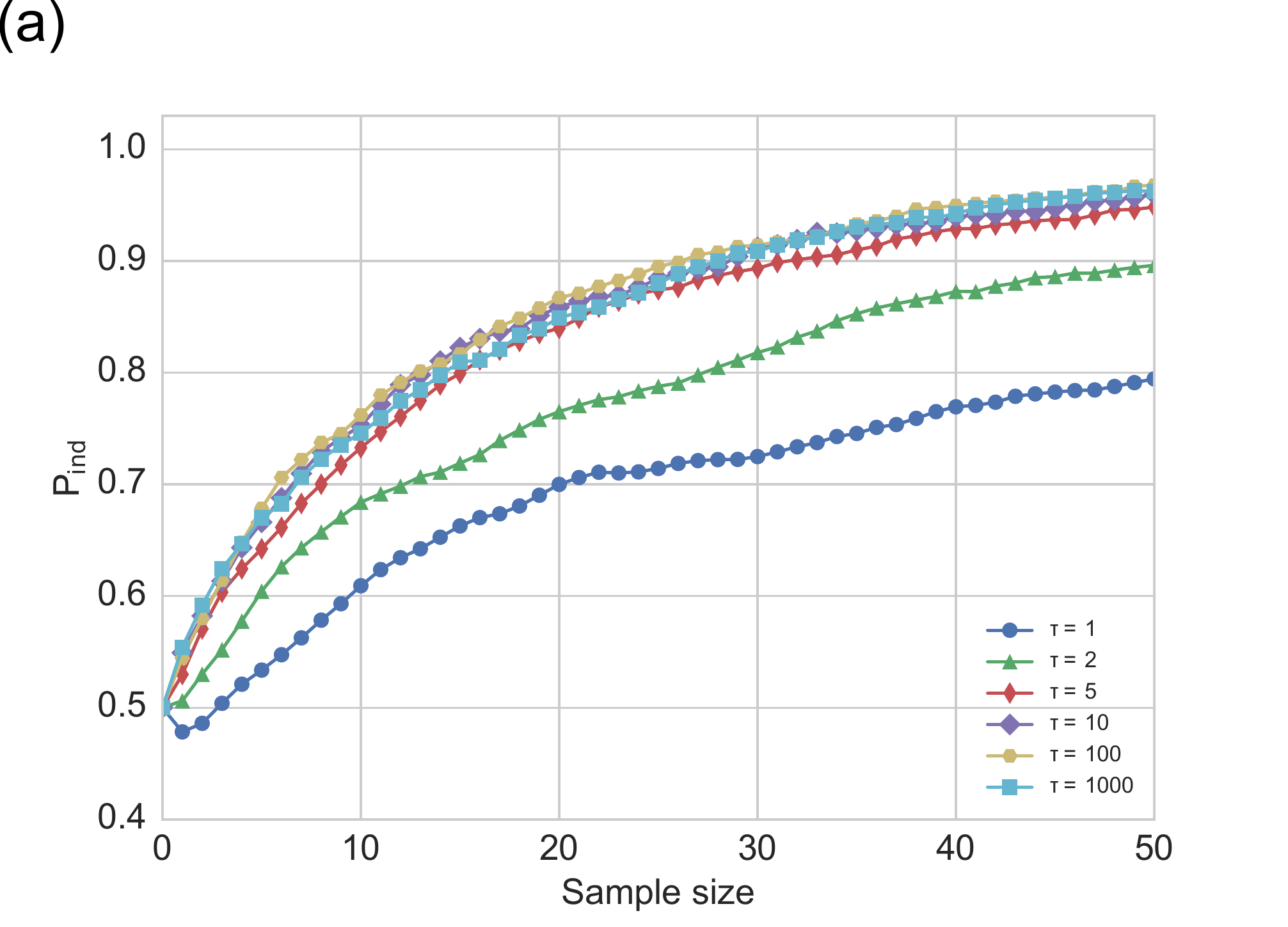}}
	\subfloat{\includegraphics[width=0.5\columnwidth]{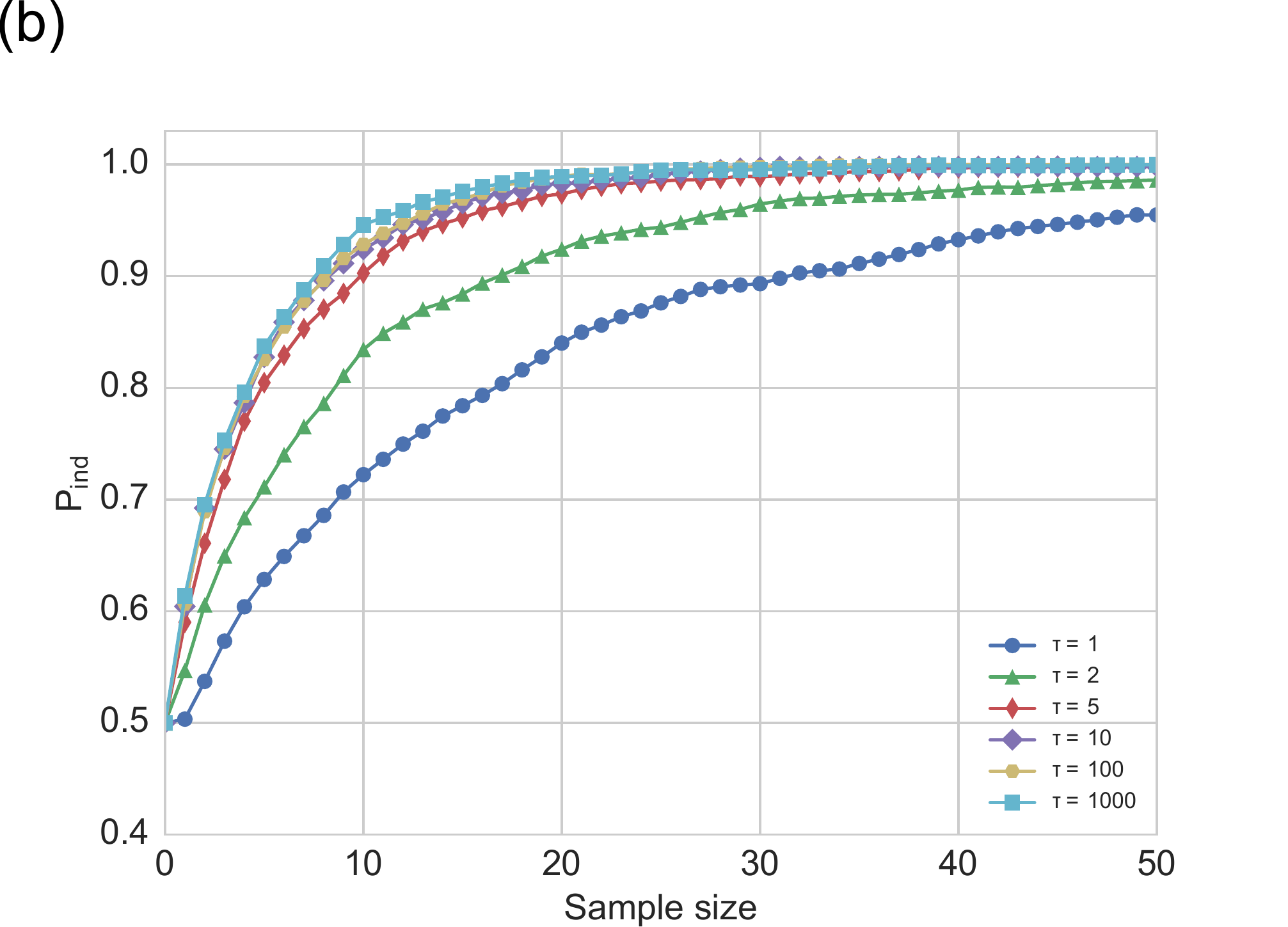}}
	\linebreak
	\subfloat{\includegraphics[width=0.5\columnwidth]{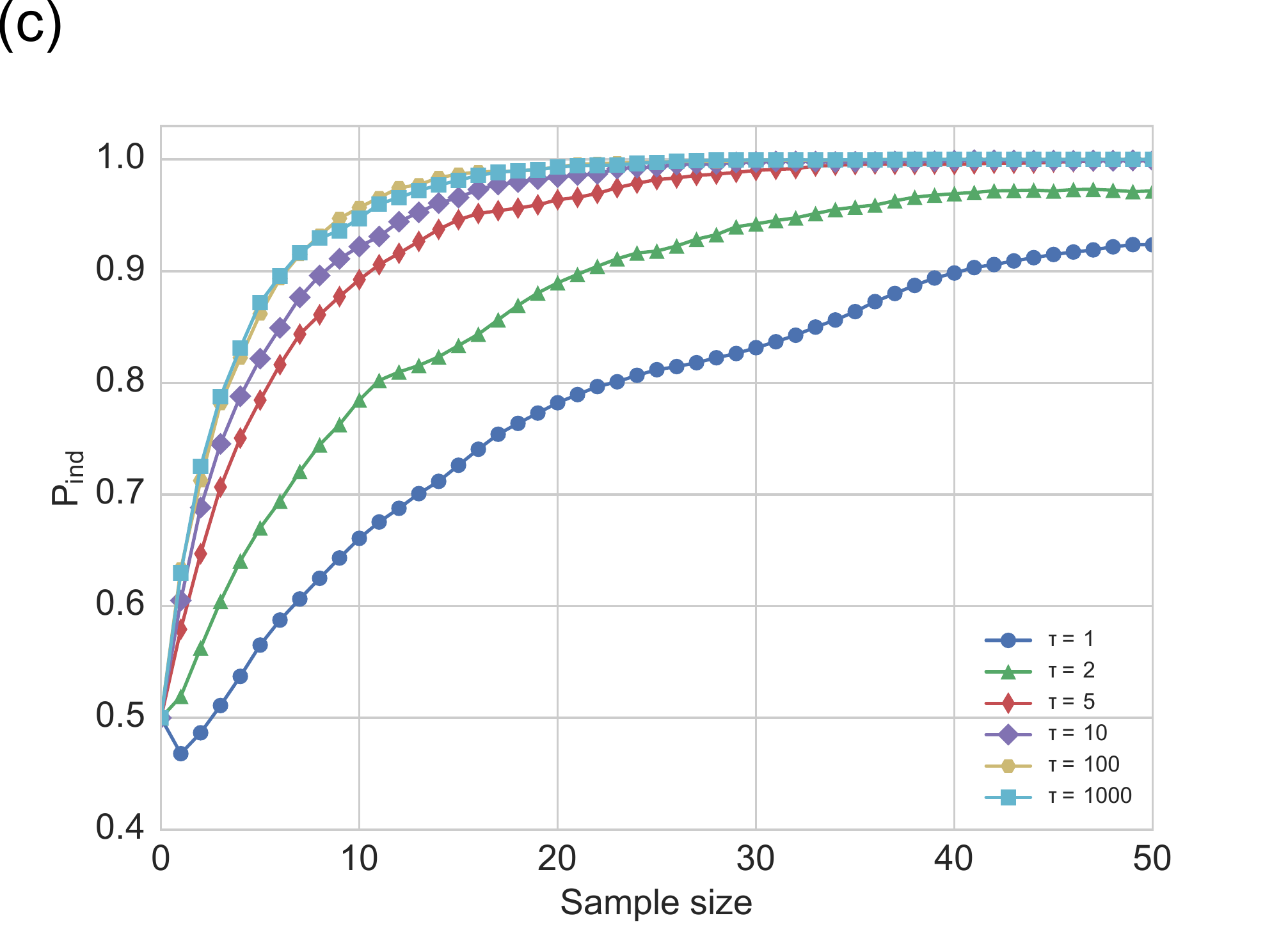}}
	\subfloat{\includegraphics[width=0.5\columnwidth]{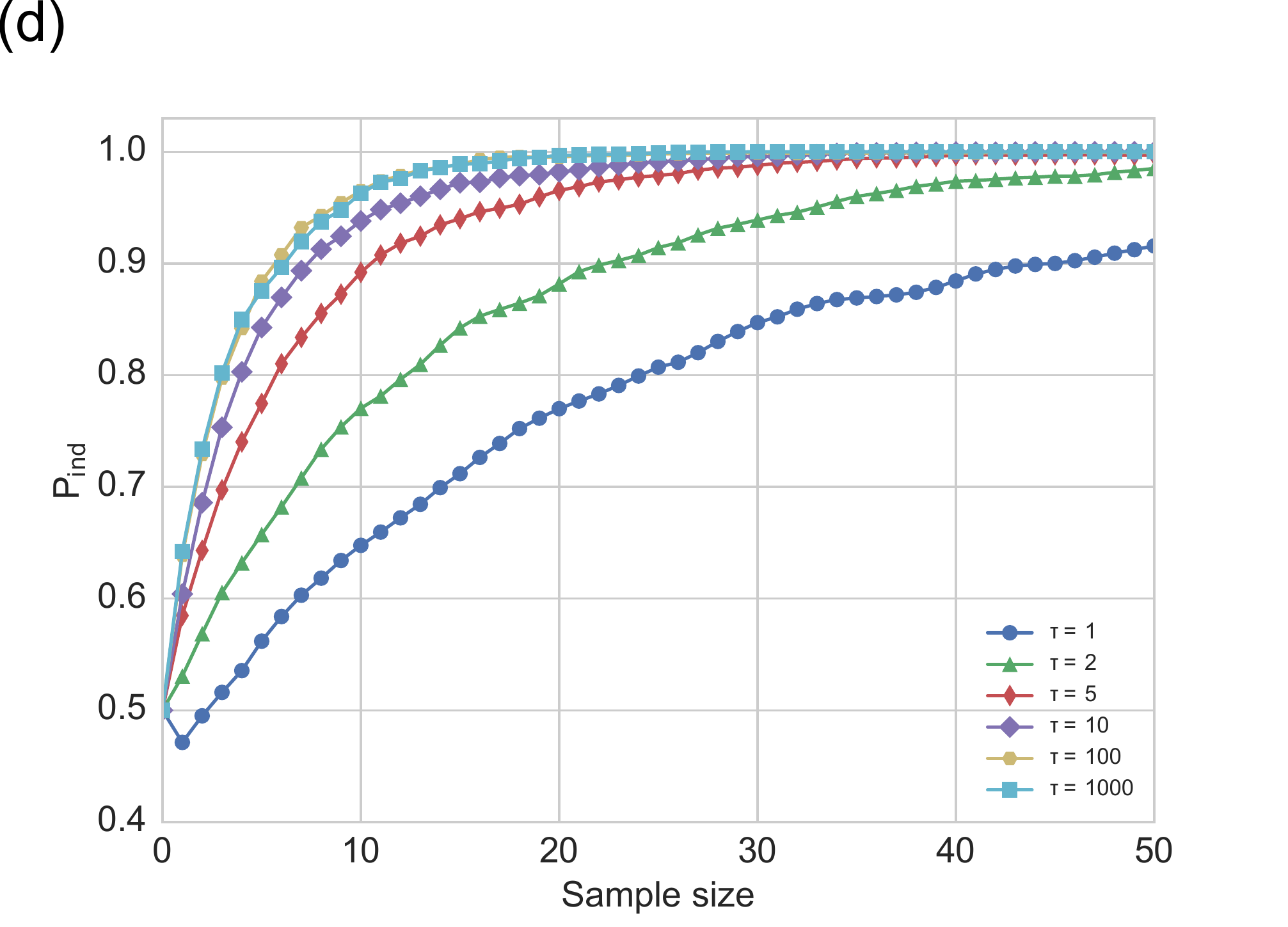}}
	\caption{Results of likelihood ratio tests performed on samples generated using MIS with different values of $\tau$.
		Here $\tau$ is shorthand for both $\tau_{\mathrm{burn}}$ and $\tau_{\mathrm{thin}}$.
		All data points are the mean value of $P_\mathrm{ind}$ over $1000$ independently generated samples.
		Confidence intervals have been omitted for ease of viewing.
		(a): Data for a Haar-random instance of the \bs problem with $n=3$ and $m=9$. 
		(b): Similarly for $n=7$ and $m=49$.
		(c): $n=12$ and $m=144$.
		(d): $n=20$ and $m=400$.}
	\label{fig:main-burnin-thin}
\end{figure}

\begin{figure}[tp]
	\captionsetup[subfloat]{farskip=0pt,captionskip=0pt}
	\centering
	\subfloat{\includegraphics[width=0.35\columnwidth]{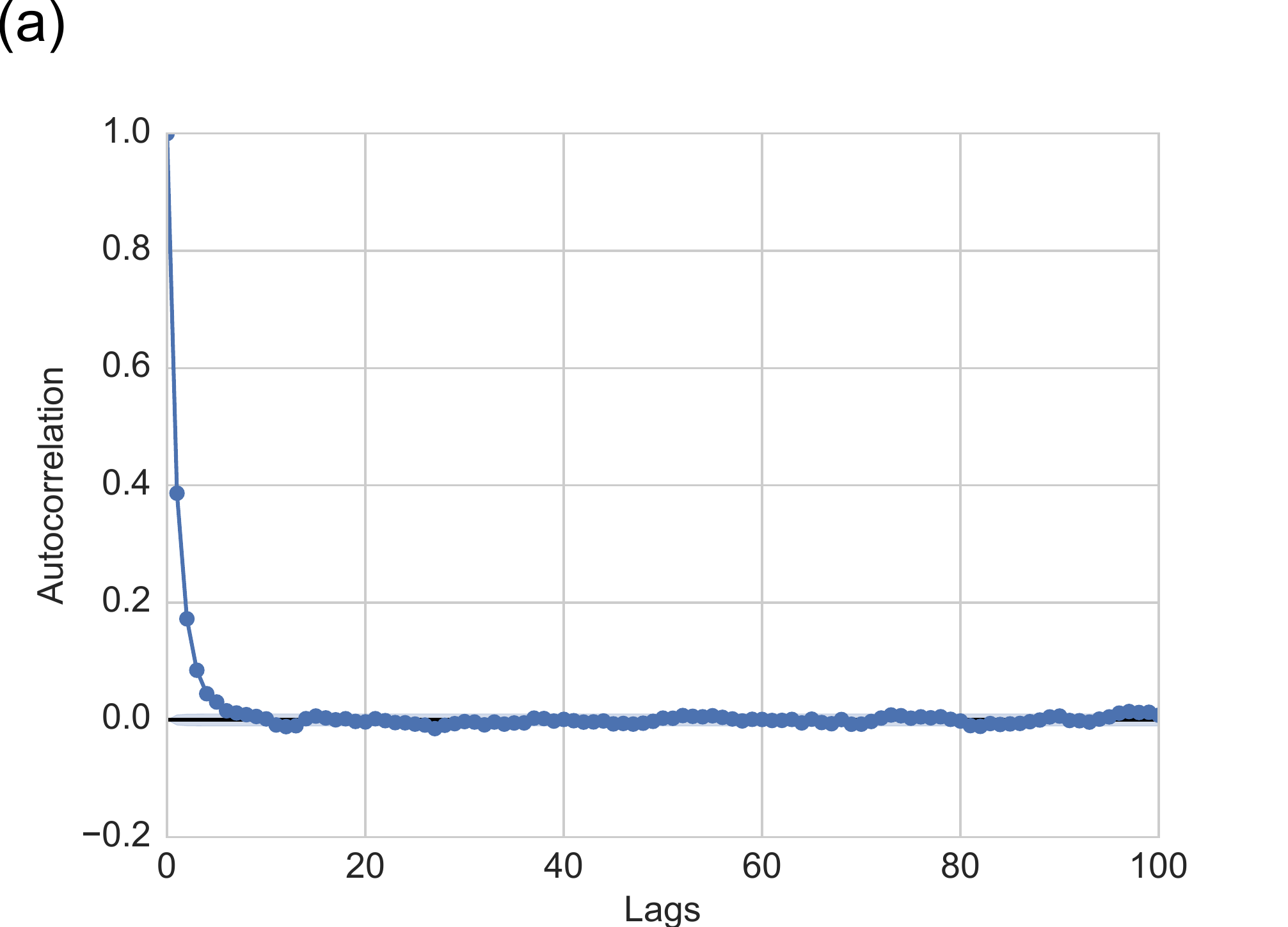}}
	\subfloat{\includegraphics[width=0.35\columnwidth]{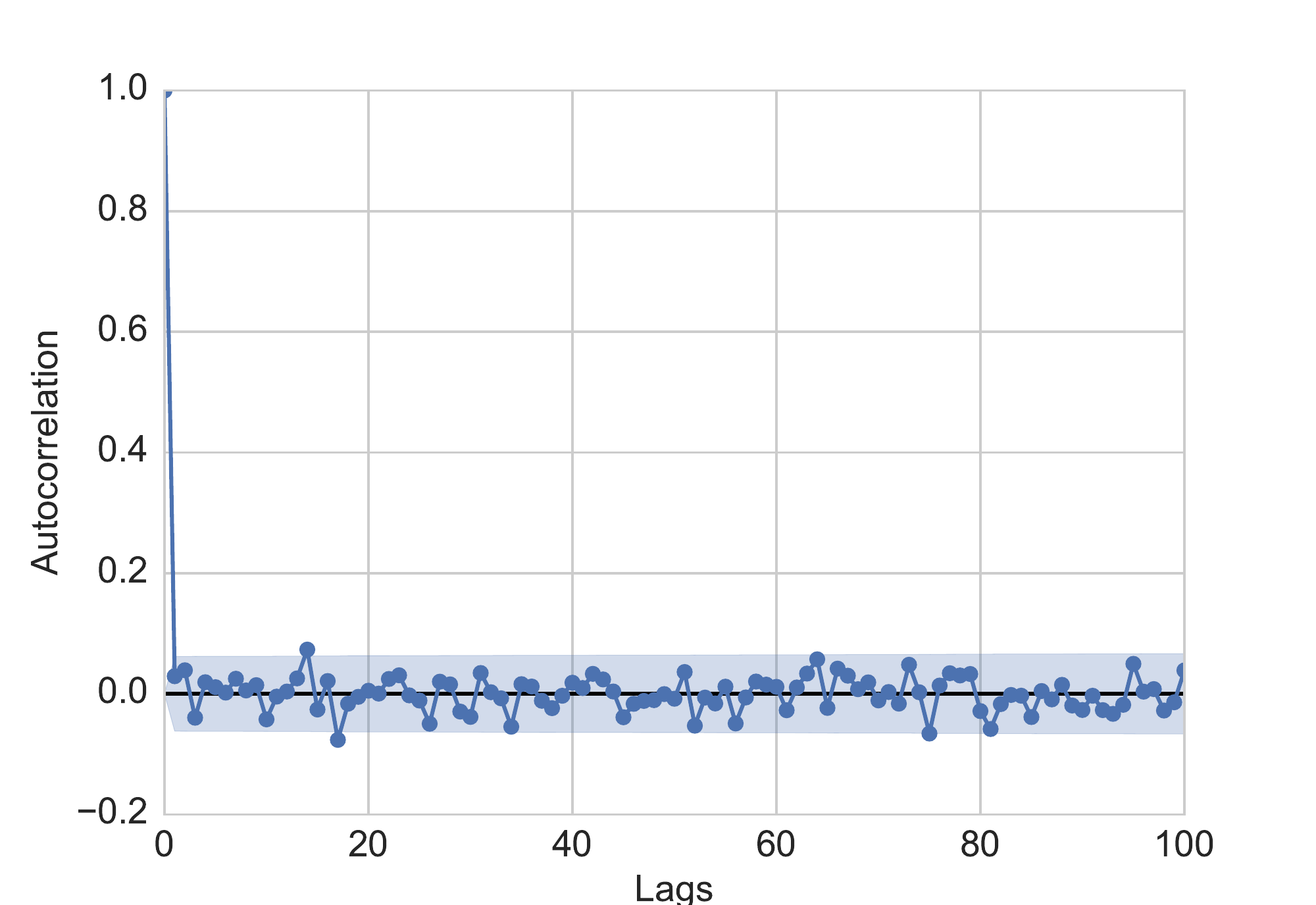}}
	\linebreak
	\subfloat{\includegraphics[width=0.35\columnwidth]{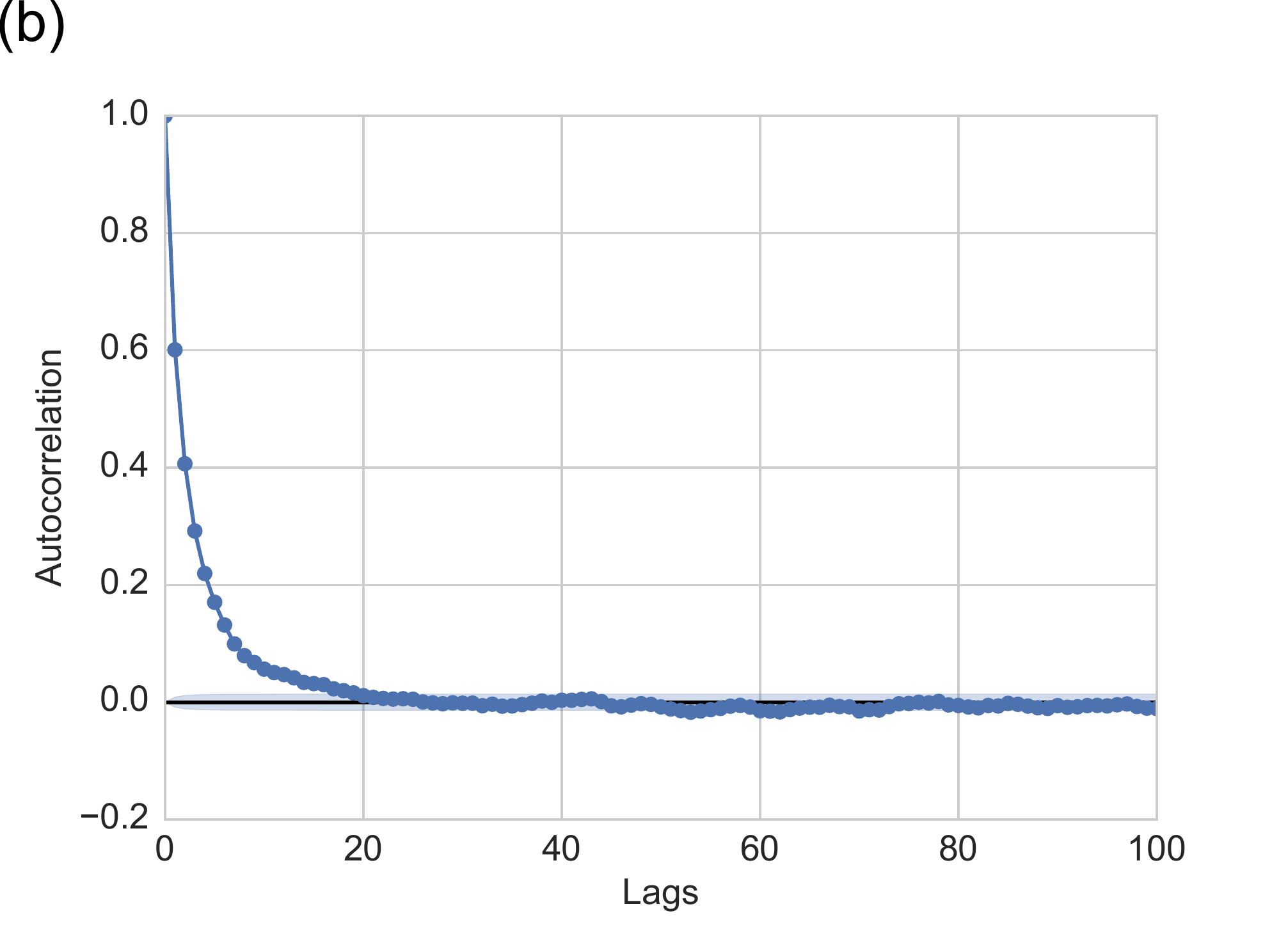}}
	\subfloat{\includegraphics[width=0.35\columnwidth]{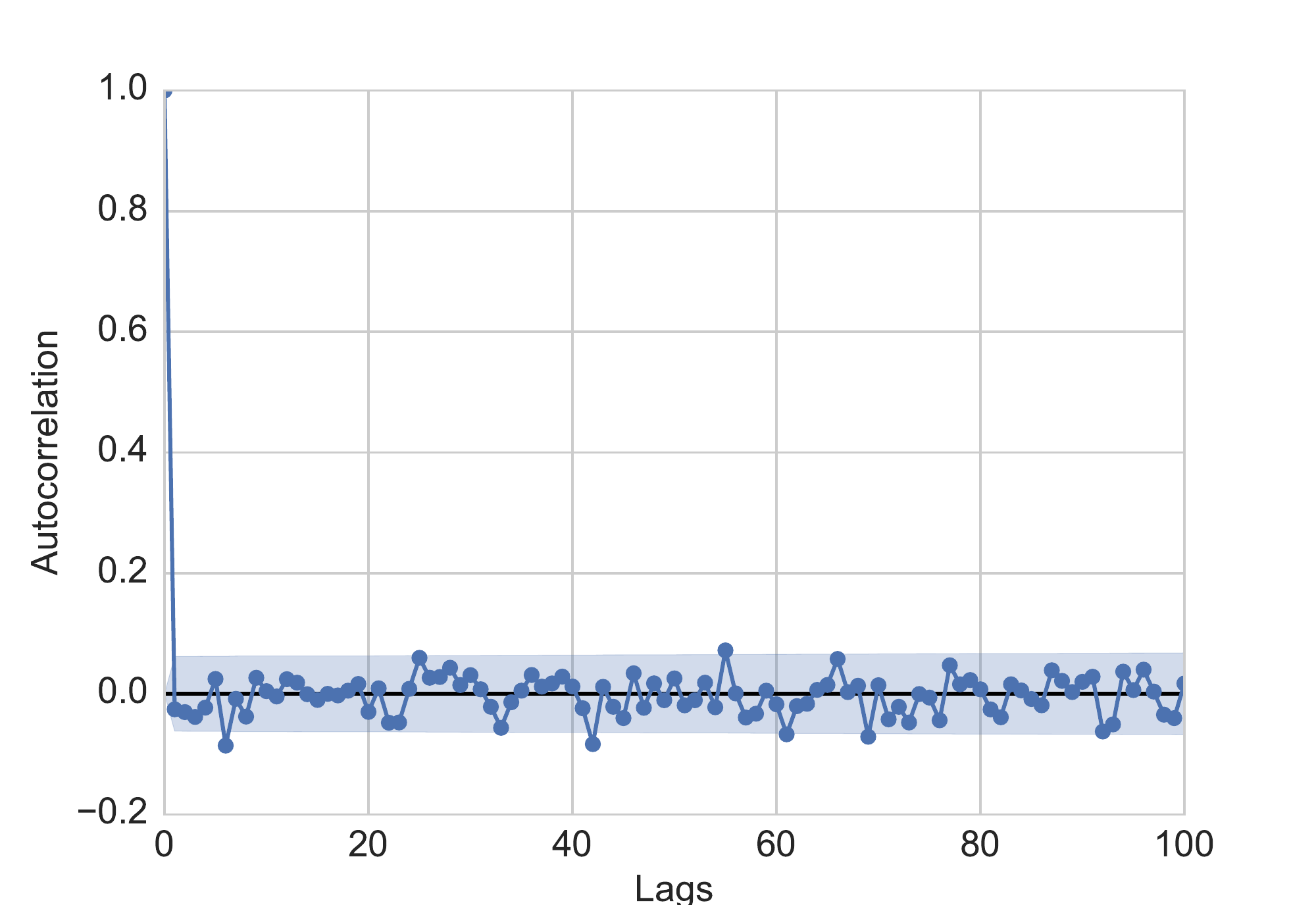}}
	\linebreak
	\subfloat{\includegraphics[width=0.35\columnwidth]{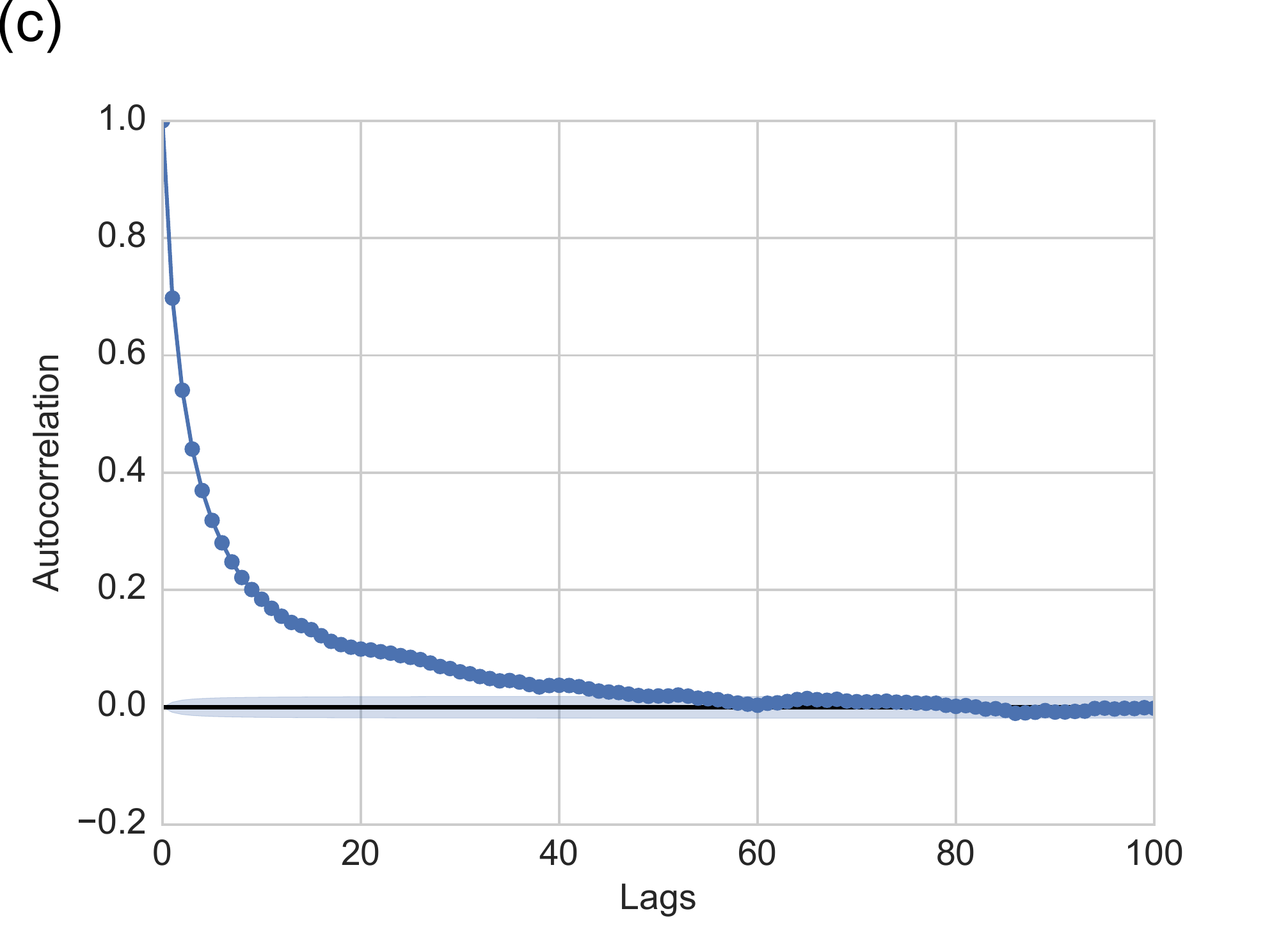}}
	\subfloat{\includegraphics[width=0.35\columnwidth]{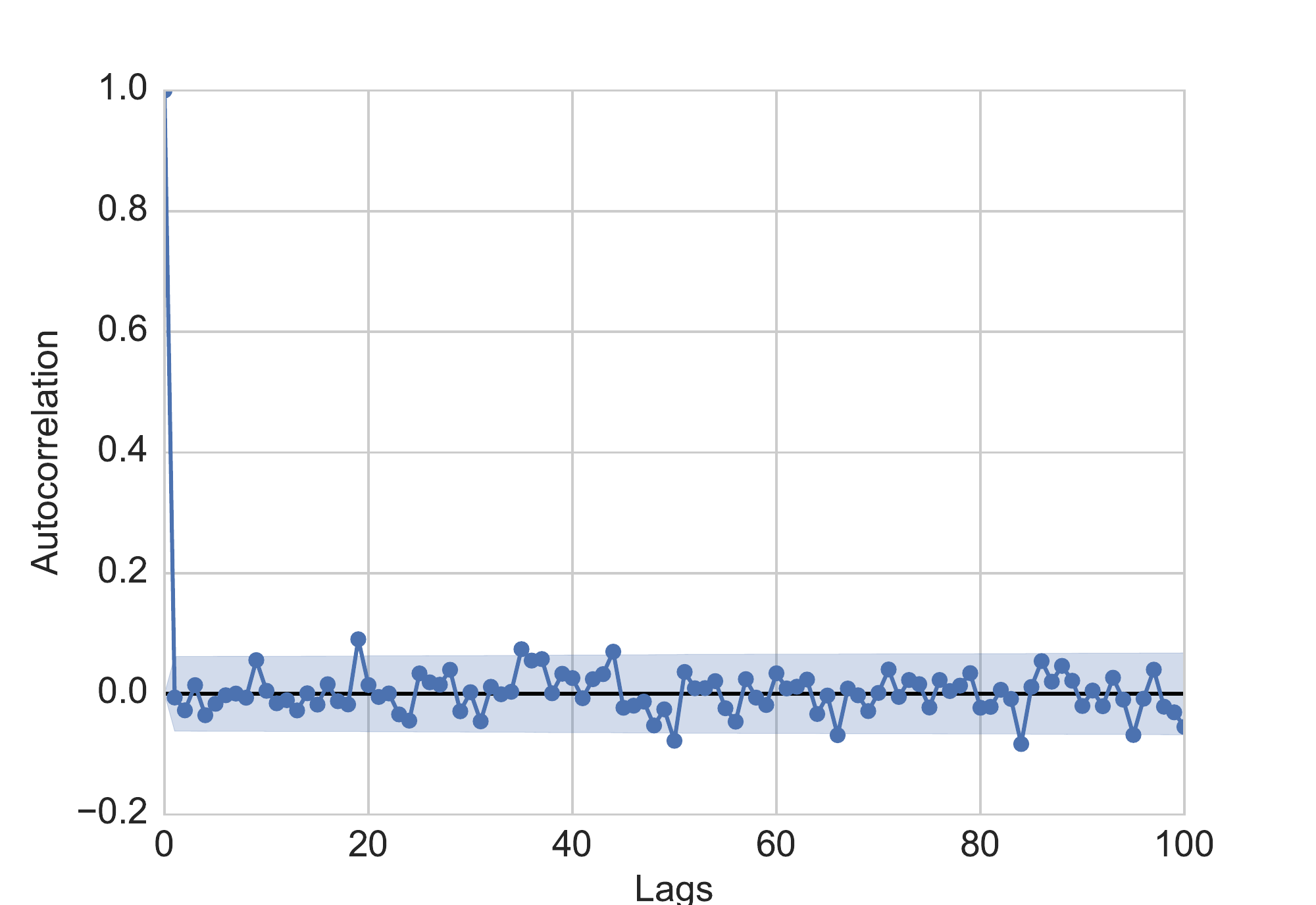}}
	\linebreak
	\subfloat{\includegraphics[width=0.35\columnwidth]{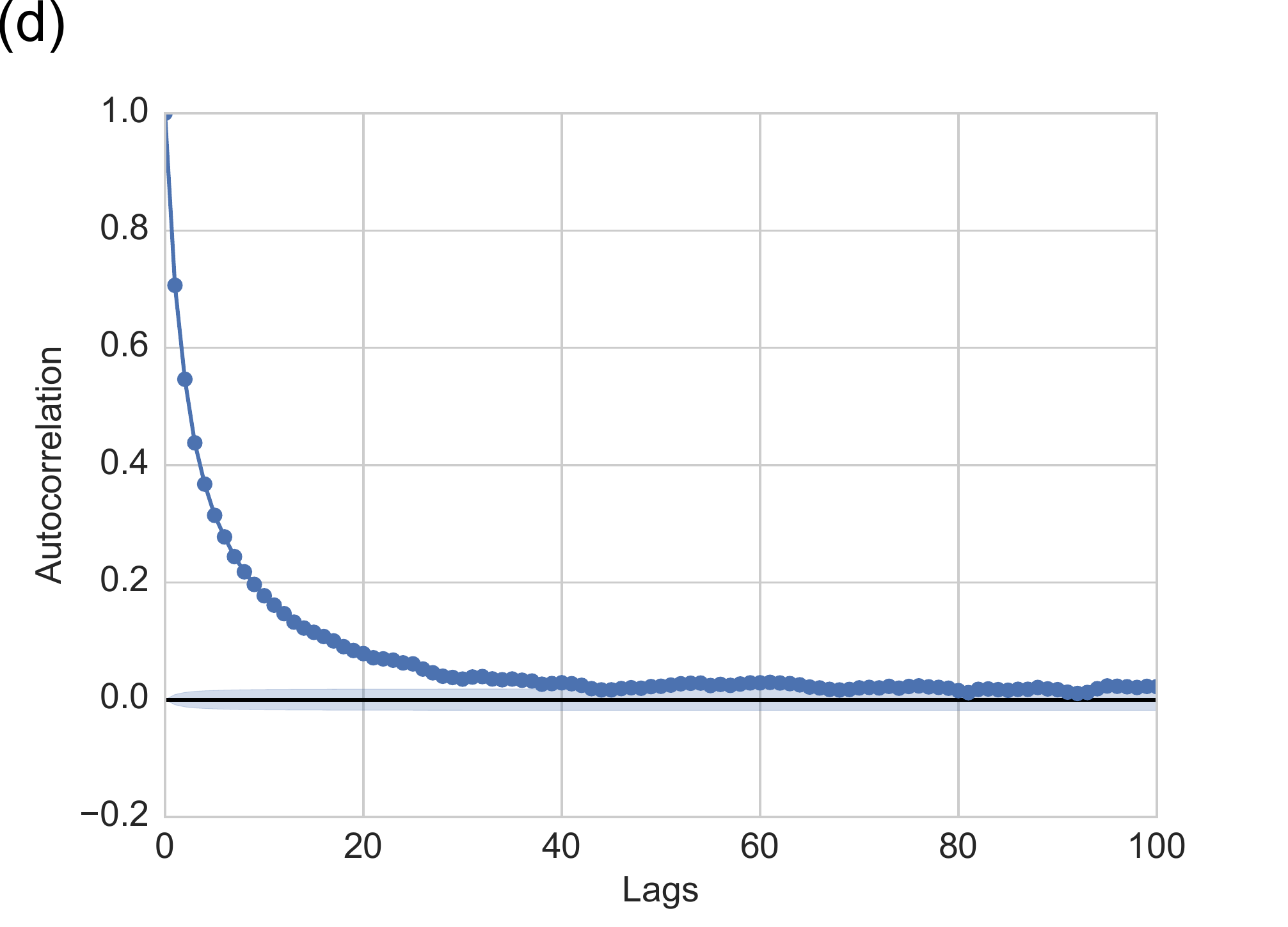}}
	\subfloat{\includegraphics[width=0.35\columnwidth]{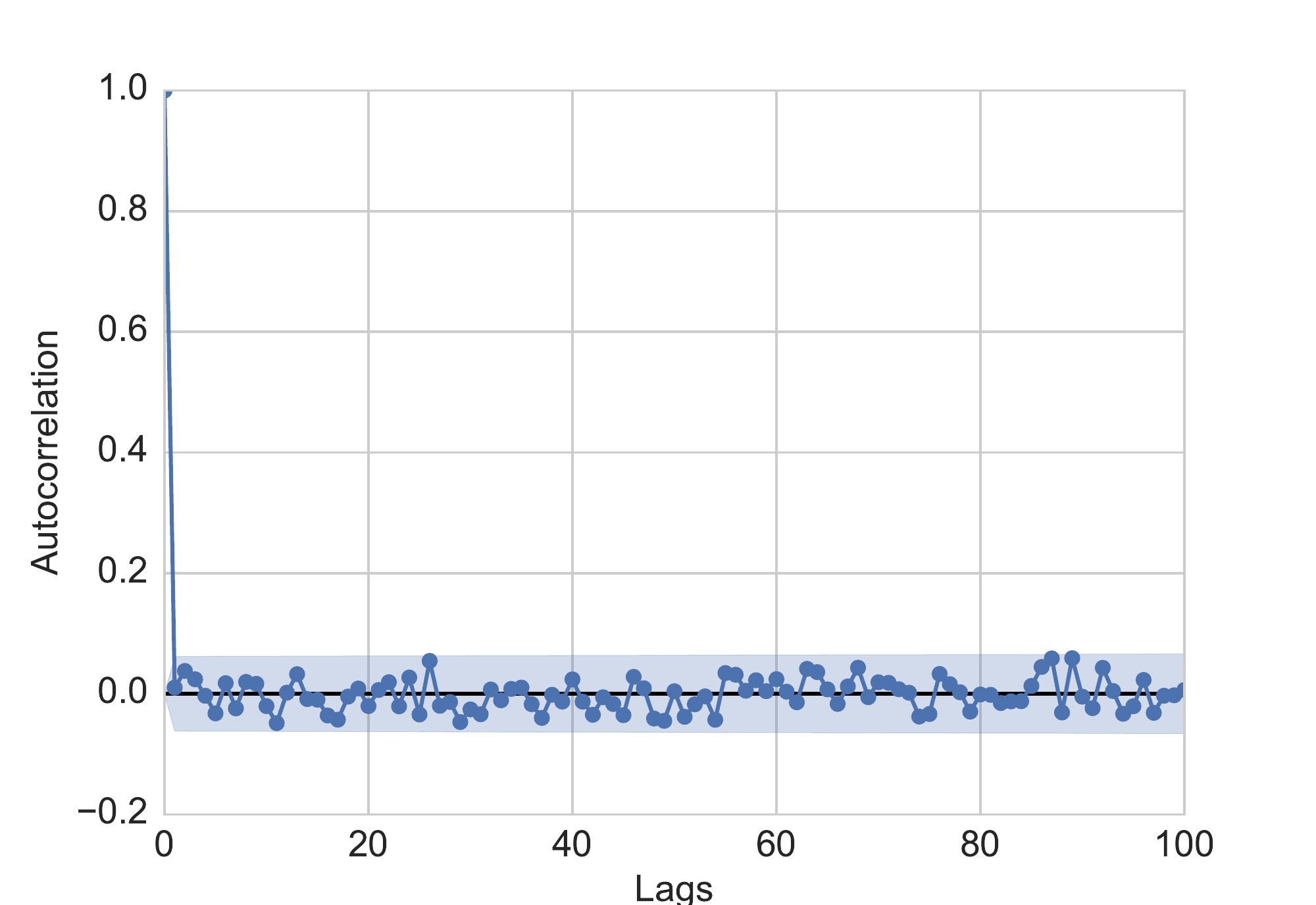}}
	\linebreak
	\subfloat{\includegraphics[width=0.35\columnwidth]{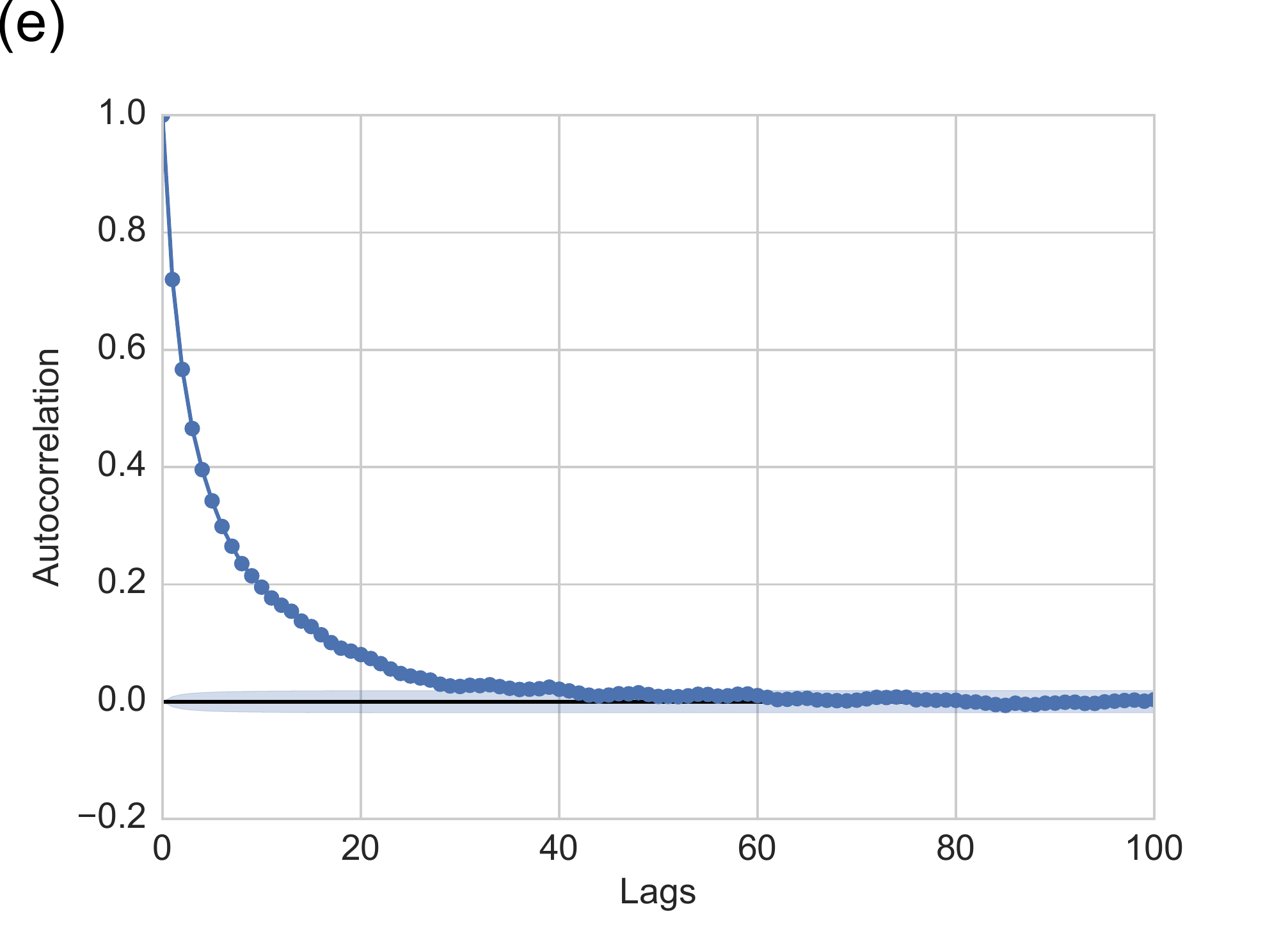}}
	\subfloat{\includegraphics[width=0.35\columnwidth]{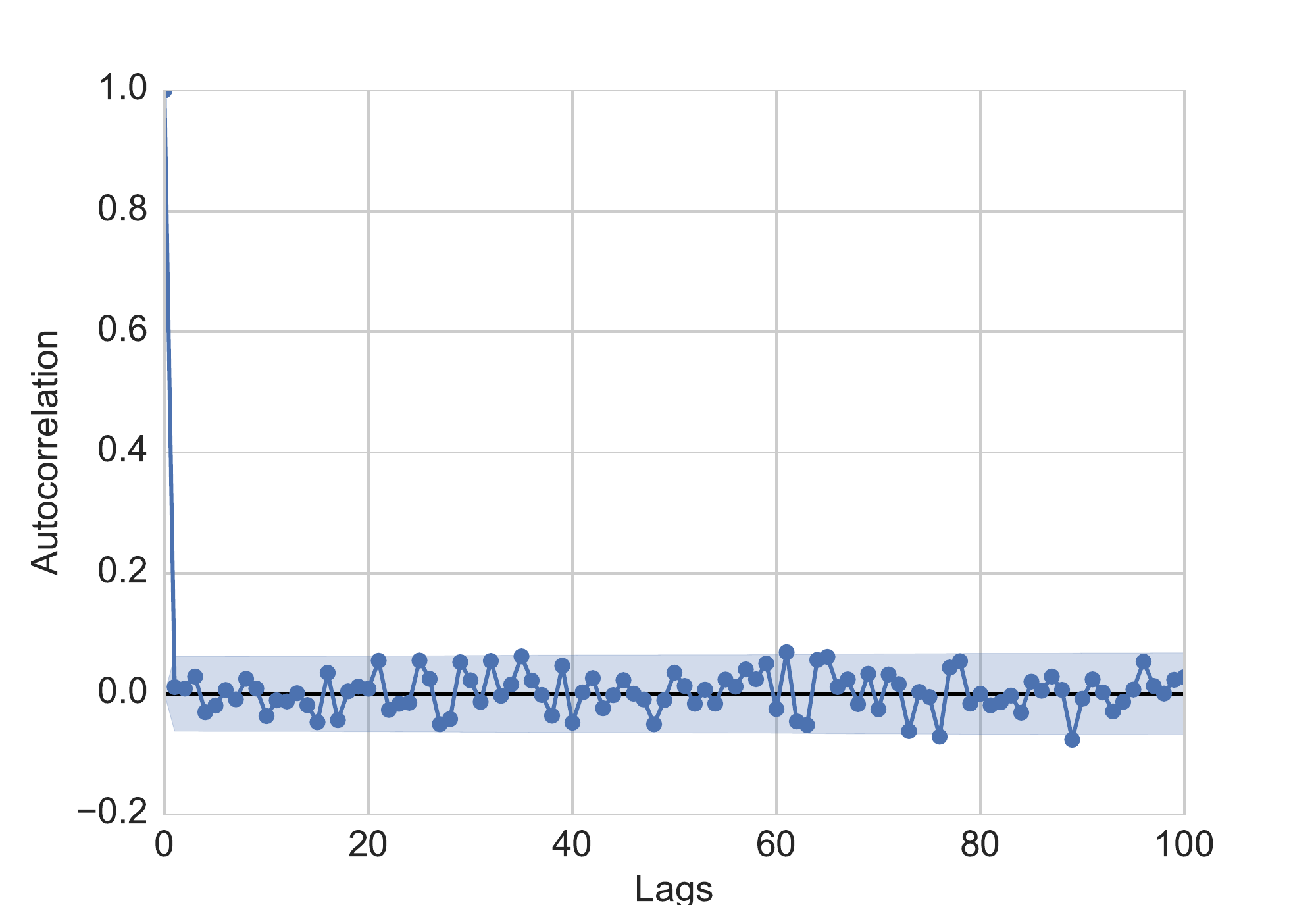}}
	\caption{Examples of the autocorrelation of $-\log(|\mathrm{Per}(A_S)|^2)$ present at lags up to 100 in Markov chains before (left) and after (right) applying our burn in and thinning procedures.
		The un-thinned sample size was 50000 (reduced from 100100 by taking the first 50000 tuples), and the thinned sample size was 1000.
		Shaded regions represent a $95\%$ confidence interval.
		(a): Data for a Haar-random instance of the \bs problem with $n=3$ and $m=9$.
		(b): Similarly for $n=7$ and $m=49$.
		(c): $n=12$ and $m=144$.
		(d): $n=20$ and $m=400$.
		(e): $n=25$ and $m=625$.
	}
	\label{fig:autocorrelation}
\end{figure}

How can we confirm that our samplers are sampling from an approximately correct distribution? 
Boson sampling is believed to be computationally hard to verify~\cite{Aaronson2011,Gogolin2013,Aaronson2013}, so, to date, experimental implementations of the problem have relied upon circumstantial evidence to support the claim that they are, in fact, sampling from the target distribution. 
This, coupled with the fact that reliability of MCMC methods is notoriously difficult to certify, requires us to proceed with caution.

However, the fact that we have 3 completely different samplers, each of which we can ask to provide us with a sample for the same instance of the problem, allows us to gather mutually supportive evidence for them all sampling from the correct target distribution; if we cannot reject that all pairs of samples come from the same underlying distribution, we can have some confidence that they are \emph{all} in fact sampling from the target distribution.

A natural quantity to use to compare the distributions on matrices that we consider is $|\mathrm{Per}(A_S)|^2$ of the sampled matrix $A_S$, which is equal to the probability of the sampled matrix in the \bs distribution; for convenience we actually use $\log |\mathrm{Per}(A_S)|^2$.

We generated samples of size 20000 of tuples (assigning each of them an index of an integer value between 0 and $\binom{m}{n}-1$), together with their corresponding $-\log(|\mathrm{Per}(A_S)|^2)$, using various samplers and for problems of size $n=3, 7, 12, 20$. 
The data are plotted in fig.\ \ref{fig:sample-distributions}.

We expect that the distribution of $-\log(|\mathrm{Per}(A_S)|^2)$ differs between samples of indistinguishable bosons and distinguishable particles.
This feature manifests itself clearly for each attempt and each method of attempted \bs; the distinguishable particle values (yellow) are distributed visibly differently to those of other sampling methods (all other colours).
More rigorously, we performed 2-sample bootstrap Kolmogorov-Smirnov (KS) tests~\cite{Praestgaard1995,Sekhon2011} comparing the distribution of $-\log(|\mathrm{Per}(A_S)|^2)$ values 
for tuples obtained by sampling from the distinguishable particle distribution with those obtained using our classical \bs methods.
The result was that we could reject the null hypothesis that this feature of the samples was distributed identically at the 0.1\% significance level (i.e.\ $p<0.001$ for all samplers and all problem sizes considered).
This result was found to be consistent over 5 repetitions of taking a sample for each input unitary.

Not only do we see that our sampling methods give significantly differently distributed $-\log(|\mathrm{Per}(A_S)|^2)$ to the distinguishable particle sampler, we see \emph{no} significant difference between these distributions for each of our sampling methods.
Applying the same 2-sample bootstrap KS tests to compare these distributions returned that we cannot reject the null hypothesis that they are distributed identically at any reasonable significance level.
We see this as evidence that all of our samplers are sampling from the target \bs distribution.

Due to some sampling methods being more efficient than others, we witness problem sizes beyond which certain methods become unreasonably slow. 
As a rough guide, on a personal computer these problem sizes are $n=7$ for brute force sampling and $n=12$ for rejection sampling.
Thus, in the case of $n=20$, we compare one MIS sampler (with $\tau_{\mathrm{burn}}=\tau_{\mathrm{thin}}=100$) with another MIS sampler (with $\tau_{\mathrm{burn}}=\tau_{\mathrm{thin}}=1000$).

In table\ \ref{table:p-values} we present the $p$-values obtained when performing 2-sample bootstrap Kolmogorov-Smirnov tests on the values $-\log(|\mathrm{Per}(A_S)|^2)$ for sets of $20000$ tuples obtained with various samplers.
Independent samples were generated $5$ times in order to repeat the tests.
The null hypothesis in this test is that the values are identically distributed.
As $p$-values are uniformly distributed when the null hypothesis is true, we see no significance in any relatively small values in table \ref{table:p-values}; they are to be expected when quoting a large number of $p$-values.

\begin{table}[h]
	
	\begin{tabular}{|l|l|l|l|l|l|l|l|}
		\hline
		n & Sampler 1 & Sampler 2 & \multicolumn{5}{|c|}{2-sample KS test $p$-values} \\
		\hline
		3 & MIS & na\"ive brute-force & 0.651 & 0.34 & 0.653 & 0.368 & 0.089 \\
		& MIS & rejection & 0.209 & 0.415 & 0.679 & 0.053 & 0.167 \\
		& rejection & na\"ive brute-force & 0.03 & 0.588 & 0.39 & 0.277 & 0.331 \\
		\hline
		7 & MIS & na\"ive brute-force & 0.448 & 0.606 & 0.858 & 0.734 & 0.468 \\
		& MIS & rejection & 0.628 & 0.615 & 0.246 & 0.581 & 0.775 \\
		& rejection & na\"ive brute-force & 0.27 & 0.252 & 0.287 & 0.954 & 0.263 \\
		\hline
		12 & MIS & rejection & 0.628 & 0.919 & 0.998 & 0.77 & 0.519 \\
		\hline
		20 & MIS ($\tau=100$) & MIS ($\tau=1000$) & 0.548 & 0.258 & 0.832 & 0.483 & 0.585 \\
		\hline
	\end{tabular}
	\caption{$p$-value data for various problem sizes and sampler combinations.
		Here, $\tau$ is shorthand for both $\tau_{\mathrm{burn}}$ and $\tau_{\mathrm{thin}}$.}
	\label{table:p-values}
\end{table}

Note that the assessment of MCMC algorithms is non-trivial: in particular, slow convergence to the target distribution and autocorrelation within the chain can result in an erroneous sample being output.
We therefore also performed a modified version of the likelihood ratio test described by Bentivegna et al.~\cite{Bentivegna2014}. Following the notation of~\cite{Bentivegna2014}, the test works as follows. We define two hypotheses: $\mathcal{Q}$, the indistinguishable boson hypothesis, and $\mathcal{R}$, an alternative hypothesis.
In our case, the alternative hypothesis $\mathcal{R}$ is the distinguishable particle hypothesis.
Let $q_x$ be the probability of seeing the sampled event $x$ according to hypothesis $\mathcal{Q}$, and $r_x$ be the corresponding probability under hypothesis $\mathcal{R}$. 
Assigning equal priors to each hypothesis, Bayes' theorem tells us that
\begin{equation}\label{bayes-lr}
\frac{P\left(\mathcal{Q}|N_{\mathrm{events}}\right)}{P\left(\mathcal{R}|N_{\mathrm{events}}\right)} = \prod_{x=1}^{N_{\mathrm{events}}} \left(\frac{q_x}{r_x}\right) = \mathcal{X},
\end{equation}
where, for example, $P\left(\mathcal{Q}|N_{\mathrm{events}}\right)$ is the probability of hypothesis $\mathcal{Q}$ being correct given that a sample of size $N_{\mathrm{events}}$ with events $\left\{k_x\right\}$ was obtained. 
A large value of $\mathcal{X}$ would correspond with us having a high degree of confidence that hypothesis $\mathcal{Q}$ provides a better description of the observed data than hypothesis $\mathcal{R}$. 

It is convenient to rewrite equation~\eqref{bayes-lr} in terms of the probability assigned to hypothesis $\mathcal{Q}$,
\begin{equation}\label{lr-pind}
P_{\mathrm{ind}} \equiv P\left(\mathcal{Q}|N_{\mathrm{events}}\right) = \frac{1}{\mathcal{X}+1} \prod_{x=1}^{N_{\mathrm{events}}} \left(\frac{q_x}{r_x}\right)
\end{equation}
which is now normalised such that $ P\left(\mathcal{Q}|N_{\mathrm{events}}\right) +  P\left(\mathcal{R}|N_{\mathrm{events}}\right) = 1$.

One, perhaps subtle, point here is that $q_x$ and $r_x$ do not simply correspond to $\left| \mathrm{Per}\left(A_x\right)\right|^2$ and $\mathrm{Per}\left(\left| A_x\right|^2\right)$ respectively. 
The reason for this is that we are restricted to the CFS, and the probability of a sample being collision-free differs between indistinguishable bosons and distinguishable particles.
So $q_x$ and $r_x$ must be normalised independently, such that they independently sum to 1 over all events $x$. Unfortunately, doing this exactly would require summing all probabilities in the CFS for the specific instance of the problem being considered -- a computationally daunting task, as previously discussed. 
To approximate the normalisation of $q_x$ efficiently, here we instead average the probability of the output being collision-free over the Haar measure, as reported in~\cite{Aaronson2011,Arkhipov2012}:
\begin{equation}\label{cfs-probability}
P_{\mathrm{CFS}} \approx \left.\binom{m}{n} \middle/ \binom{m+n-1}{n}\right. .
\end{equation}
For the hypothesis $\mathcal{R}$, we can efficiently sample output tuples, and so the ratio of collision-free tuples to tuples with collisions in a large sample provides an approximate normalisation for $r_x$.

For each problem size, we use this likelihood ratio test to assess the performance of MIS samplers with different $\tau_{\mathrm{burn}}$ and $\tau_{\mathrm{thin}}$. 
The reasoning is that, as the proposal distribution for the sampler is the distinguishable particle distribution, we might expect that if the chain has not converged to the target distribution, this will manifest itself as samples looking more like they are from the distinguishable particle distribution than they should.
Also, we expect that samplers with larger $\tau_{\mathrm{burn}}$ and $\tau_{\mathrm{thin}}$ are more likely to sample from the target distribution. 
Because of this, we expect to be able to observe a point at which increasing $\tau_{\mathrm{burn}}$ and $\tau_{\mathrm{thin}}$ has, on average, no effect on the outcome of a likelihood ratio test between hypotheses $\mathcal{Q}$ and $\mathcal{R}$.

The data corresponding to this for $n=3,7,12,20$ are presented in fig.\ \ref{fig:main-burnin-thin}. We can see that, indeed, for a given problem size, increasing $\tau_{\mathrm{burn}}$ and $\tau_{\mathrm{thin}}$ beyond a point does not improve the likelihood ratio test results. Not only this, but we observe that this point gradually increases in size with the problem size $n$. 
For example, with problem size $n=3$, $\tau_{\mathrm{burn}} = \tau_{\mathrm{thin}} = 5$ provides results comparable with $\tau_{\mathrm{burn}} = \tau_{\mathrm{thin}} = 1000$. 
However, with problem size $n=20$, $\tau_{\mathrm{burn}} = \tau_{\mathrm{thin}} = 5$ provides markedly worse samples on average than $\tau_{\mathrm{burn}} = \tau_{\mathrm{thin}} = 1000$. 
In fact, $\tau_{\mathrm{burn}} = \tau_{\mathrm{thin}} = 100$ seems to be the point at which increasing these values provides no benefit.

Additionally, in fig.\ \ref{fig:autocorrelation} we see that autocorrelation between $-\log(|\mathrm{Per}(A_S)|^2)$ for 100 lags in an un-thinned chain is negligible for random instances of the problem for $n=3,7,12,20,25$.
This is further supported by the autocorrelation being negligible at \emph{all} lags for chains generated by applying our thinning procedure to the un-thinned chains.

Based on this, we claim that $\tau_{\mathrm{burn}} = \tau_{\mathrm{thin}} = 100$ is sufficient for MIS for problems up to size $n=20$, and is likely to be sufficient for values of $n$ greater than 20. 
The consequence of this is that a sample from the \bs distribution can be output, using a classical computer, in roughly the time that it takes to compute 100 $n \times n$ real valued matrix permanents and 100 $n \times n$ complex valued matrix permanents. 
For $n=20$ and $m=400$, this equates to a computational saving of a factor of $\sim 10^{31}$ when compared with brute force sampling.


\subsection{Moving to larger $n$}

Although MIS is efficient in the number of matrix permanent computations required, it still requires the inefficient computation of matrix permanents by Ryser's algorithm~\cite{ryser63}.
As such, gathering a $20000$ tuple sample as was obtained in fig.\ \ref{fig:sample-distributions} is challenging for $n>20$ using readily available hardware.
We were able to perform \bs using MIS for $n=25$ and $m=625$, with a sample size of $1000$ tuples for plotting the distributions of indexed tuples and $-\log(|\mathrm{Per}(A_S)|^2)$ of these sampled tuples, alongside a sample from the distinguishable particle distribution.
We also performed a likelihood-ratio test to compute the probability of the indistinguishable boson hypothesis against the distinguishable particle hypothesis, averaged over 100 samples of 25 tuples for MIS with different values of $\tau$.

In addition to this, we performed boson sampling using MIS for $n=30$ and $m=900$, obtaining a $250$ tuple sample with $\tau = 100$, and compared the convergence of $P_\mathrm{ind}$ with other values of $\tau$.

\begin{figure}[tp]
	\captionsetup[subfloat]{farskip=0pt,captionskip=0pt}
	\centering
	\subfloat{\includegraphics[width=0.5\columnwidth]{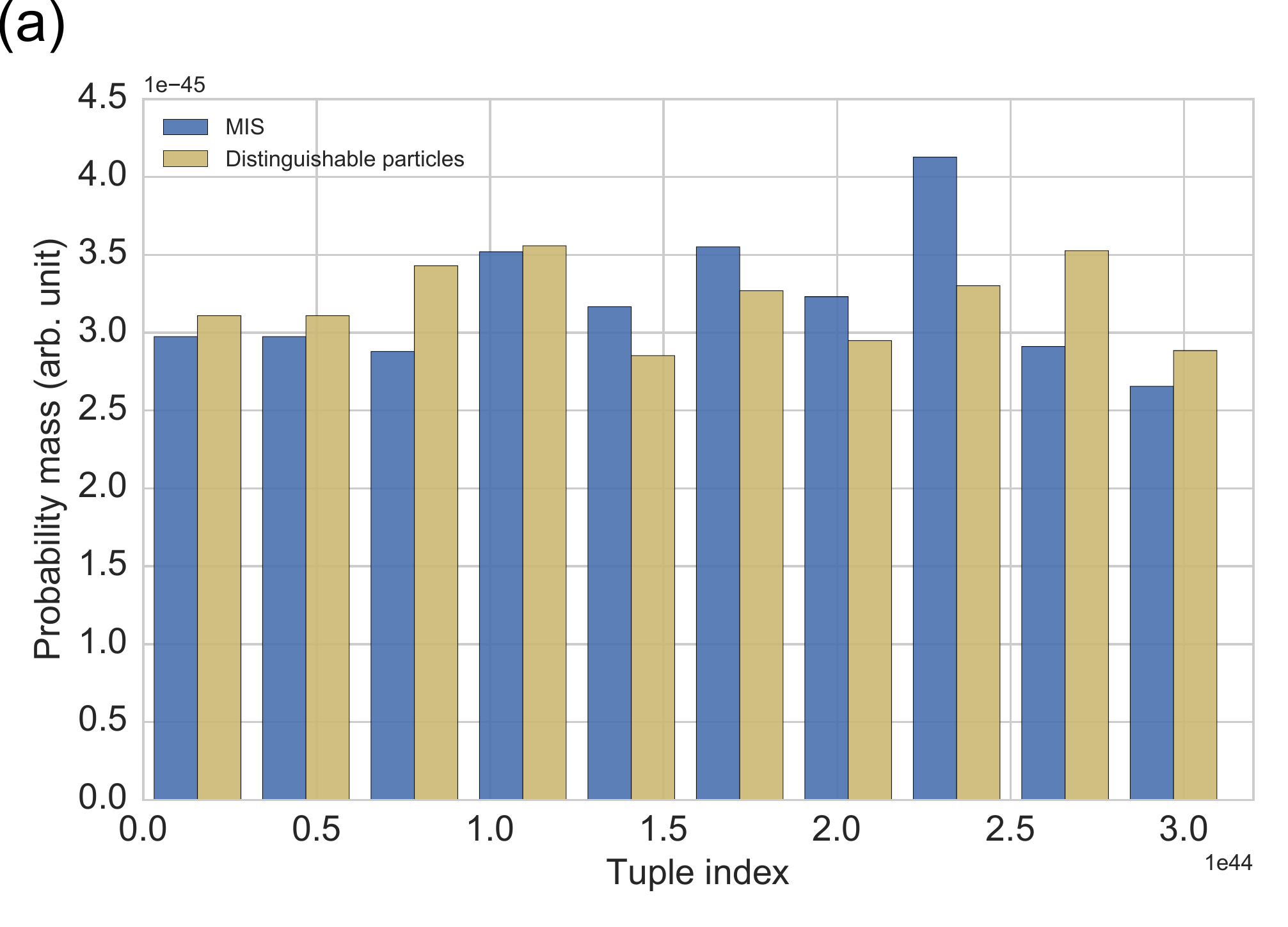}}
	\subfloat{\includegraphics[width=0.5\columnwidth]{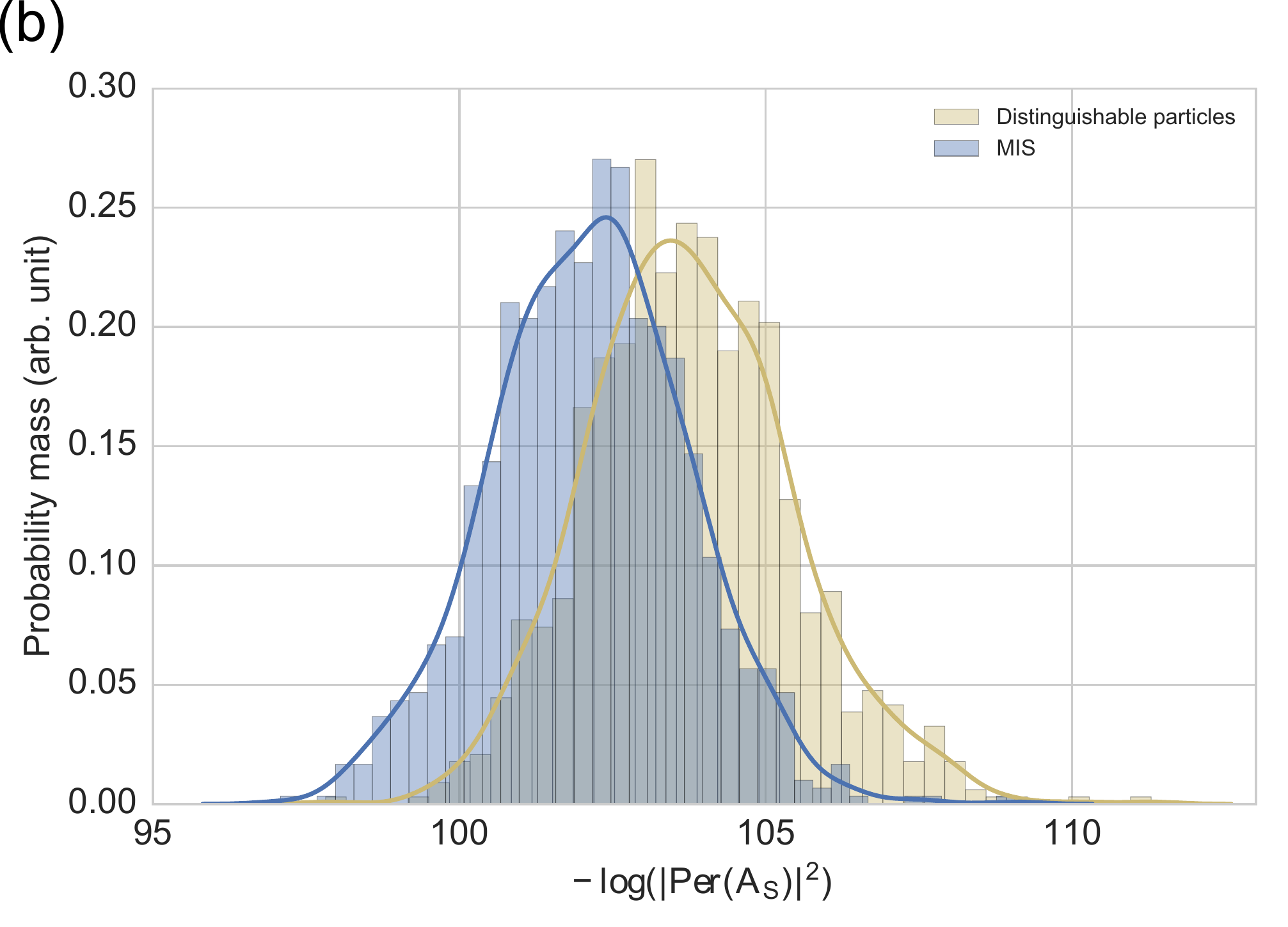}}
	\linebreak
	\subfloat{\includegraphics[width=0.5\columnwidth]{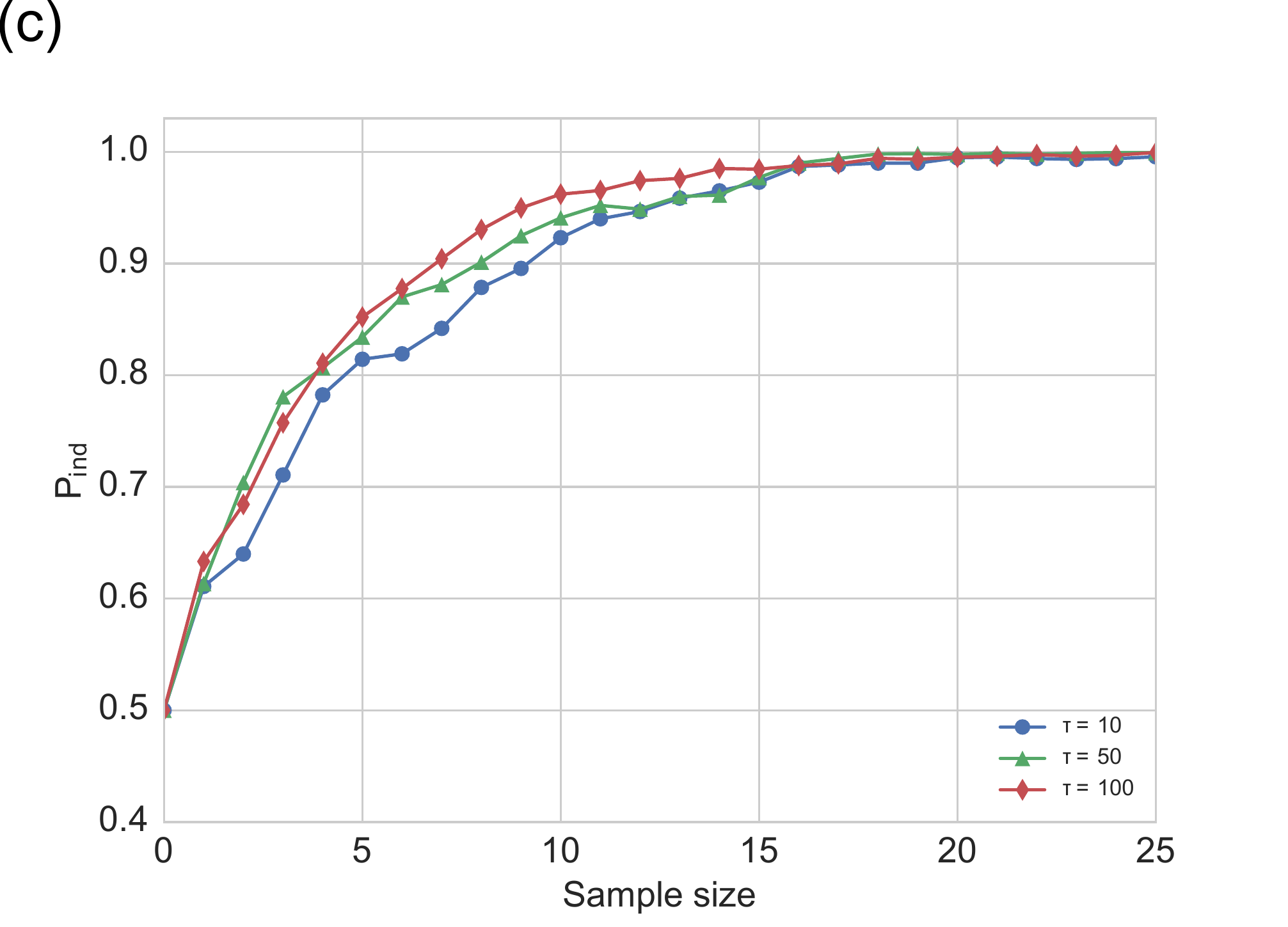}}
	\subfloat{\includegraphics[width=0.5\columnwidth]{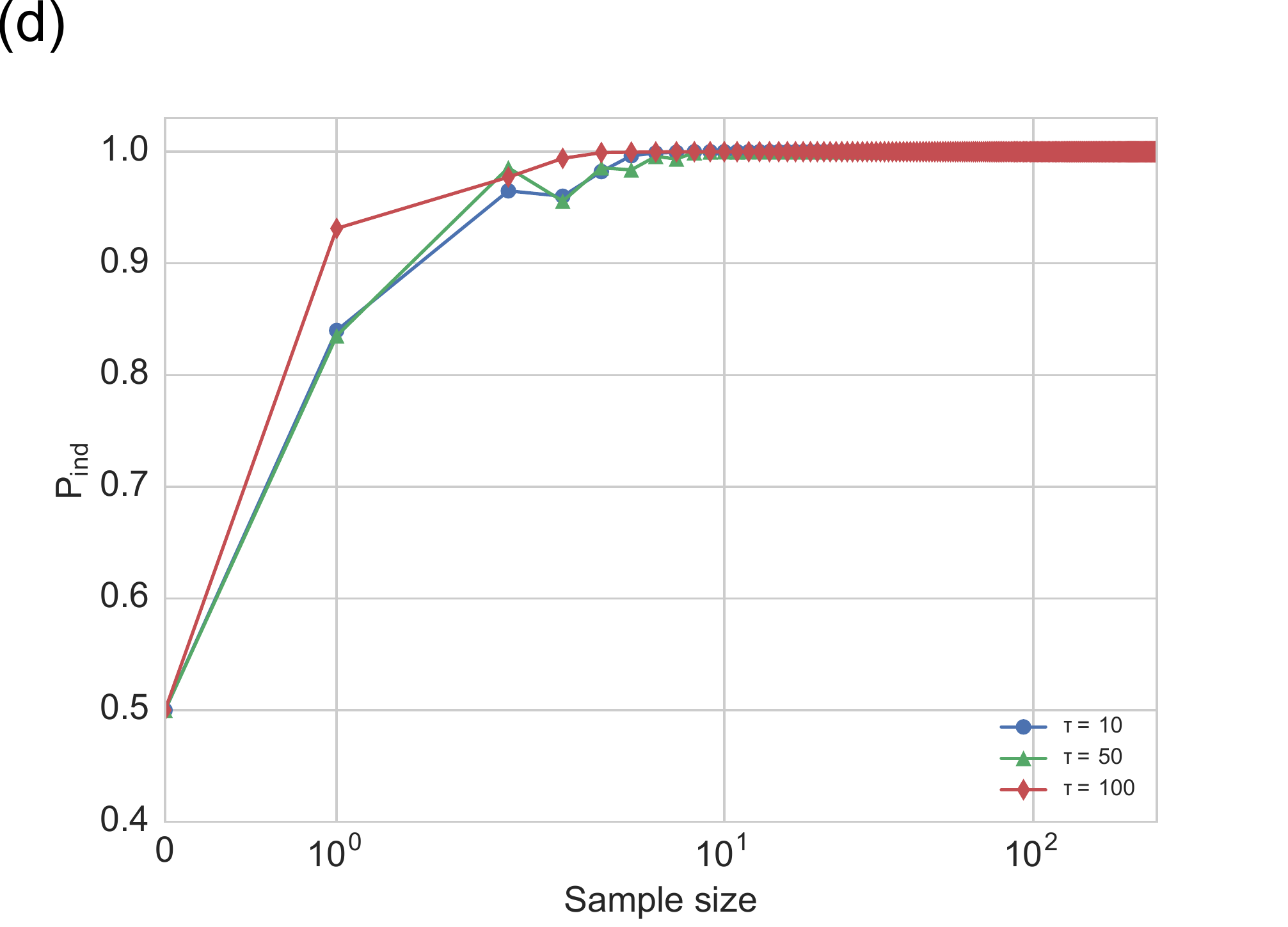}}
	\caption{(a): Distribution of 1000 indexed and binned tuples obtained using MIS (blue) and a sampler sampling from the distinguishable particle distribution (yellow) for a Haar-random instance of the \bs problem with $n=25$ and $m=625$.
		(b): Distribution of $-\log(|\mathrm{Per}(A_S)|^2)$ for each tuple $S$ sampled.
		Solid lines are obtained via kernel density estimation, and are included as a visual aid.
		(c): Results of likelihood ratio tests performed on samples generated using MIS for a Haar-random instance of the \bs problem with $n=25$ and $m=625$.
		(d): Results of a likelihood ratio test performed on a sample generated using MIS for a Haar-random instance of \bs with $n=30$ and $m=900$.}
	\label{fig:p25}
\end{figure}

All of the data for $n>20$ was taken on a cluster of 4 servers\footnote{\label{servers} 2 dual-socket Intel Xeon CPU E5-2697A v4 @ 2.60GHz and 2 dual-socket Intel Xeon CPU E5-2680 v3 @ 2.50GHz
} at the University of Bristol, allowing for $122$ MIS instances (and therefore $122$ Markov chains) run in parallel (vs a maximum of $4$ on our personal computer).
The data are presented in fig.\ \ref{fig:p25}.
For $n=25,30$, we notice no change in the performance of MIS beyond having to compute permanents of larger matrices.

\section{Comparison with experimental implementations}
\label{sec:comparison}

Limiting ourselves to the standard \bs problem for now, we will consider how the time to obtain a single tuple sample scales with experimental parameters for a physical \bs device.

Parameters which have an effect on the rate at which samples can be generated experimentally are: the $n$-photon generation rate $R(n)$, the single photon transmission probability $\eta$, and the Haar-averaged probability of the $n$-photon state being in the CFS given by eqn.\ (\ref{cfs-probability}), which we take to be an approximation of the relative size of the CFS for any \bs input $U$.

It is important to notice here that in fact $\eta$ depends on $n$.
This is because the loss is composed of fixed loss $1-\eta_\mathrm{f}$ (source efficiency, detection loss, insertion loss etc.) and loss within the circuit, $1-\eta_\mathrm{c}$.
The circuit depth -- and therefore the amount of material (or number of components) each photon must propagate through -- necessarily grows with the size of the problem, as non-trivial interference between all photons and across all modes must occur.

Circuits generally considered for \bs have depth $d$ which scales at least linearly with $m$ \cite{Reck1994, Clements2016} and hence at least quadratically with $n$.
However, Aaronson and Arkhipov have shown that \bs can be implemented using a different circuit construction with depth $d = \mathcal{O}(n \log m)$~\cite{Aaronson2011} if the input modes are fixed, and recently a circuit and predictions of performance with a linear mode scaling of $m=4n$ have been reported~\cite{Wang2016}.

This might suggest that $d=\mathcal{O}(n)$ could also be sufficient, although note that when $m=\mathcal{O}(n)$ the probability of a sample being collision-free is exponentially small in $n$ (see eqn.\ (\ref{cfs-probability})), and no theoretical argument for computational hardness is known.

Here we compare the rates with which we expect to obtain samples from a \bs experiment for the two regimes $d=n^2$, $d=4n$ where experimental attempts have been documented~\cite{Spring2013,Crespi2013,Tillmann2013,Broome2013,Carolan2014,spagnolo2014,Bentivegna15,carolan2015,He2016,Wang2016,Loredo2017}.
The transmission probability of each photon through the circuit is $\eta_\mathrm{c} = {\eta_0}^d$, where $\eta_0$ is the probability of a photon surviving per 2-mode coupling length (or component) in the interferometer.
Using these parameters, we can write the time in which we expect to obtain a single sample from a quantum device as a function of the problem size:
\begin{equation}\label{expt_rate}
q_t\left(n\right) = \left( R(n) P_\mathrm{CFS} \eta^n \right)^{-1}
\end{equation}
where $\eta = \eta_\mathrm{f} \eta_0^d$
Expanding $\eta$ for linear and quadratic mode scaling respectively, we see that
\begin{align*}
q_t^{\mathrm{lin}}\left(n\right) &= \left(R(n){\eta_\mathrm{f}}^n {\eta_0}^{4n^2}P_{\mathrm{CFS}}\right)^{-1}\\
&\sim \left(R(n) \left(\frac{4\eta_\mathrm{f}}{5}\right)^n {\eta_0}^{4n^2}\right)^{-1}\\
q_t^{\mathrm{quad}}\left(n\right) &= \left(R(n) {\eta_\mathrm{f}}^n {\eta_0}^{n^3} P_{\mathrm{CFS}}\right)^{-1}\\
&\sim \left(R(n) {\eta_\mathrm{f}}^n \left(\frac{1}{e}\right){\eta_0}^{n^3}\right)^{-1}
\end{align*}
where it should be emphasised that $\eta_{\mathrm{f}}, \eta_0 < 1$. 

Using these bounds, we see that the time required to obtain a sample scales exponentially in either $n^2$ or $n^3$.
However, classical computation of the permanent using Ryser's algorithm scales only exponentially in $n$.
Therefore, for very large $n$ and realistic loss parameters, the runtime scaling of a classical sampling algorithm would be favourable compared with the \bs experiment, even if the algorithm needed to compute exponentially many permanents to obtain one sample.

\begin{figure}[tp]
	\centering
	\captionsetup[subfloat]{farskip=0pt,captionskip=0pt}
	\subfloat{\includegraphics[width=0.5\linewidth]{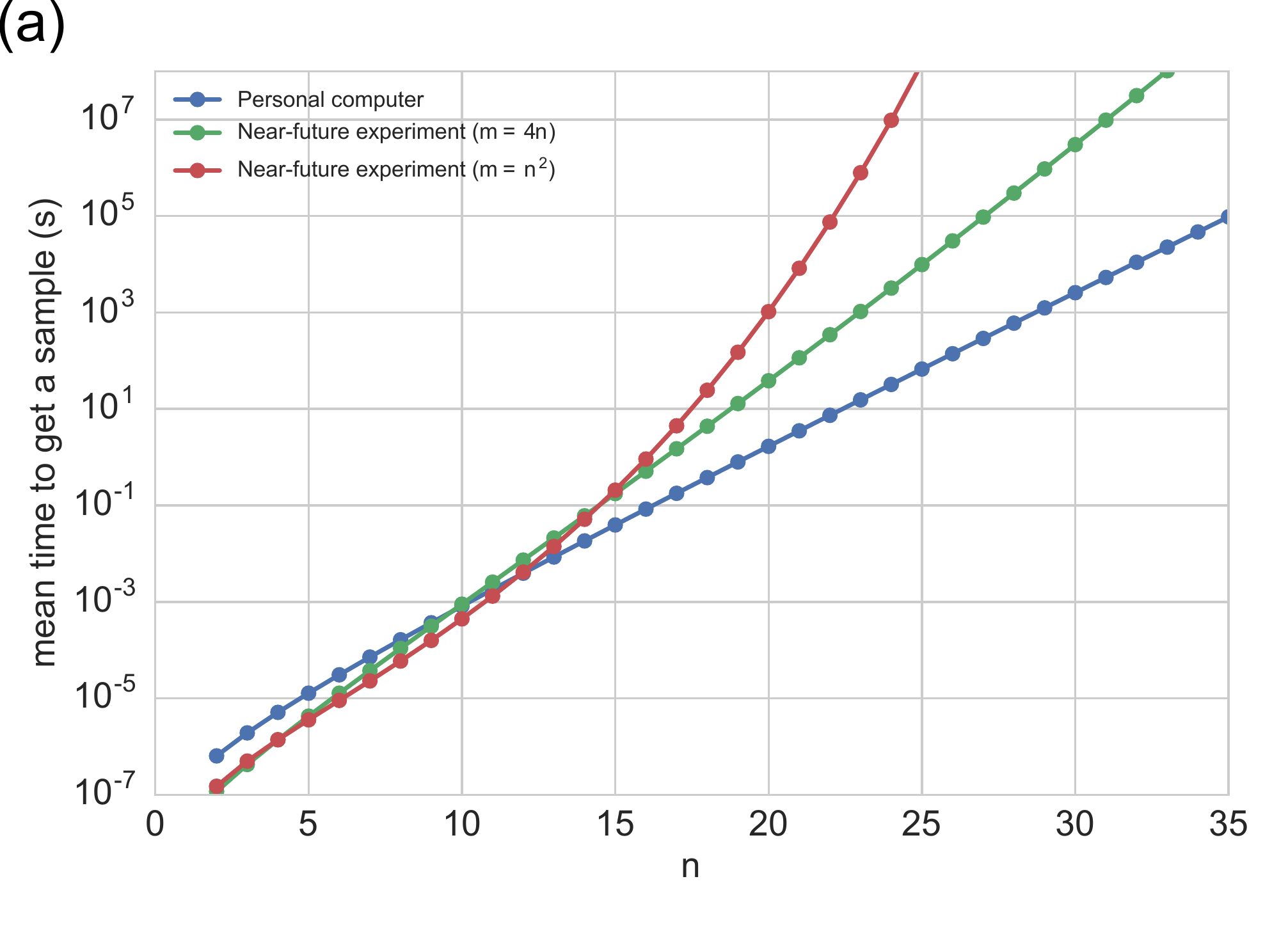}}
	\subfloat{\includegraphics[width=0.5\linewidth]{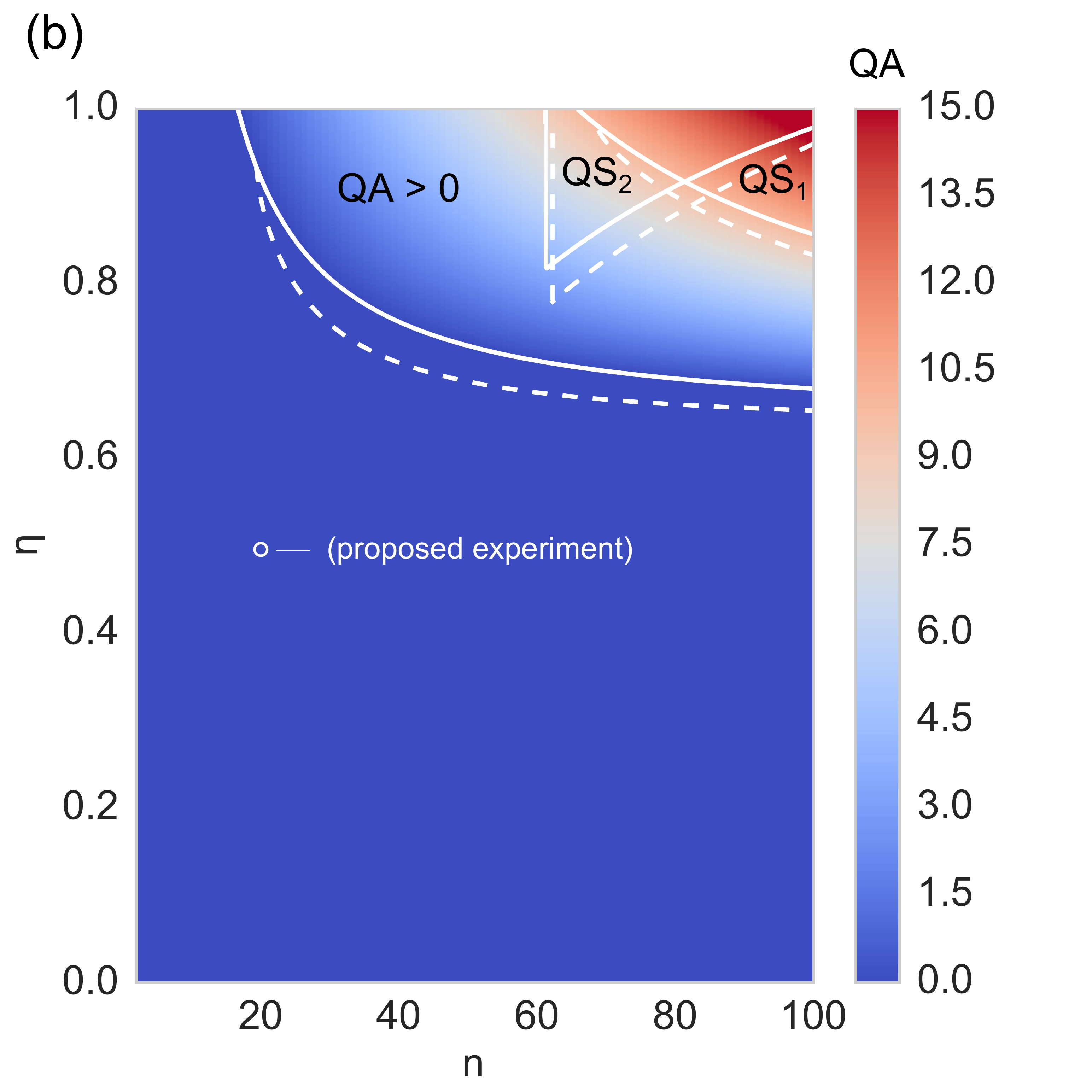}}
	\caption{(a): Comparison of average time to output a single tuple sample from the \bs distribution experimentally (red: $m=n^2$, green: $m=4n$) and using a personal computer running MIS (blue).
		Experimental points were estimated using projected future loss parameters and extrapolated from values stated in~\cite{Wang2016}.
		Exact parameter values are stated in Sec.\ \ref{sec:comparison}.
		(b): Quantum advantage, QA, as a function of $n$ and $\eta$ assuming the classical time scaling of a supercomputer and an experimental rate $R(n) = 76n^{-1}$MHz. Lines separate the regions of no quantum advantage, positive quantum advantage and quantum supremacy (as measured by criteria QS$_1$ or QS$_2$ or both). Dashed lines demonstrate adjusted regions when up to $2$ photons can be lost (optimised to maximise QA). The empty circle represents a proposed future experiment in ref.\ \cite{Wang2016}}
	
	\label{fig:standard-comparison}
\end{figure}

In order to provide a meaningful comparison of our classical \bs methods to an experimental implementation, we focus on the current leading experimental approach by Wang et al.~\cite{Wang2016}.
This approach involves the use of a quantum dot based photon source, and has resulted in increased photon numbers and significantly higher rates compared to the current best SPDC based demonstrations (see Extended Data Table 1 in~\cite{Wang2016}).
In fig.\ \ref{fig:standard-comparison} (also see fig.2d in the main text), we consider the rate at which a similar experiment with projected near-future parameters~\cite{Wang2016} can generate sample values with both $m=4n$ and $m=n^2$, and compare this to the average sample value generation rate for a personal computer running the MIS algorithm for \bs.

Specifically, in~\cite{Wang2016} Wang et al. report a quantum dot based photon source with $R(n) = 76n^{-1}$MHz.
Also reported are loss parameter values predicted to be achievable in the near future:
\[ \eta_\mathrm{f} = \eta_\mathrm{QD} \eta_\mathrm{de} \eta_\mathrm{det}, \]
where $\eta_\mathrm{QD}$ is the single photon source end-user brightness ($=0.74$), $\eta_\mathrm{de}$ is the demultiplexing efficiency for each channel ($=0.845$) and $\eta_\mathrm{det}$ is the efficiency of each detector ($=0.95$).
The authors of~\cite{Wang2016} report the efficiency of their 9 mode interferometer to be $\sim99\%$, and so we approximate the efficiency of a similar interferometer scaled to $m$ modes as $\eta_\mathrm{c} = 0.99^\frac{m}{9}$ and hence $\eta_0 = 0.99^{1/9}$ as we are assuming $d=m$.

We see that, although the personal computer is outperformed for a small problem size, beyond $n\approx14$ the personal computer outperforms the experimental boson sampler. 
Clearly, the point at which the classical device becomes dominant would increase if the experiment had access to an $n$-photon state generator with greater repetition rate.
On the other hand, if the classical device were more powerful (i.e.\ could perform more floating point operations per second), it would start to become dominant at even smaller problem sizes.
More important is that the rate falls off noticeably more rapidly for the experimental sampler with these loss parameters, for both regimes of scaling the number of modes with the number of photons.

To emphasise this point, we perform the following analysis.
Assuming that our MIS sampler continues to perform equally well for larger instance sizes,
we can compare its runtime with current and future experiments.
The classical runtime for an instance of size $n$ bosons in $m=n^2$ modes can be estimated as
\[
c_t(n) = 3\times10^{-13} n^2 2^n 
\]
where $3\times10^{-15}s$ is the time scaling for computing one real and one complex permanent recently reported for the supercomputer \emph{Tianhe 2}~\cite{wu16}.
We note here that we do not assume that we can use the efficient (i.e Gray code ordered) version of Ryser's algorithm here, as we would need to parallelise the computation of the permanent to deal with large $n$ problems.
We define the \emph{quantum advantage} (QA) as the order of magnitude improvement in quantum runtime versus classical runtime,
\begin{equation}
\text{QA}(n,\eta) = \max\Big[0, \log_{10}\Big(\frac{c_t}{q_t}\Big)\Big].
\end{equation}

We now consider two possible definitions of quantum supremacy.
First, we can define supremacy as
a speedup so large that it is unlikely to be overcome
by algorithmic or hardware improvements to the classical sampler,
for which we choose a speedup of ten orders of magnitude.
Secondly, we may wish to define supremacy as the point at which
a computational task is performed in a
practical runtime on a quantum device, for which we choose under a week,
but in an impractical runtime on a classical device, for which we choose over a century.

These can be summarised as
\begin{align}
&\text{QS}_1: \text{QA} > 10 \\
&\text{QS}_2: q_t < 1 \text{week} , c_t > 100 \text{yrs}.
\end{align}
For a quadratic mode scaling and a photon source with rate $R(n) = 10$GHz (beyond the capabilities of the current best single photon sources), we plot QA against $n$ and $\eta$ in fig.1 of the main text, where $q_t^\mathrm{quad}$ is used.
In order to achieve both quantum supremacy criteria in this regime, we see that it is required to have $\eta > 0.6$ and $n>60$.
We can perform the same analysis with $q_t^\mathrm{lin}$ and $R(n) = 76n^{-1}$MHz as per~\cite{Wang2016}.
The data for this are plotted in fig.\ \ref{fig:standard-comparison}.
Although using a linear mode scaling increases $\eta$, we see that the requirements for quantum supremacy become even more stringent.
Indeed, in order to achieve both quantum supremacy criteria here we require $\eta > 0.9$ and $n>70$.

\section{Lossy \bs}
\label{sec:loss}

We have seen that loss can cause significant difficulties in the setting of the standard \bs problem, leading to a classical sampler outperforming the quantum experiment.
In this section, we explore whether modifying the problem to build loss in from the start could mitigate these issues.
One might expect that losing photons would also make the \bs problem easier classically, because the matrices whose permanents are required to be computed would be smaller.
However, it is conceivable that the quantum advantage would increase, enabling a demonstration of quantum supremacy.

A lossy variant of \bs was introduced by Aaronson and Brod~\cite{Aaronson16}.
In this variant, we assume that $n-k$ photons are lost before entering the circuit enacting the linear optical transfer matrix, so $k$ photons remain.
Probabilities in this setting are not determined directly by $\left| \mathrm{Per}\left( A_S\right) \right| ^{2}$ for some $m$-tuple $S$, but by the average of this quantity over all possible ways of losing $n-k$ photons from $n$ photons:
\begin{equation}\label{temperament}
\Pr(S) = \frac{1}{\left|\Lambda_{n,k}\right|}\sum_{T \in \Lambda_{n,k}}\left| \mathrm{Per}\left( A_{S,T}\right) \right| ^{2}
\end{equation}
where $\Lambda_{n,k}$ is the set of $k$-subsets of $\{1,\dots,n\}$ and $A_{S,T}$ is the $k \times k$ submatrix of $A$ obtained by taking columns of $A$ according to $T$ and rows of $A$ according to $S$, which remains a subset of $\{1,\dots,m\}$.
Note that once again we restrict to the collision-free subspace, making the assumption that the probability of a collision is low enough that this does not significantly affect the probabilities.

It was shown by Aaronson and Brod~\cite{Aaronson16} that, if at most a constant number of photons in total are lost before entering the circuit, the lossy \bs problem remains hard, under the same assumptions as the original \bs problem.
However, it was left as an open problem whether the same holds true in the more realistic setting where a constant fraction of photons are lost.
Another open problem in~\cite{Aaronson16} was to generalise the loss model to include loss within and after the linear optical circuit.
In~\cite{Berry2010} it is shown that uniform (i.e. mode-independent) loss channels can commute with linear optics.
Here we prove the slightly stronger result, that as long as the overall transfer matrix is proportional to a unitary, loss can always be considered at the input even if the physical loss channels, wherever they occur, are not uniform (i.e are mode-dependent).


Consider a boson sampling device consisting of an ideal unitary linear optical transformation $U$ on a set of $m$ optical modes which is preceded or succeeded by path-independent loss. This loss can be modelled by considering a set of $m$ additional virtual ancilla modes such that the optical transfer matrix on all $2m$ modes remains unitary. A uniform transmission probability of $\eta$ can then be described by beamsplitters coupling each mode to its corresponding ancilla, resulting in the transfer matrix

\begin{equation}
L = \begin{pmatrix}
\sqrt{\eta}  & \sqrt{1-\eta} \\
\sqrt{1-\eta}& -\sqrt{\eta} 
\end{pmatrix} \otimes \mathbb{1}_m
\end{equation}
and so including the interferometer, the full transfer matrices for input and output losses are
\begin{align}
M_I &= (U \oplus \mathbb{1}_m)L =   \begin{pmatrix}
\sqrt{\eta}U & \sqrt{1-\eta} U \\
\sqrt{1-\eta} \mathbb{1} & -\sqrt{\eta}\mathbb{1}
\end{pmatrix} \\
M_O &= L(U \oplus \mathbb{1}_m) =   \begin{pmatrix}
\sqrt{\eta}U & \sqrt{1-\eta} \mathbb{1} \\
\sqrt{1-\eta} U & -\sqrt{\eta}\mathbb{1}
\end{pmatrix} .
\end{align}

Any m-mode optical state can be expressed in a coherent state basis~\cite{gerry2005}:
\begin{equation}
\rho_m = \int \lambda(\boldsymbol{\alpha},\boldsymbol{\beta}) |\boldsymbol{\alpha}\rangle\!\langle\boldsymbol{\beta}| d^m\boldsymbol{\alpha}d^m\boldsymbol{\beta}
\end{equation}
where $|\boldsymbol{\alpha}\rangle= |(\alpha_1,\alpha_2,...,\alpha_m)^T\rangle \equiv \bigotimes_{i=1}^m |\alpha_i\rangle$ is an $m$-mode coherent state and $\langle \boldsymbol{\beta}| = \langle (\beta_1,\beta_2,...\beta_m)^T| = \bigotimes_{i=1}^m\langle\beta_i|$. A coherent state evolves under a transfer matrix $T$ as
\begin{equation}
\mathcal{U}(T)|\boldsymbol{\alpha}\rangle = \bigotimes_{i=1}^m \left|\left. \sum_j T_{ij}  \alpha_j \right\rangle\right. = |T\boldsymbol{\alpha}\rangle
\end{equation}
It can then be shown that when the initial state contains vacuum in all ancilla modes, $\rho = \rho_m \otimes |\boldsymbol{0}\rangle\!\langle\boldsymbol{0}|_\text{an}$, the same
state is produced in the $m$ system modes under the transformations $M_O$ and $M_I$  
\begin{equation}
\begin{split}
\rho_O & =  \Tr_\text{an}\left[\mathcal{U}(M_O)\rho\mathcal{U}^\dagger(M_O)\right]\\
& = \Tr_{\text{an}}\left[ \int d^{2m}\boldsymbol{\alpha}d^{2m}\boldsymbol{\beta} \Big|\sqrt{\eta}U\boldsymbol{\alpha}\Big\rangle\!\Big\langle\sqrt{\eta}U\boldsymbol{\beta}\Big| \otimes \big|\sqrt{1-\eta}U\boldsymbol{\alpha}\big\rangle\!\big\langle\sqrt{1-\eta}U\boldsymbol{\beta}\big|\right] \\
& = \int d^{2m}\boldsymbol{\alpha}d^{2m}\boldsymbol{\beta} \Big|\sqrt{\eta}U\boldsymbol{\alpha}\Big\rangle\!\Big\langle\sqrt{\eta}U\boldsymbol{\beta}\Big| \Tr\left[ \big|\sqrt{1-\eta}U\boldsymbol{\alpha}\big\rangle\!\big\langle\sqrt{1-\eta}U\boldsymbol{\beta}\big|\right] \\
& =\int d^{2m}\boldsymbol{\alpha}d^{2m}\boldsymbol{\beta} \Big|\sqrt{\eta}U\boldsymbol{\alpha}\Big\rangle\!\Big\langle\sqrt{\eta}U\boldsymbol{\beta}\Big| \Tr\left[\mathcal{U}(U) \big|\sqrt{1-\eta}\boldsymbol{\alpha}\big\rangle\!\big\langle\sqrt{1-\eta}\boldsymbol{\beta}\big|\mathcal{U}^\dagger(U)\right] \\
& =\int d^{2m}\boldsymbol{\alpha}d^{2m}\boldsymbol{\beta} \Big|\sqrt{\eta}U\boldsymbol{\alpha}\Big\rangle\!\Big\langle\sqrt{\eta}U\boldsymbol{\beta}\Big| \Tr\left[ \big|\sqrt{1-\eta}\boldsymbol{\alpha}\big\rangle\!\big\langle\sqrt{1-\eta}\boldsymbol{\beta}\big|\right] \\
& = \rho_I.
\end{split}
\end{equation}

More generally, wherever loss occurs in the experiment, the overall transfer matrix $K$ on the system modes can be efficiently characterised \cite{Laing2012,Rahimi-Keshari2013}. Since path-dependent loss is usually small in experiments, and can be mitigated by interferometer design \cite{Clements2016}, the matrix $K/||K||_2\approx U$. The matrix $K$ can then be embedded into a larger unitary matrix acting on additional modes as before. We note all unitary dilations of $K$, $M_K \in U(m+p)$ where $p \geq m$, can be parameterised using the Cosine-Sine decomposition as

$$ M_K = \begin{pmatrix}
A & 0 \\
0 & X
\end{pmatrix}
\left(
\begin{array}{c|cc}
\cos(\Theta) & -\sin(\Theta) & 0 \\
\hline \sin(\Theta) & \cos(\Theta) & 0 \\
0 & 0 & \mathbb{1}_{p-m}
\end{array} 
\right)
\begin{pmatrix}
B^{\dag} & 0 \\
0 & Y
\end{pmatrix}
$$
where $K = A\cos{(\Theta)}B^{\dag}$, with $A,B \in U(m)$ and $\cos{(\Theta)} = \text{diag}(\cos{\theta_1},...,\cos{\theta_m})$ with $\theta_1\leq \theta_2 \leq .. \leq \theta_m$, is a singular value decomposition of $K$ and $X,Y \in U(p)$. In fact, all unitary dilations are related by the choice of $X$ and $Y$ \cite{Horn1991,allen2006}. Since $\mathcal{U}(Y) \ket{\mathbf{0}} = \ket{\mathbf{0}}$ and the choice of $X$ does not affect $\rho_K$ using the cyclic property of the trace as above, setting $\eta = ||K||_2^2$, we see that $\rho_K = \rho_I$. Moreover, we have shown that all unitary dilations of a transfer matrix produce the same output state and therefore any boson sampling experiment with overall path-independent losses is equivalent to introducing uniform loss channels with transmission probability $\eta$ at the input, followed by the ideal unitary evolution.

Our MIS method can readily be adapted to deal with loss at the input, by inserting an initial step for each tuple to be output, which generates a uniformly random input subset $T$.
This would be followed by the usual MIS method with permanents of $k \times k$ submatrices computed.
We described a similar adaptation in Sec.~\ref{sec:mis} to deal with scattershot \bs.
The core part of the classical sampling procedure for both the lossy and scattershot variants therefore follows precisely that of standard \bs.  
From our analysis of the required burn in period for MIS, we can see that the performance of our sampler will be similar to the standard \bs case. 
That is, it is likely that lossy and scattershot \bs is no more difficult classically than standard \bs.

%
The modified MIS sampler can then be compared with experimental performance.
Using the fixed loss regime of~\cite{Aaronson16}, we adjusted the quantum and classical runtimes and optimised QA for the case where up to 2 photons were allowed to be lost.
This resulted in the dashed lines in fig.2d) and e) of the main text, and in fig.\ \ref{fig:standard-comparison}.
We see that allowing this loss relaxes the requirements for quantum supremacy, but does not significantly change the experimental regimes required.

In current and future experiments, where $\eta \ll 1$, more photons than this are usually lost.
Again using expected loss parameter values from~\cite{Wang2016}, we find that the optimal number of photons to lose (in terms of performance enhancement relative to the classical sampler) increases with $n$.
The estimated experimental time to produce a tuple when $n=20$, $k=12$ and $m=80$ is $11.8\mu$s, where this time encompasses the expected number of repetitions required to detect exactly $k$ photons.
With the same parameter values and problem size, the estimated MIS runtime on a personal computer is $3.9$ms.
This amounts to a factor of $333$ performance enhancement for the experiment over the personal computer.

However, the MIS sampler is not the only way in which one could attempt to sample from the lossy \bs distribution: indeed, it could even be the case that enough loss could render the \bs distribution easy to sample from classically.
For example, it was shown in~\cite{bremner16b} that a class of quantum circuits whose output probability distributions are  likely to be hard to sample from classically becomes easy in the presence of noise.
We also remark that Rahimi-Keshari, Ralph and Caves~\cite{rahimikeshari16} have proven classical simulability of boson sampling under various physically motivated models of errors (e.g.\ loss, mode mismatch, and dark counts), and that Aaronson and Brod have speculated~\cite{Aaronson16} that each probability in the small-$k$ lossy \bs distribution tends to the product of the squared 2-norms of the rows of $A_S$ -- an efficiently computable quantity.

Lossy \bs also suffers from an additional difficulty when one considers performing likelihood based verification techniques.
In particular, computing the likelihood of a sample according to the indistinguishable boson hypothesis requires computing eqn.\ (\ref{temperament}) for each tuple $S$ in a sample.
The sum over all $T \in \Lambda_{n,k}$  necessitates a factor of $\binom{n}{k}$ slowdown in computing the likelihood relative to a standard (i.e.\ lossless) $k$-photon \bs experiment.
%
%
The worst case for this is when $k=n/2$ photons survive, amounting to $\sim2^n / \sqrt{\pi n/2}$ matrix permanent computations per likelihood calculation.
To avoid this exponential slowdown, one would be forced to use a nonstandard verification technique~\cite{Aaronson2013, Walschaers2016}, or devise a more efficient means of computing (\ref{temperament}) than na\"ive evaluation of the sum.

\section{Discussion}

We finish by discussing the limitations of our techniques, and prospects for future experimental and classical improvements.

\subsection{Limitations}

Our sampling methods have some limitations that we now set out.  
Our rejection sampling method which is applicable for $n$ up to approximately $12$, is guaranteed to give independent samples.  
Our approach to improving the efficiency of the method requires us first to estimate the maximum probability $\mu$ of the distribution. 
If this estimate is very poor, that is if there is a large proportion of the probability mass greater than $\mu$, then there is a risk we will not sample from the \bs distribution correctly.   
From our computational experiments, we can see that this appears to be very unlikely to occur at least in the range we are able to test.  

For our MIS method,  we have to be concerned both with how long to wait until convergence and the related question of potential dependence between samples. 
It is unfortunately not possible to prove that the samples we take are entirely independent of each other or indeed of the initial state.  
Although this question of dependence in \bs is not one that has arisen previously, it is desirable
for samples to be taken independently.   

We have mitigated this problem in two ways. 
The first is by our choice of MIS as a sampling method, which means that samples from our proposal distribution are taken independently. The second is by our use of a thinning procedure. 
In fig.\ \ref{fig:autocorrelation} we see that this thinning procedure successfully reduces the auto-correlation
of the computed probabilities of the sampled submatrices to near zero.
In addition to this, in fig.\ \ref{fig:main-burnin-thin} we see that extending the thinning procedure beyond a certain point causes no improvement in the results of a likelihood ratio test as outlined in~\cite{Bentivegna2014}. 
However, it is still possible that there remains dependence on some other detailed property of the sampled submatrices which is not exposed by looking at the computed probabilities or likelihood ratios.

\subsection{Prospects for experimental \bs}

The fact that producing \bs machines requires low loss rates has been known to be a hurdle since the conception of the problem~\cite{Aaronson2011}.
However, the extent of this hurdle has not been well understood, not least due to the lack of an optimised classical \bs algorithm.

Our results suggest that with loss parameters and photon numbers expected in the near future, no straightforward or significant quantum supremacy is achievable via \bs.
This is \emph{not} to say that it is impossible to build a \bs machine which demonstrates quantum supremacy.
However, it is absolutely necessary to further decrease loss rates, or develop efficient fault-tolerance techniques targeted at linear optics in order to achieve this.
A machine performing \bs with tens of photons in thousands of modes is likely required if we are to witness an experiment significantly outperforming a classical simulator. 
We note the engineering feat required to build an interferometer of this size alone may prohibit its realisation in the near future.

One approach to mitigating loss in an experiment would be to reduce circuit depth.
In fig. 1e) of the main text and fig.\ \ref{fig:standard-comparison} we have assumed a circuit depth of $d=m$ to infer $\eta$ for the specific experimental points plotted (solid and open white points).
Using more exotic structures such as the 3D platform used in~\cite{Crespi2016}, it may be possible to reduce this scaling, and hence increase $\eta$.

In addition to the problems caused by photon loss, it is also necessary to build reconfigurable linear optical circuits capable of performing high precision transformations on modes~\cite{leverrier15,kalai14,Arkhipov2015}, to generate highly indistinguishable photons~\cite{Tichy2015,Shchesnovich14} and to limit detector noise and higher order photon number terms in the input state~\cite{rahimikeshari16} in order to perform a classically intractable task.
We note that although there exist methods to aid in achieving these things -- such as the spectral filtering of single photons or adding additional components to perform higher fidelity beamsplitter transformations~\cite{Miller2015,Wilkes2016} -- they typically result in an increased amount of photon loss.

There have also been proposals for performing boson sampling using other hardware platforms, such as trapped ions \cite{Shen2014} and superconducting qubits \cite{Peropadre2016}. Although currently experimentally untested, it is possible these approaches could provide better scalability. However, coherence and gate operation times are likely to limit the rates and sizes of implementations possible in these architectures, e.g. in ref. \cite{Peropadre2016} current estimates suggest up to 20 photons.

\subsection{Prospects for classical \bs}

Our current approach to simulating a \bs experiment assumes a Fock state input of the form given in eqn.\ (\ref{input_state}). 
Another proposal for \bs type experiments is based on Gaussian input states~\cite{lund14,Hamilton2016}.
In this case, Hamilton et al.~\cite{Hamilton2016} showed that the probabilities in the output distribution are related to the hafnian~\cite{minc78} of a submatrix, rather than the permanent.
It has been shown by Bj\"orklund that hafnians can be computed using an algorithm with runtime matching Ryser's algorithm for the permanent, up to polynomial factors~\cite{bjorklund12}.
We are hopeful that the techniques we have presented in this paper can be carried over to Gaussian \bs, potentially allowing for the classical simulation of experiments designed to generate molecular vibronic spectra~\cite{Huh15}.

In terms of performance, it is possible that improvements can be made in a number of ways.
Firstly, it is likely that MIS with the distinguishable particle proposal distribution is not optimal for classical \bs.
Although computing matrix permanents will remain the bottleneck, requiring fewer of these computations per output tuple could result in up to a $\sim 100$ times speed-up.
Secondly, the implementation could be scaled up to run on high performance computing hardware.
This would require careful parallelisation of both MIS and Ryser's (or Balasubramanian--Bax/Franklin--Glynn) algorithm.
We remark that parallelisation of the most computationally expensive parts of MIS is relatively trivial.
As the proposal distribution is independent of the current state, we can distribute the permanent computations across multiple processors.
In addition to this, it may be possible to build a more realistic model of the experiment (including, for example, partial distinguishability of photons) to allow for some approximations of matrix permanents to be made, or for permanents of smaller submatrices to be computed with some probability.


\section{Conclusion}

We have shown that \bs can be simulated classically for a range of parameters that was previously considered likely to be computationally intractable~\cite{Aaronson2011,preskill12,Bentivegna15,Latmiral2016}.
Our results pose a challenge for both experimentalists and theorists.
The challenge for experimentalists is to build low-loss, large-scale linear optical circuits going substantially beyond current technology, while the challenge for theorists is either to develop efficient linear-optical loss-tolerance schemes, or to find some way to distinguish our classical \bs simulators from the real \bs distribution.

Note that one way of implementing a large-scale loss-free \bs experiment is simply to simulate \bs on a fault-tolerant universal quantum computer, perhaps based on some other technology than optics.
However, in the near term this is unlikely to be a more efficient means of demonstrating quantum supremacy than an approach targeted directly at the underlying hardware~\cite{boixo16,bremner16,lund17,BermejoVega2017}.

We close by highlighting another interpretation of our results: as providing an additional tool for classical verification of future \bs experiments.  Due to the computational difficulty of simulating the \bs distribution, existing verification procedures have not been designed to take advantage of statistical information about the true boson sampling distribution. Indeed for $n > 7$ this information has simply not been available within a realistic time frame.  Our new efficient computational sampling procedure allows us for the first time to test directly whether an  experiment is sampling from the correct distribution.  Setting the null hypothesis to be that we are sampling from the \bs distribution, we can sample directly from the true distribution using our MIS procedure and then perform standard statistical hypothesis tests on the results of any \bs experiment.
MIS can also be used to certify lossy versions of boson sampling, where previous methods based on computing likelihoods are inefficient, as discussed in Section \ref{sec:loss}.
This will potentially give a new level of confidence that was not previously available in the accuracy of any experimental design and implementation.

\end{document}